\documentclass[reprint,onecolumn,prd,amsmath,amssymb,aps,superscriptaddress]{revtex4-1}
\usepackage{graphicx}
\usepackage{dcolumn}
\usepackage{bm}
\usepackage{lipsum}  
\usepackage{hyperref}
\usepackage{verbatim}
\usepackage{mathrsfs}

\newcommand{\lcdm}{\Lambda\text{CDM}}
\newcommand{\dm}{\mathrm{DM}}
\newcommand{\ev}{\text{eV}}
\newcommand{\kev}{\text{keV}}
\newcommand{\mev}{\text{MeV}}
\newcommand{\gev}{\text{GeV}}
\newcommand{\tev}{\text{TeV}}
\newcommand{\beq}{\begin{eqnarray}}
\newcommand{\eeq}{\end{eqnarray}}
\newcommand{\sv}{\langle \sigma v \rangle}
\newcommand{\mpl}{M_{\mathrm{Pl}}}
\newcommand{\msol}{{\rm M}_{\odot}}
\newcommand{\mpbh}{M_{\mathrm{PBH}}}
\newcommand{\lrf}[2]{\left(\frac{#1}{#2}\right)}

\def\corrauthornb{\footnote{Corresponding author: \href{mailto:nbozorgnia@ualberta.ca}{nbozorgnia@ualberta.ca}}}
\def\corrauthordm{\footnote{Corresponding author: \href{mailto:dmorri@triumf.ca}{dmorri@triumf.ca}}}

\begin{document}
\def\lnp{Lec.\ Notes in Physics}
\def\cpc{Comp.\ Phys.\ Comm.}
\def\jpg{J. Phys. G}
\def\ijmpa{Int.\ J.\ Mod.\ Phys.\ A}
\def\epjc{Eur.\ Phys.\ J.\ C}
\def\nima{Nuc.\ Inst.\ Methods A}
\def\nimb{Nuc.\ Inst.\ Methods B}
\def\njp{New J.\ Phys.}
\def\rmp{Rev.\ Mod.\ Phys.}
\def\app{Astropart.\ Phys.}
\def\aj{AJ}%
\def\actaa{Acta Astron.}%
\def\araa{ARA\&A}%
\def\arnps{Ann.~Rev.~Nucl.~\& Part.~Sci.}%
\def\apj{ApJ}%
\def\apjl{ApJ}%
\def\apjs{ApJS}%
\def\ao{Appl.\ Opt.}%
\def\apss{Ap\&SS}%
\def\aap{A\&A}%
\def\aapr{A\&A~Rev.}%
\def\aaps{A\&AS}%
\def\azh{AZh}%
\def\pos{PoS}%
\def\baas{BAAS}%
\def\bac{Bull.\ Astr.\ Inst.\ Czechosl.}%
\def\caa{Chinese Astron.\ Astrophys.}%
\def\cjaa{Chinese J.\ Astron.\ Astrophys.}%
\def\icarus{Icarus}%
\def\jhep{JHEP}%
\def\jcap{JCAP}%
\def\jinst{JINST}%
\def\jpsj{J.\ Phys.\ Soc.\ Japan}%
\def\jrasc{JRASC}%
\def\canjphys{Can.~J.~Phys.}
\def\apphys{Astropart.~Phys.}
\def\mnras{MNRAS}%
\def\memras{MmRAS}%
\def\na{New A}%
\def\nar{New A Rev.}%
\def\pasa{PASA}%
\def\pra{Phys.\ Rev.\ A}%
\def\prb{Phys.\ Rev.\ B}%
\def\prc{Phys.\ Rev.\ C}%
\def\prd{Phys.\ Rev.\ D}%
\def\pre{Phys.\ Rev.\ E}%
\def\prx{Phys.\ Rev.\ X}%
\def\prl{Phys.\ Rev.\ Lett.}%
\def\pasp{PASP}%
\def\pasj{PASJ}%
\def\qjras{QJRAS}%
\def\rmxaa{Rev. Mexicana Astron. Astrofis.}%
\def\skytel{S\&T}%
\def\solphys{Sol.\ Phys.}%
\def\sovast{Soviet~Ast.}%
\def\ssr{Space~Sci.\ Rev.}%
\def\zap{ZAp}%
\def\nat{Nature}%
\def\science{Science}%
\def\sci{\science}%
\def\iaucirc{IAU~Circ.}%
\def\aplett{Astrophys.\ Lett.}%
\def\apspr{Astrophys.\ Space~Phys.\ Res.}%
\def\bain{Bull.\ Astron.\ Inst.\ Netherlands}%
\def\fcp{Fund.\ Cosmic~Phys.}%
\def\gca{Geochim.\ Cosmochim.\ Acta}%
\def\grl{Geophys.\ Res.\ Lett.}%
\def\jcp{J.\ Chem.\ Phys.}%
\def\jgr{J.\ Geophys.\ Res.}%
\def\jqsrt{J.\ Quant.\ Spec.\ Radiat.\ Transf.}%
\def\memsai{Mem.\ Soc.\ Astron.\ Italiana}%
\def\nphysa{Nucl.\ Phys.\ A}%
\def\nphysb{Nucl.\ Phys.\ B}%
\def\physrep{Phys.\ Rep.}%
\def\physscr{Phys.\ Scr}%
\def\planss{Planet.\ Space~Sci.}%
\def\procspie{Proc.\ SPIE}%
\def\repprogphys{Rep.\ Prog.\ Phys.}%
\def\jpcrd{J. Phys. Chem. Ref. Data}%
\def\jphysb{J. Phys. B}%
\def\jphysd{J. Phys. D}%
\def\jphysconfseries{J. Phys. Conf. Series}%
\def\physrev{\pr}
\def\pr{Phys. Rev.}%
\def\josa{J. Opt. Soc. Amer. (1917-1983)}%
\def\josab{J. Opt. Soc. Amer. B}%
\def\pla{Phys. Lett. A}%
\def\plb{Phys. Lett. B}%
\def\os{Opt. Spectrosc. (Russ.)}%
\def\jas{J. Appl. Spectrosc.}%
\def\annp{Ann. Phys.}%
\def\sa{Spectrochim. Acta}%
\def\prsoca{Proc. R. Soc. London Ser. A}%
\def\zphysa{Z. Phys. A}%
\def\zphysb{Z. Phys. B}%
\def\zphysc{Z. Phys. C}%
\def\zphysd{Z. Phys. D}%
\def\zphyse{Z. Phys. E}%
\def\zphys{Z. Phys.}%
\def\adndt{Atom. Data Nuc. Data Tables}%
\def\jmolspec{J. Mol. Spectrosc.}%
\def\aphysb{Appl. Phys. B}%
\def\nim{Nuc. Inst. Meth.}%
\def\jphysique{J. Phys. (Paris)}%
\def\epjp{Eur.~Phys.~J.~Plus}%
\def\epjc{Eur.~Phys.~J.~C}%
\def\epl{Europhys.~Lett}%
\def\njp{New J.~Phys.}
\let\astap=\aap
\let\apjlett=\apjl
\let\apjsupp=\apjs
\let\applopt=\ao

\title{Dark Matter Candidates and Searches}

\author{Nassim Bozorgnia \corrauthornb{}}
\affiliation{Department of Physics, University of Alberta, Edmonton, AB T6G 2E1, Canada}
\affiliation{Theoretical Physics Institute, University of Alberta, Edmonton, AB T6G 2E1, Canada}

\author{Joseph Bramante}
\affiliation{Arthur B. McDonald Canadian Astroparticle Physics Research Institute, Queen's University, Kingston, ON K7L 3N6, Canada}
\affiliation{Department  of  Physics, Engineering Physics and Astronomy, Queen's  University,  Kingston,  ON  K7L  3N6, Canada}
\affiliation{Perimeter  Institute  for  Theoretical  Physics,  Waterloo,  ON  N2J  2W9, Canada}
\author{James M. Cline}
\affiliation{McGill  University,  Department  of  Physics, Montr\'eal, QC H3A 2T8,  Canada}
\author{David Curtin}
\affiliation{Department of Physics, University of Toronto, Toronto, ON M5S 1A7, Canada}
\author{David McKeen}
\affiliation{TRIUMF, 4004 Wesbrook Mall, Vancouver, BC V6T 2A3, Canada}
\affiliation{Department of Physics and Astronomy, University of Victoria, Victoria, BC V8P 5C2, Canada}
\author{David E. Morrissey 
\corrauthordm{}
}
\affiliation{TRIUMF, 4004 Wesbrook Mall, Vancouver, BC V6T 2A3, Canada}
\affiliation{Department of Physics and Astronomy, University of Victoria, Victoria, BC V8P 5C2, Canada}
\author{Adam Ritz}
\affiliation{Department of Physics and Astronomy, University of Victoria, Victoria, BC V8P 5C2, Canada}
\author{Simon Viel}
\affiliation{Department  of  Physics,  Carleton  University,  Ottawa,  ON  K1S  5B6, Canada}
\affiliation{Arthur B. McDonald Canadian Astroparticle Physics Research Institute, Queen's University, Kingston, ON K7L 3N6, Canada}
\author{Aaron C. Vincent}
\affiliation{Arthur B. McDonald Canadian Astroparticle Physics Research Institute, Queen's University, Kingston, ON K7L 3N6, Canada}
\affiliation{Department  of  Physics, Engineering Physics and Astronomy, Queen's  University,  Kingston,  ON  K7L  3N6, Canada}
\affiliation{Perimeter  Institute  for  Theoretical  Physics,  Waterloo,  ON  N2J  2W9, Canada}
\author{Yue Zhang}
\affiliation{Department  of  Physics,  Carleton  University,  Ottawa,  ON  K1S  5B6, Canada}
\affiliation{Arthur B. McDonald Canadian Astroparticle Physics Research Institute, Queen's University, Kingston, ON K7L 3N6, Canada}
%
%
\begin{abstract}
Astrophysical observations suggest that most of the matter in the cosmos consists of a new form that has not been observed on Earth. The nature and origin of this mysterious dark matter are among the most pressing questions in fundamental science. In this review we summarize the current state of dark matter research from two perspectives. First, we provide an overview of the leading theoretical proposals for dark matter. And second, we describe how these proposals have driven a broad and diverse global search program for dark matter involving direct laboratory searches and astrophysical observations. This review is based on a Green Paper on dark matter prepared as part of the 2020 Astroparticle Community Planning initiative undertaken by the Canadian Subatomic Physics community but has been significantly updated to reflect recent advances. 
\end{abstract}
\maketitle
\section{Introduction
\label{sec:intro}}

Over the last century, the fields of particle physics and cosmology have developed enormously in conjunction with each other, revolutionizing our understanding of the universe. Laboratory and collider data have guided the development of the Standard Model~(SM) of particle physics that describes all known fundamental particles and the forces between them.  Similarly, astronomical observations have led to a standard $\Lambda$CDM model of cosmology consisting of dark energy~($\Lambda$), dark matter~(DM), and the particles of the SM, together with initial conditions consistent with inflation.

The greatest puzzle at the interface of these two fields is the nature of DM.  None of the particles predicted by the SM has the right properties to make up DM. Furthermore, many proposals for the identity of DM predict observable signals in current or future experiments, suggesting that it may be within tantalizingly close reach.

Evidence for DM comes from a large and diverse set of astrophysics observations. Measurements of the cosmic microwave background~(CMB) together with other cosmological data provide very strong support for the $\lcdm$ model in which DM plays an essential role, making up about 25\% of the total energy density of the universe today and over 80\% of the matter in it~\cite{Planck:2018vyg}. At shorter distances, additional non-luminous matter is needed to account for the motions of galaxies within galaxy clusters~\cite{Zwicky:1933gu,Zwicky:1937zza} as well as the rotational velocities of stars within individual galaxies~\cite{Rubin:1980zd,Persic:1995ru}. Determinations of matter distributions from gravitational lensing also point to the presence of DM~\cite{Kaiser:1992ps,Bartelmann:1999yn,Hoekstra:2008db,Kilbinger:2014cea}.  In addition, the observed light element abundances relative to hydrogen strongly support models where baryonic matter only constitutes a small fraction of the present energy density of the universe~\cite{Peebles:1991}. It is remarkable that the DM hypothesis of a new, massive, non-luminous particle can account for all these data.

Despite the strong evidence for DM, very little is known about its identity. Observational support for DM relies on the gravitational influence it has on regular matter or radiation. Since gravity acts universally, the observed effects of DM provide very little information about its detailed nature. The best we have been able to do is to constrain what DM is not. For example, DM must be \emph{dark}, meaning that it does not interact too readily with ordinary matter, and it needs to be \emph{matter}, implying that it dilutes and gravitates like non-relativistic matter in the early universe. Observations of certain astrophysical structures imply further that the DM must not interact too strongly with itself, and have a mass greater than about $m_{\dm} \gtrsim 10^{-19}\,\ev$ if it is a boson~\cite{Dalal:2022rmp, Turner:1983he,Hui:2016ltb} and $m_{\dm} \gtrsim 1\,\kev$ if it is a fermion~\cite{Tremaine:1979we,Boyarsky:2008ju}.

Theoretical studies have identified a wide range of candidates for DM as well as mechanisms for their creation in the early universe and potential signals they could induce in experiments~\cite{Jungman:1995df,Bertone:2004pz,Feng:2010gw,Bergstrom:2012fi,Lisanti:2016jxe,Arcadi:2017kky,Battaglieri:2017aum,Roszkowski:2017nbc,Lin:2019uvt}. A candidate that has received particular attention is the Weakly Interacting Massive Particle~(WIMP), consisting of a new elementary particle that connects to the SM exclusively through the weak force~\cite{Hut:1977zn,Lee:1977ua,Jungman:1995df,Bertone:2004pz}. WIMPs arise in many extensions of the SM that address the electroweak hierarchy puzzle, and they can very naturally develop the observed DM relic abundance through the mechanism of thermal freeze-out.  If DM consists of WIMPs, they could potentially be detected through their scattering with ordinary matter, their direct creation at high-energy colliders, or the SM particles they would produce by annihilating in our galaxy and beyond. However, WIMPs are only one example of what DM might be, with the many other possibilities including axions, primordial black holes, and supermassive non-thermal relics, each of which can produce very different signals~\cite{Feng:2010gw}.

Solving the mystery of DM will therefore require new laboratory measurements and astronomical observations that are sensitive to its detailed non-gravitational properties.  Indeed, dedicated searches for DM have been ongoing for the past 30 years. These searches can be classified broadly into three categories. First, direct searches aim to detect relic DM species through their effects on ultra-low background laboratory detectors. Second, indirect DM searches look for the SM particles produced by DM annihilation in astrophysical environments with high DM density. And third, collider searches aim to create DM (or related particles) and study their signals in high-energy collisions. While these approaches are very different from one another, they are also complementary and have the combined potential to provide a detailed understanding of the nature of DM.

In this review, we attempt to give an overview of the current state of research on DM. Our approach is to focus on two complimentary perspectives. In the first, presented in Section~\ref{sec:candidates} we discuss the leading theoretical proposals for what DM might be and how it was created. Our second approach in Section~\ref{sec:approaches} is to summarize the broad range of experimental searches for DM and connect them to bounds on DM theories. Finally, Section~\ref{sec:conclusions} is reserved for our conclusions.
This review is based on a Green Paper on DM prepared as part of the 2020 Astroparticle Community Planning initiative undertaken by the Canadian Subatomic Physics community.
We refer the reader to other articles in this journal edition for detailed accounts of dark matter search efforts by members of the community.

\section{Dark Matter Candidates\label{sec:candidates}}

Candidates for dark matter range over many order magnitude in mass and how strongly they interact with ordinary matter. We present here an overview of theoretically motivated DM candidates, loosely organized by how their density was created in the early universe.

\subsection{WIMPs and WIMP-Like Dark Matter}

Of the many DM candidates proposed so far, weakly interacting massive particles~(WIMPs) have been studied and searched for in the greatest depth. The defining feature of WIMPs is that they interact with the SM through the weak force. This allows WIMPs to be created by and come to thermodynamic equilibrium with SM particles in the very early universe. These interactions also lead to WIMP signals, such as scattering with ordinary particles in the laboratory, visible annihilation in the cosmos, and direct production at colliders. Many of these attractive features of WIMPs also carry over to WIMP-like DM that interacts with the SM in other ways.

\subsubsection{The WIMP Paradigm}

Motivation for the WIMP paradigm comes from two main features.  First, WIMPs can obtain the observed DM relic abundance in a very simple and natural way through the process of thermal freeze-out in the early universe~\cite{Hut:1977zn,Lee:1977ua}. This occurs for WIMP masses near the weak scale, as characterized by the masses of the $W^{\pm}$ and $Z^0$ vector bosons. And second, WIMPs with weak-scale masses arise in many proposed extensions of the SM that address the naturalness of the electroweak scale, such as supersymmetry~(SUSY)~\cite{Jungman:1995df,Bertone:2004pz,Feng:2010gw}. Together, these two features of WIMPs are sometimes called the \emph{WIMP Miracle}.

Thermal freeze-out is an attractive mechanism for the formation of the relic DM density because it is simple and generic~\cite{Kolb:1990vq}. If the early universe was hot enough, with temperatures approaching at least the weak scale, the weak interactions of WIMPs would have allowed them to be created, scattered, and destroyed by collisions involving SM particles, eventually reaching a thermodynamic equilibrium with the SM. As the universe expanded and cooled, these processes would have slowed relative to the expansion rate of the universe. Eventually, they would have become too slow to maintain the equilibrium density of WIMPs, leaving the number of WIMPs per comoving volume nearly constant. After this point, called freeze-out, the remaining WIMPs would have been diluted further by the cosmological expansion to produce the DM relic abundance seen today.

For a typical WIMP candidate, $\chi$, freeze-out is found to occur when the cosmological plasma temperature is roughly $T \sim m_{\chi}/25$, where $m_{\chi}$ is the WIMP mass~\cite{Kolb:1990vq}. After accounting for the subsequent cosmological dilution, the current WIMP relic density due to thermal freeze-out is given to a good approximation by~\cite{Kolb:1990vq,Steigman:2012nb}
\beq
\Omega_{\chi}h^2  \ \equiv \ \frac{\rho_{\chi}}{\rho_c/h^2}  \ \simeq \
(0.119)\,\frac{3\times 10^{-26}\,\text{cm}^3/\text{s}}{\sv} \ ,
\label{eq:omfo}
\eeq
where $\sv$ is the thermally-averaged annihilation cross section, $\rho_c = 3H_0^2/8\pi G$ is the critical density of the universe, and $h = H_0/100\,\text{km}\,\text{s}^{-1}\text{Mpc}^{-1} \simeq 0.7$ is the normalized Hubble rate. For comparison, cosmological data finds a DM relic density of $\Omega_{\chi}h^2 = 0.1193\pm 0.0009$~\cite{Planck:2018vyg}.

A notable feature of Eq.~\eqref{eq:omfo} is that the relic density depends on the properties of the WIMP nearly exclusively through $\sv$. This quantity corresponds to the cross section for $\chi+\chi \to \text{SM}+\text{SM}$, multiplied by the relative speed $v$ of the initial states and averaged over the thermal distributions of their momenta~\cite{Gondolo:1990dk,Edsjo:1997bg}.  In many cases, $\sv$ takes the parametric form
\beq
\sv \ \sim \ \frac{g_{\rm eff}^4\,m_{\chi}^2}{\max\{m_{\chi}^4,m_W^4\}}\,f(v) \ ,
\label{eq:sv}
\eeq
where $g_{\rm eff}$ is the effective dimensionless coupling strength of $\chi$ to the SM, $m_{W}$ is the mass of the W vector boson, and $f(v)$ is a function of the DM relative velocity. For perturbative annihilation dominated by the angular momentum partial wave $\ell$, one has $f(v) \sim v^{2\ell}$, so that $\ell = 0$ corresponds to $s$-wave, $\ell = 1$ to $p$-wave, and so on. Applying Eq.~\eqref{eq:sv} to couplings in the range $g_{\rm eff} \sim 10^{-3}$ -- $10^{-1}$ expected for a WIMP, the result of Eq.~\eqref{eq:omfo} leads to a mass range of $m_{\chi} \sim 2\,\gev$ -- $10\,\tev$.

Candidates for WIMP DM arise in many extensions of the SM that stabilize the weak scale against quantum corrections. The archetypal example is SUSY with soft breaking near the weak scale~\cite{Martin:1997ns}. In SUSY, every particle of the SM is predicted to have a superpartner differing in spin by half a unit. If the theory has an additional symmetry like $R$-parity, the lightest superpartner particle~(LSP) can be stable. If the LSP is also neutral under electromagnetism and the strong force, it is a candidate for WIMP DM~\cite{Goldberg:1983nd,Ellis:1983ew}. In the minimal supersymmetric extension of the SM~(MSSM), this candidate is the lightest neutralino, a mixture of the superpartners of the photon, the $Z^0$ vector boson, and the neutral Higgs bosons. Neutralinos as WIMP DM have been studied and searched for extensively, and they remain a promising possibility~\cite{Jungman:1995df,Bertone:2004pz,Feng:2010gw,Bramante:2015una,Roszkowski:2017nbc,Leane:2018kjk,Bottaro:2021snn}.

WIMPs also emerge in many other theories of new physics motivated by electroweak naturalness, including Universal Extra Dimensions~\cite{Servant:2002aq,Cheng:2002ej}, Little/Composite Higgs with Parity~\cite{Cheng:2003ju,Hubisz:2004ft}, Twin Higgs~\cite{Dolle:2007ce,Garcia:2015loa}, and more.  Note, however, that WIMPs need not be connected to the electroweak hierarchy problem. Some more general examples include the Inert Higgs Doublet~\cite{Barbieri:2006dq,Cao:2007rm} and other new electroweak multiplets~\cite{Cirelli:2005uq}.

The direct connection of WIMPs to the SM that allows them to develop the correct relic density through thermal freeze-out also makes them potentially observable in experiments. WIMP DM can scatter with nuclei and electrons in laboratory detectors through the exchange of weak vector or Higgs bosons, motivating direct detection searches for DM~\cite{Goodman:1984dc,Drukier:1986tm}. In regions of our galaxy and beyond where DM is particularly dense, WIMPs can find each other, annihilate, and produce SM particles such as gamma-rays, neutrinos, and cosmic rays, motivating indirect searches for WIMPs through these products~\cite{Gunn:1978gr,Zeldovich:1980st}. And WIMPs can be created in high-energy colliders such as the Large Hadron Collider~(LHC) leading to distinctive signals of missing energy~\cite{Haber:1984rc}. These features have motivated a broad range of searches for WIMP DM that will be discussed in detail in Section~\ref{sec:approaches} below.

\subsubsection{Generalizing WIMPs}

The very attractive freeze-out mechanism for WIMP DM can also operate for other DM candidates or it can be modified by a more complicated cosmological evolution. This motivates the search for WIMP-like signals over a much larger range of masses than the $m_{\chi} \sim 2\,\gev$--$10\,\tev$ expected for minimal WIMPs.

Thermal freeze-out does not rely specifically on the weak force. It can also proceed if the DM connects to the SM through a new force carrier, such as an exotic scalar or vector boson~\cite{Boehm:2003hm,Pospelov:2007mp}. Generalizing the results of Eqs.~(\ref{eq:omfo},\,\ref{eq:sv}) to a mediator $\phi$ with mass $m_{\phi}$, SM coupling $g_\text{SM}$, and DM coupling $g_{\chi}$, the resulting relic density scales approximately as $\Omega_{\chi}h^2 \propto \max\{m_{\chi}^4,m_{\phi}^4\}/g_\text{SM}^2g_{\chi}^2m_{\chi}^2$~\cite{Feng:2008ya}. Thus, more general effective couplings allow for a much wider range of WIMP-like DM masses from thermal freeze-out. This range can extend from $m_{\chi}\sim 5\,\kev$--$100\,\tev$, where the lower bound comes from the DM being sufficiently non-relativistic~\cite{Viel:2013apy,Irsic:2017ixq,DES:2020fxi} and the upper limit from unitarity~\cite{Griest:1989wd,Smirnov:2019ngs}. In practice, however, the lower bound on the mass of thermal WIMP-like DM tends to be closer to $m_{\chi}\gtrsim 10\,\mev$, since the annihilation of lighter candidates injects significant energy into electromagnetic species or neutrinos after neutrino decoupling, thereby changing the effective temperature of neutrinos relative to photons~\cite{Boehm:2012gr,Boehm:2013jpa,Nollett:2013pwa,Nollett:2014lwa}. Just like for WIMPs, WIMP-like candidates can be observable in direct, indirect, and collider searches through their interactions with the SM. 

The relic density of WIMP or WIMP-like DM can also be modified by non-minimal cosmological evolution during or after freeze-out. A simple example is the decay of a very massive particle after thermal DM freeze-out. Such decays can inject entropy into the cosmological plasma which has the effect of diluting the abundance of DM and allowing larger DM masses above $m_{\chi} > 100\,\tev$~\cite{Gelmini:2006pw,Gelmini:2006pq}. Such decays can also increase the final DM density relative to minimal freeze-out if the DM or any of its precursors are created as products of the decay~\cite{Moroi:1999zb,Arcadi:2011ev}.

Together, these considerations motivate generalizing searches for WIMP DM to a broader range of masses and couplings. While WIMPs are a target that should certainly be looked for, it is important to search broadly as well.

\subsection{Non-WIMP Dark Matter}

DM can also be created and evolve in ways that are very different from the thermal freeze-out processes typically expected for WIMPs.  In this section we discuss a necessarily incomplete list of non-WIMP DM candidates with significantly different creation mechanisms. Many of them have been investigated in detail for the new ways in which they address the relic density, or for the novel signals they produce that broaden the scope of DM searches.

\subsubsection{Axions and Other Ultralight Bosons}

Ultralight bosonic fields with masses $m_a\ll \kev$ can generate the observed relic density through a non-thermal mechanism that is fundamentally different from thermal WIMPs~\cite{Turner:1983he,Antypas:2022asj,Adams:2022pbo}. Such a field is generically expected to be be displaced from the minimum of its potential in the early universe. In conjunction with inflation, this displacement can be nearly uniform over cosmologically large distances. As the universe cools, the field will begin to oscillate coherently around the minimum of its potential provided it is thermally decoupled from the cosmological plasma (and itself). The energy of the oscillations gravitates just like a relic abundance of non-relativistic massive particles and can play the role of DM. This class of DM candidates only works for bosonic fields due to the large occupation numbers required, and is sometimes referred to as wave-like DM. 

The most popular realization of ultralight bosonic DM is the (QCD) axion. This new scalar particle was first proposed to address the strong $CP$ problem of quantum chromodynamics~(QCD)~\cite{Peccei:1977hh, Peccei:1977ur, Weinberg:1977ma, Wilczek:1977pj}, a major naturalness puzzle of the SM. In the SM Lagrangian, the symmetries of the theory allow a so-called {theta}-term of the form 
\beq
\mathscr{L}_{\rm SM}  \supset  - \Theta\,\frac{\alpha_s}{8\pi}\,G_{\mu\nu}\widetilde G^{\mu\nu} \ ,
\label{eq:theta}
\eeq
where $G_{\mu\nu}$ is the gluon field strength tensor and $\widetilde G$ is its dual. This term violates both parity~($P$) and time reversal~($T$) discrete symmetries, and generates a contribution to the electric dipole moment of the neutron proportional to $\Theta$.  To be consistent with existing bounds, this forces $|\Theta| \lesssim 10^{-10}$ in the absence of extreme fine tuning~\cite{Shindler:2015aqa}.  Since the natural range is $\Theta \in [-\pi,\pi)$, the strong $CP$ problem is the puzzle of why the apparent value of $\Theta$ is so small. To address this puzzle, Peccei and Quinn proposed a new $U(1)$ global symmetry that is anomalous with respect to strong interaction and spontaneously broken at a high energy scale $f_a$~\cite{Peccei:1977hh, Peccei:1977ur}. At lower energies below $f_a$, the theory has a light  pseudo-Nambu-Goldstone field $a(x)$, the axion, that develops a coupling to the gluon of the same form as Eq.~\eqref{eq:theta} but with $\Theta \to \theta(x) = a(x)/f_a$. This coupling generates a potential for the axion field through QCD confinement such that $\theta(x) \to 0$ at the minimum, thus providing an elegant solution to the strong $CP$ problem. The axion potential also leads to an axion mass, given by~\cite{Gorghetto:2018ocs}
\beq
m_a \ \simeq \ 6.0\times 10^{-6}\,\mathrm{eV}\,\left(\frac{10^{12}\,\gev}{f_a}\right) \ .
\label{axionrelicmass}
\eeq
Note that the mass is inversely proportional to the symmetry breaking energy $f_a$.

Axions can make up the DM through a version of the ultralight boson mechanism discussed above~\cite{Preskill:1982cy, Dine:1982ah, Abbott:1982af}.  Interestingly, the realization of the mechanism is closely connected to the axion solution to the strong $CP$ problem. Prior to the QCD phase transition in the early universe, the axion has a flat potential and can take any field value within its allowed range $\theta = a/f_a \in [-\pi,\pi)$. During QCD phase transition at temperature $T \sim 150\,\mev$, a non-trivial potential for the axion is generated. Since the potential was previously flat, the axion field generically begins to evolve with a displacement from the minimum of the new potential. This vacuum misalignment leads to axion field oscillations whose energy density acts as DM. The axion relic density produced this way is~\cite{Marsh:2015xka,Ballesteros:2016xej}
\beq
\Omega_ah^2 = 0.12\,\left(\frac{f_a}{9\times 10^{11}\,\gev}\right)^{1.165}\,\left[\bar{\theta}_i^2 + (\delta\theta_i)^2 \right]\,F \ ,
\label{eq:axrd}
\eeq
where $\bar{\theta}_i = a(t_i)/f_a$ is the mean initial misalignment angle (averaged over the Hubble volume), $\delta\theta_i$ is its root-mean-square fluctuation, and $F$ is a factor that accounts for anharmonicities in the potential but is on the order of unity for $f_a \lesssim \mpl/10$.

There is a close interplay between Peccei-Quinn~(PQ) $U(1)$ symmetry breaking and cosmic inflation~\cite{Preskill:1982cy,Dine:1982ah,Abbott:1982af,Turner:1990uz,Marsh:2015xka,Ballesteros:2016xej,OHare:2024nmr}. If the PQ symmetry was broken after inflation completes, the universe today is comprised of many previously disconnected patches, each with an independent initial field displacement $\theta_i$. The axion relic density is then proportional to the average of these angles squared, corresponding to the fluctuation term in Eq.~\eqref{eq:axrd} with $(\delta\theta_i)^2 \sim 1$. In addition to axions, the phase transition can also produce topological defects such as domain walls and cosmic strings which can be a further source of axions and lead to interesting (and potentially dangerous) effects themselves~\cite{Sikivie:1982qv,Vilenkin:1982ks}.

Alternatively, the PQ symmetry could have been broken before inflation. The presence of a massless axion field during inflation leads to isocurvature perturbations~\cite{Seckel:1985tj,Turner:1990uz}. To satisfy CMB constraints on such perturbations, the scale of inflation cannot be too high. In this scenario, the observable universe today exists in a patch with a single value of $\theta_i = \bar{\theta}_i \gg |\delta\theta_i|$. This allows for the possibility of obtaining the correct relic density with a smaller $\theta_i$ and a correspondingly larger $f_a$. Assuming the vacuum misalignment mechanism is the dominant source of the axion DM relic density, the allowed range of the QCD axion mass lies between $m_a = 10^{-12}\,$eV -- $10^{-4}$\,eV, with the corresponding PQ symmetry breaking scale between $f_a \sim 10^{12}\,\gev$ -- $\mpl$~\cite{Visinelli:2014twa}.

Experimental signatures of the axion depend on the ultraviolet~(UV) completion of the theory. The two benchmark UV completions are the KSVZ~\cite{Kim:1979if, Shifman:1979if} and DSFZ models~\cite{Dine:1981rt, Zhitnitsky:1980he}. They make qualitatively similar but quantitatively different predictions for the axion couplings to SM particles. All of the couplings are controlled by the PQ symmetry breaking scale $f_a$, and are thus connected to the axion mass. Of particular importance is the axion-photon coupling through the operator
\beq
\mathscr{L} \supset -\frac{g_{a\gamma\gamma}}{4}\,a\,F_{\mu\nu}\widetilde{F}^{\mu\nu} \ ,
\label{eq:agg}
\eeq
where $F_{\mu\nu}$ is the photon field strength and $g_{a\gamma\gamma} \sim \alpha/(2\pi\,f_a)$ for both the KSVZ and DFSZ axion models. This coupling and others related to it can modify the evolution of stars by helping them lose energy more efficiently, and these considerations typically push $f_a\gtrsim 10^9\,\gev$~\cite{Raffelt:2006cw,Kim:2008hd}. On the other side, larger values of $f_a \gtrsim 10^{12}\,\gev$ tend to generate too much axion DM unless the initial displacement angle is much smaller than unity or there is a dilution of their density~\cite{Dimopoulos:1988pw,Fox:2004kb}. The coupling of Eq.~\eqref{eq:agg} can also produce signals in direct laboratory searches based on electromagnetic cavities~\cite{Sikivie:1983ip,Asztalos:2009yp,Graham:2015ouw}.

From a broader perspective, the QCD axion is a well-motivated representative of a much wider class of ultralight boson DM candidates~\cite{Hui:2016ltb}.  Light pseudoscalars that only couple to the SM through higher-dimensional operators suppressed by powers of a large energy scale $f$ arise in many theories~\cite{Svrcek:2006yi,Arvanitaki:2009fg,Jaeckel:2010ni}. Their masses need not be related to QCD confinement, and they are generically called axion-like particles~(ALPs). Similarly, new $U(1)$ vector bosons with feeble interactions with the SM, often called dark photons, can have naturally tiny masses by gauge invariance~\cite{Redondo:2008ec,Goodsell:2009xc,Nelson:2011sf,Graham:2015rva}. These more general ultralight bosons typically have masses well below the keV scale and can obtain a relic density through misalignment or other mechanisms. If they are make up the DM, they can be as light as $\sim 10^{-19}\ev$; for masses below this the particle Compton wavelength suppresses cosmic structure unacceptably at small scales~\cite{Dalal:2022rmp, Hu:2000ke}. As a result, ultralight boson DM with mass close to this lower bound is sometimes called fuzzy DM. Ultralight bosons can also produce a range of signals in high-precision laboratory experiments~\cite{Graham:2015ouw,Safronova:2017xyt}.

\subsubsection{Dark Photons as Dark Matter}

Light $U(1)$ gauge bosons,  known as dark photons, are a widely considered extension of the standard model.  They can get mass either through the Higgs mechanism, implying the existence of a larger scalar sector, typically at the same scale, or more simply by the Stueckelberg mechanism, which does not require further light degrees of freedom. For a recent review, see Ref.~\cite{Cline:2024qzv}.

Dark photons are frequently proposed as mediators between DM and the SM, but they can comprise the DM in their own right in a specific form of ultralight bosonic DM. Dark photons with mass below $2m_e$ have a highly-suppressed decay rate into either $3\gamma$ or $2\nu$, and naturally comprise a non-thermal dark matter candidate \cite{Pospelov:2008jk,Redondo:2008ec}. The possibility of freeze-in production in the early universe was excluded through indirect constraints and direct detection searches for absorption \cite{XENON100:2014csq,An:2014twa}, but other non-thermal production mechanisms still allow for a viable parameter range. 

Unlike scalars or pseudoscalars, misalignment does not easily lead to a significant relic density for vectors; due to their conformal invariance, any energy density initially present due to misalignment is diluted by the Hubble expansion faster than for pressureless matter.  This can be overcome by adding a nonminimal coupling $R A_\mu' A'^\mu$ to gravity \cite{Golovnev:2008cf,Arias:2012az}, but it results in ghosts over some range of momenta \cite{Karciauskas:2010as}, with their typical instabilities.   Fixing the ghost problem requires further model-building complications~\cite{Nakayama:2019rhg}.

A simpler production mechanism is to use the generic fluctuations of the $A'$ field during inflation to generate the relic density \cite{Graham:2015rva}.  The desired abundance is achieved for inflationary Hubble rate $H_i$ if the $A'$ mass is
\beq
    m_{A'} = 6\times 10^{-6}\left(\frac{10^{14}\,{\rm GeV}}{H_i}\right)^4 \ ,
\eeq
reminiscent of Eq.\ (\ref{axionrelicmass})
for the axion.  However, it has been shown \cite{Reece:2018zvv} that this scenario is disfavored by the weak gravity conjecture unless $m_{A'} \gtrsim 0.3\,$eV. Other inflationary production mechanisms include Refs.~\cite{Kolb:2020fwh,Capanelli:2024nkf}.

Misalignment as an origin for dark photons can be rescued if $A'$ couples to an axion 
\cite{Agrawal:2018vin,Co:2018lka} or a dark Higgs  $H'$ \cite{Dror:2018pdh}, which is initially produced by misalignment, and subsequently decays into dark photons.  Such couplings are generic, for example $a F'_{\mu\nu}\tilde F'^{\mu\nu}$ for the axion, or the gauge coupling to $H'$ if the latter gives mass to the dark photon.

The above discussion focused on light dark photons in the non-WIMP regime.  For $m_{A'}\gtrsim 1$\,MeV, the classic freeze-out mechanism could give rise to a thermal $A'$ abundance from a dark sector in which $A'$ is the lightest annihilation product, 
for example $H'H'\to A'A'$, and is sufficiently stable.  The latter criterion would require sufficiently small kinetic mixing $\epsilon$ with the standard model photon to suppress $A'\to 3\gamma$ and $A'\to e^+e^-$.  Such small $\epsilon$ can alternatively lead to the desired relic abundance through freeze-in \cite{Fradette:2014sza}.

\subsubsection{Sterile Neutrinos and Freeze-In Dark Matter}

Sterile neutrinos can provide a simple DM candidate with a possible connection to neutrino masses.  This can arise from a new SM gauge singlet fermion $\nu_s$ with a small mixing with a linear combination of the active neutrino states $\nu_a$. DM comes from the mostly singlet neutrino mass eigenstate defined by $\nu_4 = \nu_s \cos\theta + \nu_a \sin\theta$, where $\theta$ is the mixing parameter. To act as a fermionic DM species, the sterile neutrino $\nu_4$ must satisfy the Tremaine-Gunn bound~\cite{Tremaine:1979we, Gorbunov:2008ka, Boyarsky:2008ju} which requires it to be heavier than about a keV. 

An acceptable relic density of sterile neutrinos can be generated in a number of ways, but a particularly attractive mechanism is that of Dodelson and Widrow~\cite{Dodelson:1993je}. At high temperatures in the early universe, the active neutrinos reach chemical equilibrium through their weak interactions with the rest of the SM plasma but, for sufficiently small mixing angles, the singlet neutrinos do not. Once an active neutrino state $\nu_a$ is created in the thermal plasma, it starts to oscillate into the gauge singlet $\nu_s$ driven by their mixing. This can allow a small component of the $\nu_s$ state to develop and remain intact before another weak interaction destroys the active neutrino part. The process is repeated many times until the weak interaction falls out of equilibrium and a relic density of singlet neutrinos is built.  The dominant production is found to occur at temperatures near $T\sim 100\,\mev$.  Assuming a vanishing initial density of sterile neutrinos, the correct relic density can be generated by this mechanism for a range of masses and mixing angles~\cite{Abazajian:2001nj,Drewes:2016upu,Abazajian:2017tcc,Dasgupta:2021ies}.

Within this minimal setup, the DM $\nu_4$ is not absolutely stable; it can decay into three neutrinos, or a neutrino plus a photon. The latter mode serves as an important indirect detection channel since the photon from the decay is monochromatic. Many experiments are searching for such a signal, and they have set stringent upper limits on the mixing angle $\theta$ as a function of DM mass. Interestingly, an unidentified X-ray line excess of this sort at $E_{\gamma} = 3.55\,\kev$ was discovered recently by two groups~\cite{Boyarsky:2014jta, Bulbul:2014sua}. However, whether this excess is due to DM or unresolved astrophysical backgrounds is a topic of intense debate~\cite{Dessert:2018qih,Abazajian:2020unr,Boyarsky:2020hqb,Dessert:2023fen}. Measurements of this type, together with lower limits on the sterile neutrino mass from Lyman-$\alpha$ observations and structure formation, also put very strong constraints on the parameter region consistent with sterile neutrino DM production through the Dodelson-Widrow mechanism~\cite{Drewes:2016upu,Abazajian:2017tcc,Dasgupta:2021ies}. Potential solutions to this tension include other production mechanisms for the sterile neutrino~\cite{Shaposhnikov:2006xi,Asaka:2006ek,Bezrukov:2009th,Nemevsek:2012cd,Roland:2014vba,Dror:2020jzy}, new self-interactions between the active neutrinos that enhance sterile neutrino creation~\cite{deGouvea:2019phk} and that could produce striking new missing transverse momentum signals at accelerator neutrino experiments such as DUNE~\cite{Kelly:2019wow}, and the existence of a large lepton number asymmetry in the universe~\cite{Shi:1998km}.

More generally, sterile neutrinos are a prototypical example of the freeze-in mechanism of DM creation~\cite{McDonald:2001vt,Hall:2009bx,Bernal:2017kxu}. This mechanism operates when the connection between DM and the SM plasma is too weak to bring the two sectors into full thermodynamic equilibrium with each other. Even so, a sub-equilibrium density of DM can be created through reactions initiated by SM particles alone. For example, if allowed, reactions of the form $\text{SM}+\text{SM} \to \chi+\chi$ will populate a DM species $\chi$ and can produce the dominant relic density. DM created this way is sometimes called a FIMP, standing for Freeze-In Massive Particle.

Freeze-in creation of DM is usually dominated by reactions near a specific temperature. If the DM connection to the SM comes from a renormalizable operator, production is typically peaked at temperatures near the DM mass, $T\sim m_{\chi}$.  In contrast, when the connection is due to a non-renormalizable operator, DM production from freeze-in tends to be greatest at much higher temperatures, such as the reheating temperature after inflation or the highest temperature at which the non-renormalizable operator is well-defined~\cite{Elahi:2014fsa}. These two regimes are referred to as infrared and ultraviolet freeze-in, respectively. After freeze-in production, DM may undergo further number changing reactions such as annihilation, decay, and self-thermalization~\cite{Cheung:2010gj,Chu:2011be}. An important feature of both types of freeze-in is that they allow for particle DM masses well below $m_{\chi} \lesssim 10\,\mev$, which is challenging to obtain with freeze-out. This motivates extending particle DM searches through direct detection down to $m_{\chi} \sim \kev$ masses~\cite{Krnjaic:2017tio,Knapen:2017xzo,Dvorkin:2019zdi,Dvorkin:2020xga}.

\subsubsection{Asymmetric Dark Matter}

Asymmetric dark matter~(ADM) generates the DM abundance through a mechanism that is analogous to the baryon asymmetry of the universe~\cite{Kaplan:2009ag,Davoudiasl:2012uw, Petraki:2013wwa, Zurek:2013wia}. It has been well established that the relic density of baryons is asymmetric, consisting nearly entirely of baryons with close to no antibaryons. Similarly, in ADM scenarios the DM particle $\chi$ has a distinct antiparticle $\bar{\chi}$ and it is the particle that dominates the relic total DM density. Some realizations of ADM also connect the density of DM to that of baryons, which are numerically similar with $\Omega_{\rm DM} \sim 5\,\Omega_b$.

An asymmetry between the densities of DM $\chi$ and its antiparticle $\bar{\chi}$ can be created through a variety of mechanisms in the early universe in analogy with the generation of the baryon asymmetry~\cite{Riotto:1998bt,Cline:2006ts,Davidson:2008bu}. Assuming such an asymmetry has been produced, the subsequent cosmology undergoes a freeze-out process similar to WIMP-like DM but with an important twist. At high temperatures above the DM mass the ADM is fully thermalized, and both DM particles and antiparticles are present with nearly equal abundances. As the temperature cools, particles and antiparticles annihilate with each other until these reactions become too slow to maintain thermodynamic equilibrium. If the annihilation reactions are efficient, this freezeout will only occur after nearly all the DM antiparticles have annihilated away. In this regime the DM relic density is set by the DM asymmetry rather than the annihilation cross section, provided it is large enough~\cite{Graesser:2011wi, Gelmini:2013awa}. In the end, the DM relic abundance consists of particles, with only an exponentially smaller density of antiparticles left over.

Some models of ADM also connect the origin of the densities of DM and baryons to each other.  This can be realized through an approximate symmetry that relates the particle content of the DM sector to the SM and similar asymmetry production mechanisms in both~\cite{Foot:2014mia}. Alternatively, an asymmetry can be generated in one sector first and then transferred partially to the other through a portal operator~\cite{Kaplan:2009ag}. With such portals, the charge of DM must typically connect in a non-trivial way with baryon or lepton number in the SM. The number densities of DM and baryons are typically similar to each other in all these scenarios, and thus the favoured mass scale for ADM connected to the baryon asymmetry is $m_{\chi} \sim \gev$ up to a few exceptions.

Viable realizations of ADM suggest a range of possible observable signals. For ADM to be dominated by its asymmetry, the annihilation rate in the early universe must be considerably higher than what is needed to produce the correct relic through standard (symmetric) freeze-out~\cite{Graesser:2011wi,Iminniyaz:2011yp}. The rates needed are usually too large to be achieved through the weak interactions, and thus ADM motivates the existence of a new force carrier below the weak scale that couples directly to DM. If the new mediator also couples to the SM, it can mediate ADM scattering with matter in direct detection searches and possibly also lead to new features in the recoil energy spectrum~\cite{DelNobile:2015uua}. The coupling between DM and the light mediator could also lead to strong self interaction among DM particles~\cite{Buckley:2009in}. In contrast to WIMP-like DM, however, signals of ADM in indirect detection can be suppressed since the DM particles have no antiparticles to annihilate with, although some exceptions exist~\cite{Buckley:2011ye, Cirelli:2011ac, Tulin:2012re}. The suppression of annihilation also leads to interesting effects in stars. For example, ADM can be captured in neutron stars, build up a large density, and lead to formation of a black hole that destroys the star~\cite{Kouvaris:2010vv, McDermott:2011jp, Bramante:2014zca}

\subsubsection{Primordial Black Holes}

Dark matter could be made up of primordial black holes~(PBHs) formed in the early universe in regions with large upward fluctuations in the local matter density~\cite{Zeldovich:1967lct,Hawking:1971ei}. They are arguably the simplest candidate that does not directly require a new particle beyond the SM. However, if PBHs are to make up the entire DM abundance, new physics is typically needed to produce large enough density fluctuations. These can arise from enhanced fluctuations during inflation~\cite{Carr:1975qj,Ivanov:1994pa,Yokoyama:1998qw} or at a later stage from strongly first-order phase transitions~\cite{Crawford:1982yz,Hawking:1982ga,Kodama:1982sf}.

A broad set of astrophysical observations constrain the fraction $f_{\mathrm{PBH}}$ of the total DM abundance consisting of PBHs~\cite{Carr:2020gox,Carr:2020xqk,Green:2020jor,Escriva:2022duf}. Only in a limited mass range can PBHs make up the entire DM density. For masses below $\mpbh \lesssim 10^{-17}\,\msol$, PBHs emit enough energetic, visible particles through Hawking radiation~\cite{Hawking:1974sw} to exclude $f_{\mathrm{PBH}} = 1$ from the effects this would have on BBN and extragalactic photon observations. Similarly, PBHs as all the DM are strongly excluded for masses $\mpbh \gtrsim 100\,\msol$ from the distortions they would induce in the CMB from the accretion of matter on them in the early universe. Between these two extremes, PBHs can make up a significant fraction of the DM density, and even all of it within certain windows. Over this range, $f_{\mathrm{PBH}}$ still faces constraints from femtolensing of distant gamma-ray bursts for $\mpbh \in [10^{-17}, 10^{-14}]\,\msol$, capture and destruction of neutron stars for $\mpbh \in [10^{-15}, 10^{-8}]\,\msol$, the survival of white dwarfs in $\mpbh \in [10^{-14}, 10^{-13}]\,\msol$, and the microlensing of stars and supernovae over $\mpbh \in [10^{-11}, 10]\,\msol$.

The recent discovery of gravitational waves from binary black hole mergers by the LIGO collaboration have given further motivation for PBHs with $\mpbh \sim 30\,\msol$ making up at least some of the DM. For example, PBHs if in this mass range comprise an order-one fraction of the DM relic abundance, their classical gravitational-capture cross section leads to roughly the same observed merger rate as seen by LIGO~\cite{Bird:2016dcv,Clesse:2017bsw}. Recent proposals have been made to cover the relevant $\mpbh \sim 30\,\msol$ mass window by using the CHIME experiment to search for the lensing of fast radio bursts by such PBHs~\cite{Munoz:2016tmg}.

\subsubsection{And More\ldots}

Given the unknown nature of DM, it is important to  consider a broad range of possibilities for what it might be. Most of the DM candidates discussed above consist of a single DM species, sometimes accompanied by a new force mediator. However, we know that ordinary matter is comprised of many particles and force mediators that give rise to a rich variety of phenomena (\emph{e.g.} us). DM could also be part of a larger dark sector with a range of observable signals. 

Working in direct analogy with ordinary matter, DM could itself be a bound state. The simplest realization is atomic DM, consisting of a bound state of a dark photon and a dark electron with opposite charges under a dark $U(1)$ gauge force~\cite{Goldberg:1986nk,Kaplan:2009de}. Despite their names, these new states are not directly charged under the SM, and the fermion masses and the dark fine structure constant $\alpha_D$ are unknown parameters of the theory. Depending on their values, dark recombination may occur relatively late in the history of the cosmos leading to dark baryon acoustic oscillations and limiting the smallest DM structures today~\cite{CyrRacine:2012fz,Cyr-Racine:2013fsa}. Other realizations of composite DM include ADM bound by a new scalar force mediator~\cite{Wise:2014jva, Wise:2014ola,Gresham:2017cvl}, as well as strongly-coupled bound states analogous to baryons~\cite{Gudnason:2006yj,Appelquist:2014jch}, pions~\cite{Hochberg:2014kqa}, and glueballs~\cite{Okun:1980mu,Faraggi:2000pv,Boddy:2014yra,Soni:2016gzf}.

A common feature of many models of bound-state DM is a significant self-interaction cross section.  Such interactions are motivated by a number of puzzles that arise when observations of cosmic structure are compared to simulations of DM dynamics. If the DM is assumed to be effectively collisionless, as expected for WIMPs, the simulations tend to predict more structure at shorter distances than what is seen~\cite{Tulin:2017ara}. In contrast, if the DM particles are able to scatter with each other, with a cross section on the order of $\sigma/m \sim 1 \,{\rm cm^2/g}$, these interactions tend to smooth out cosmic overdensities at small scales and can potentially resolve many of these structure puzzles~\cite{Tulin:2017ara,Spergel:1999mh,Ackerman:mha}. 

While DM is usually expected to be neutral under electromagnetism and the strong force, some loopholes exist. A DM species that couples to a new massless $U(1)$ vector boson that has kinetic mixing with the photon of electromagnetism will develop a millicharge, corresponding to an effective electromagnetic charge that is a small fraction $\epsilon \ll 1$ of the charge of an electron. Existing bounds on millicharged particles come from a variety of sources~\cite{Davidson:2000hf,Dubovsky:2003yn,McDermott:2010pa}, and are typically expressed as a limit on $\epsilon$ for a given mass. Recently, an apparent anomaly in the 21-cm absorption temperature of hydrogen atoms has been found by the EDGES experiment~\cite{Bowman:2018yin}. The anomaly can be explained if a small fraction of the DM density consists of millicharged particles~\cite{Berlin:2018sjs, Barkana:2018qrx, Liu:2019knx}. A number of experiments, including ArgoNeuT, DUNE, MilliQan, LDMX, SHiP, could potentially probe this scenario~\cite{Acciarri:2019jly, Harnik:2019zee, Berlin:2018bsc, Kelly:2018brz}.

Dark matter can also interact directly with the strong force if it is superheavy, with mass above $m_{\chi} \gtrsim 10^{16}\,\gev$. When the mass is this large, the number density of the DM species is low enough for it to have a scattering cross section with ordinary matter as large as $\sigma \sim 10^{-28}\,\mathrm{cm}^2$. When such a DM candidate travels though a detector, it could scatter with the nucleus target multiple times~\cite{Bramante:2018qbc, Bramante:2018tos}. Each scattering deposits an energy similar to WIMP scattering but leaves the velocity of the very heavy DM nearly unchanged. As a result, the DM leaves a mostly collinear track of nuclear recoils behind it. Such a signal could be searched for at DM detectors such as DEAP-3600, LUX/LZ, PandaX, PICO, and XENON, as well as liquid scintillator neutrino detectors like Borexino, SNO+, and JUNO.

Even the stability of DM is not absolutely necessary. Observations of the CMB and cosmic structure require the decay lifetime of DM to be greater than ten times the age of the universe~\cite{Poulin:2016nat}. Bounds on decaying DM are even stronger if the final states are SM particles, with bounds on the lifetime reaching up to $\tau \gtrsim 10^{28}\,\mathrm{s}$ for weak-scale DM decaying to states with electromagnetic or strong charges~\cite{Blanco:2018esa}.

\subsection{Dark Sectors\label{sec:dsector}}

Dark matter and neutrino mass provide strong empirical evidence for physics beyond the SM. However, rather than suggesting a specific mass scale for new physics, they arguably point to a dark (or hidden) sector, defined by the requirement of being feebly-coupled to the SM. This feature of dark sectors relaxes many existing constraints on the mass scale of new physics, and opens up a range of high luminosity or high precision experimental probes. In turn, this has allowed significant progress both theoretically and experimentally in exploring the primary signatures of dark sectors over the past decade~\cite{Alexander:2016aln,Battaglieri:2017aum,Beacham:2019nyx,Agrawal:2021dbo,Antel:2023hkf}.

\begin{figure}[t]
\centering
\includegraphics[scale=0.5]{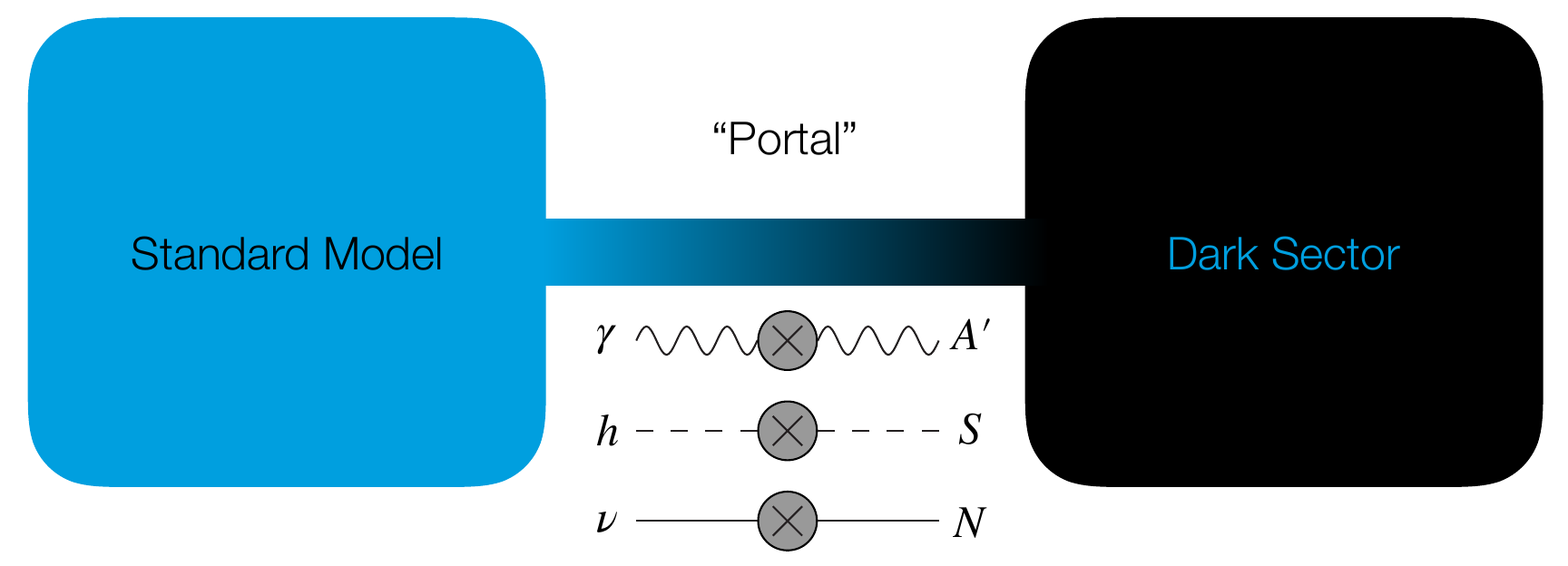}
\caption{The interaction between the SM and a dark sector is mediated by mixing neutral SM particles with those in the dark sector. Shown are the kinetic mixing (or vector), Higgs, and lepton (or neutrino) portals.}
\label{fig:portal}
\end{figure}

From a theoretical perspective, while the landscape of dark sectors is vast, it is useful to consider a general parametrization of the interactions and degrees of freedom which mediate between the SM and the dark sector. A natural assumption is that any dark sector states that are light should be SM gauge singlets. This automatically ensures weak coupling to the visible~(SM) sector, while the impact of heavier charged states is incorporated in an effective field theory expansion at or below the weak scale,
\beq
 \mathscr{L} \sim \sum_{n} \frac{c_n}{\Lambda^n} {\cal O}^{(k)}_{\rm SM} {\cal O}^{(l)}_{\rm hidden} \ ,
\eeq
where $k$ and $l$ denote operator dimensions and $n=k+l-4$. Thus lower dimension interactions, unsuppressed by the heavy scale $\Lambda$, are preferentially probed at lower energy.

The set of lowest-dimension interactions, or {\it portals}, is quite compact. Up to dimension five ($n\leq 1$), assuming SM electroweak symmetry breaking, the full list of portals is:
\begin{itemize}
\item $\mathscr{L} = -\frac{1}{2}\,\epsilon\,B^{\mu\nu}F'_{\mu\nu}$  \ -- \ dark photon $A_\mu'$ kinetically mixed with hypercharge,
\item $\mathscr{L} =(A S + \lambda S^{2})H^{\dagger}H $ \ -- \ dark scalar $S$ coupled to the Higgs,
\item $\mathscr{L} = y_N LHN $ \ -- \ dark fermion $N$ coupled via the lepton portal,
\item $\mathscr{L} =\frac{\partial_{\mu}a}{f_{a}} \overline{\psi}\gamma^{\mu}\gamma^{5}\psi$  \ -- \ axion-like pseudoscalar $a$ coupled to the axial current.
\end{itemize}
On general grounds, the couplings of these lowest dimension operators are minimally suppressed by any heavy scale, and new weakly-coupled physics would naturally manifest itself first via these interaction in any generic top-down model. Portals therefore play a primary role in dark sector phenomenology. After electroweak symmetry breaking, the portal couplings induce the mixing of neutral SM particles, such as the photon, Higgs boson, or neutrinos with new states, mediating the interactions of a dark sector with the SM. This general setup is depicted in Fig.~\ref{fig:portal}.

Focusing specifically on DM, it is notable that all the degrees of freedom introduced above as part of the primary portals are viable warm or cold DM candidates. The list includes traditional non-thermal candidates, such as axions~\cite{Preskill:1982cy, Dine:1982ah, Abbott:1982af} and sterile neutrinos~\cite{Dodelson:1993je}, but also singlet scalars and dark photons~\cite{Redondo:2008ec,Goodsell:2009xc,Nelson:2011sf,Graham:2015rva,Caputo:2021eaa}. More generally, the dark sector paradigm is also motivated by specific considerations of thermal relic cold DM candidates, which acquire a relic abundance via freeze-out in the early universe, and generalize the WIMP paradigm to a wider mass range~\cite{Boehm:2003hm,Pospelov:2007mp,Feng:2008ya,ArkaniHamed:2008qn}.  Given that the traditional WIMP annihilation rate scales as $\langle \sigma v\rangle \propto g^4 m_{\chi}^2/\max\{m_\chi^4,m_W^4\}$ for WIMP mass $m_\chi$, viability rests on the DM mass $m_{\chi}$ being above the Lee-Weinberg bound of a few GeV~\cite{Lee:1977ua}. However, thermal relic WIMP-like candidates remain a compelling DM scenario, provided that there are additional force mediators beyond the weak interactions; so-called dark forces. This again motivates a multi-component dark sector, and moreover allows the thermal relic mass range for DM to be extended below the Lee-Weinberg bound if the new dark force mediators are light.

\begin{figure}[ttt]
\begin{center}
 \includegraphics[width = .48\textwidth]{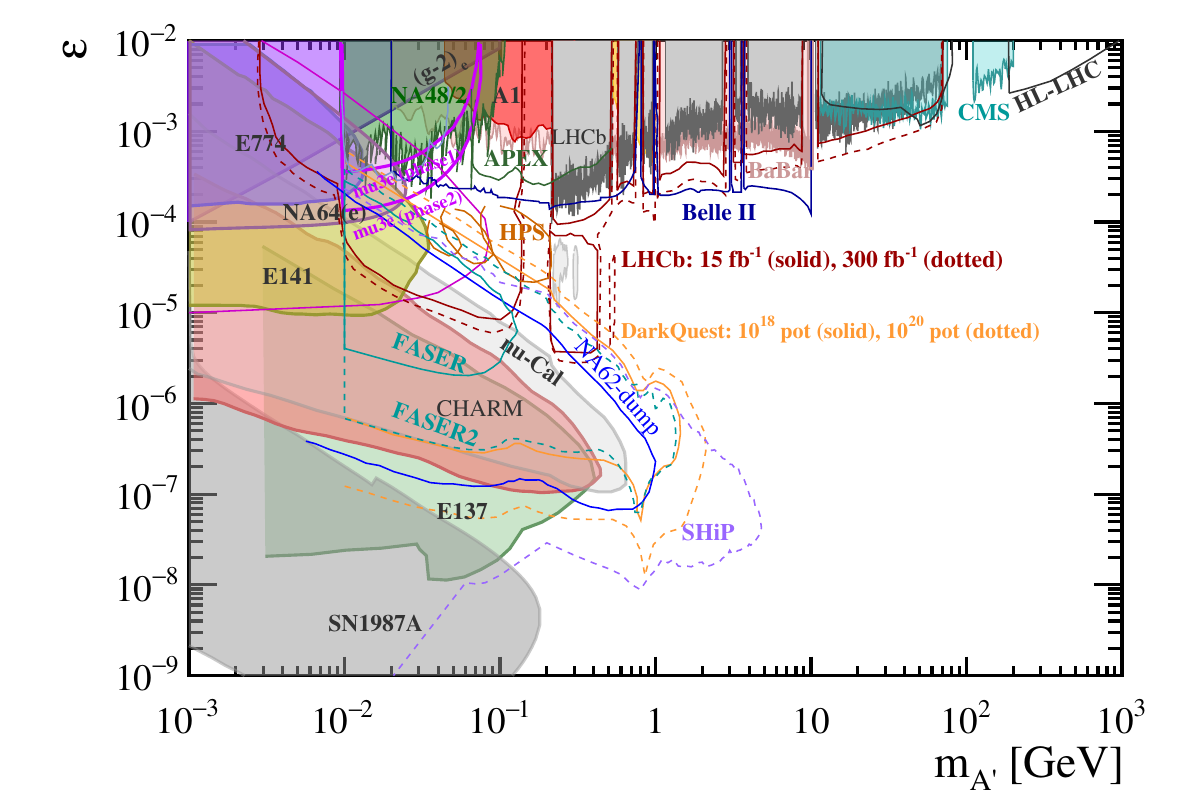}
 \includegraphics[width = .49\textwidth]{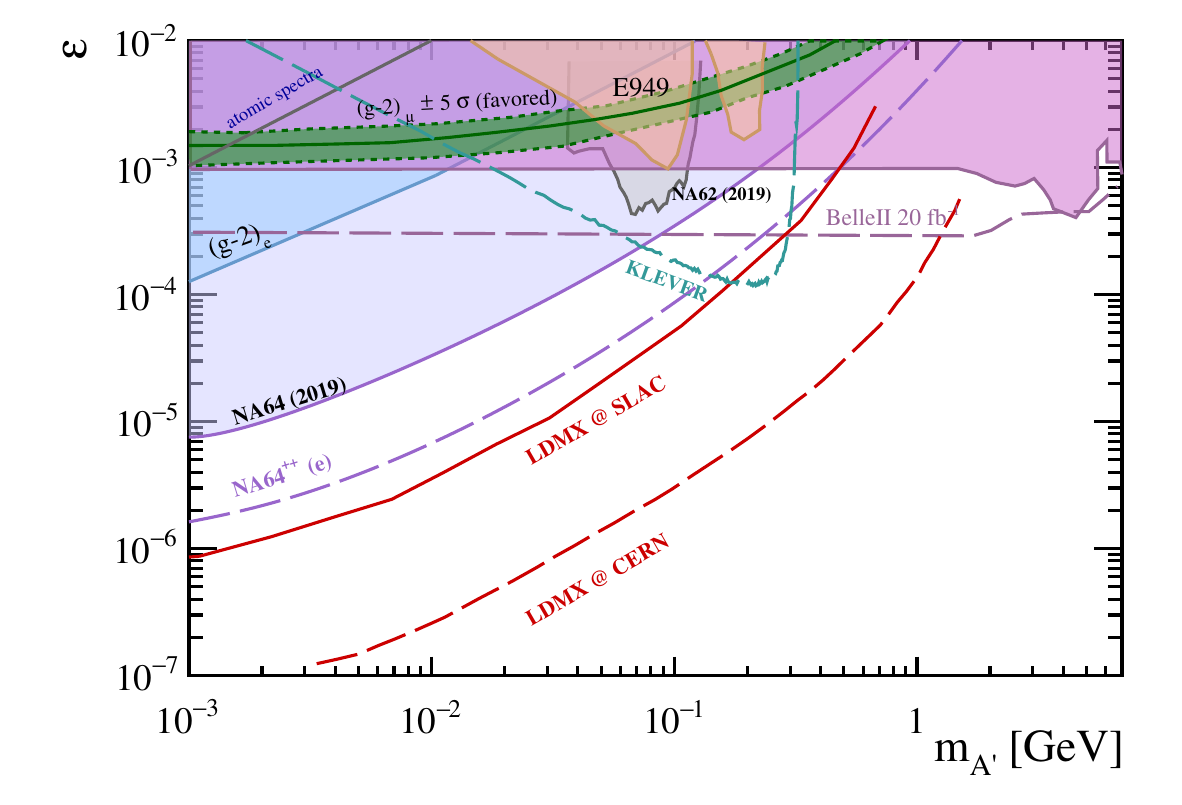}
 \caption{Existing and projected bounds on dark photons assuming visible~(left) or invisible~(right) decays in terms of the dark photon mass $m_{A'}$ and kinetic mixing parameter $\epsilon$. Figure from Refs.~\cite{Lanfranchi:2020crw,Agrawal:2021dbo}.
 } 
 \label{fig:dpbounds}
\end{center} 
\end{figure}

There has been significant work exploring dark sectors theoretically and experimentally over the past decade (see e.g.~\cite{Alexander:2016aln,Battaglieri:2017aum,Beacham:2019nyx,Agrawal:2021dbo,Antel:2023hkf}). One of the simplest hidden sectors involves states charged under a new U(1)$'$ gauge group. The corresponding gauge boson, dubbed the dark photon $A'_\mu$, is kinetically mixed with hypercharge via the vector portal above, which induces a coupling to the electromagnetic current $\mathscr{L} = -\epsilon e A'_\mu J^\mu_{\rm EM}$~\cite{Okun:1980mu,Holdom:1985ag}. This allows for experimental probes via a variety of high-luminosity electron and proton beam dump experiments and colliders~\cite{Pospelov:2008zw,Bjorken:2009mm}. Existing limits are shown in the case that decays are to visible SM final states: dileptons in the left panel of Fig.~\ref{fig:dpbounds}, and to invisible dark sector states such as light DM in the right panel of Fig.~\ref{fig:dpbounds}. These high-luminosity probes are complementary to a range of direct detection experiments searching for light DM through nuclear or electron recoils or other indirect signatures.

Dark sectors can also admit DM scenarios with a rich structure, such as self-interacting DM or multicomponent DM. Furthermore, dark sectors allow for the efficient study of self-consistent models that can match onto a wide variety of ultraviolet theories.

\section{Search Approaches and Current Bounds\label{sec:approaches}}

Having described a wide range of candidates for what DM might be, we turn next to the many approaches being pursued to identify its underlying (non-gravitational) nature. These approaches can be organized into three main categories: i) direct searches for DM in the laboratory; ii) indirect searches for astrophysical signals produced by DM; iii) creation and detection of DM in high-energy colliders. While no clear identification of DM has yet been made, these searches have significantly narrowed down the range of possibilities of what DM might be, and they offer great promise for new discoveries in the years to come.

\subsection{Direct Searches}

Direct searches are a powerful technique to unmask the DM that surrounds us through its interactions with SM particles in a controlled laboratory environment. Observations of nearby stellar velocities indicate that our solar system resides in a background of DM with a mass density of about $\rho_{\chi} \sim 0.3 ~{\rm GeV/cm^3}$, moving at speeds around $v \sim 10^{-3}$ relative to the speed of light~\cite{Bovy:2012tw,Buch:2018qdr,Benito:2019ngh,deSalas:2020hbh}. Direct DM searches aim to detect this background of DM, typically through a scattering or conversion interaction within a laboratory detector. In a scattering interaction, the DM deposits some amount of energy by recoiling  against a detector particle, usually a nucleus or electron~\cite{Lewin:1995rx}. In a conversion interaction, the mass-energy of DM is converted into visible energy in the detector. Substantial progress has been made over the last four decades in seeking out DM using these methods~\cite{Goodman:1984dc,Drukier:1986tm,Sikivie:1983ip,Jungman:1995df,Cooley:2022ufh,Akerib:2022ort}.

\subsubsection{Nuclear Scattering}

A major focus of direct searches are signals from DM scattering on nuclei in a target material~\cite{Goodman:1984dc,Drukier:1986tm}. As a struck nucleus recoils, it deposits energy in the detector leaving a potentially observable signal. This method is particularly well suited to WIMP DM based on the expected local DM velocity near $v \sim 10^{-3}$ and WIMP masses in the range of $m_\chi \sim 10\,\gev$--$10\,\tev$. In such a scattering exchange, the recoil energy $E_R$ imparted to the nucleus is
\beq
E_R = \frac{2\mu_{\chi N}^2}{m_N}\,v^2\cos^2\!\theta \ ,   
\label{eq:er}
\eeq
where $m_N$ is the mass of the nuclear target, $\mu_{\chi N} = m_{\chi}m_N/(m_\chi + m_N)$ is the DM-nucleus reduced mass, $v$ is the initial DM speed, and $\cos\theta$ is the nuclear scattering angle. These recoil energies typically lie in the $E_R \sim$ keV--MeV for WIMP DM.  Since this range  coincides with a number of nuclear and electromagnetic background processes, the most sensitive direct detection searches require extremely clean, low-background laboratories. 

The rate of DM scattering per unit recoil energy and target mass is~\cite{Goodman:1984dc}
\beq
\frac{dR}{dE_R} = \frac{1}{m_N}\left(\frac{\rho_{\chi}}{m_{\chi}}\right)\,\int_{v_{\rm min}}\!\!d^3v\,v\,f_{\rm lab}(\vec{v})\,\frac{d\sigma_N}{dE_R} \ ,
\label{eq:drder}
\eeq
where 
$d\sigma_N/dE_R$ is the differential DM-nucleus cross section, $f_{\rm lab}(\vec{v})$ is the local distribution of the DM velocity $\vec{v}$ in the lab frame, and $v_{\rm min}$ is the minimum velocity needed to produce the recoil energy $E_R$ (corresponding to $\cos\theta = 1$ in Eq.~\eqref{eq:er}).

All the dependence on the underlying theory of DM in Eq.~\eqref{eq:drder} is contained in $d\sigma_N/dE_R$. A given theory usually specifies interactions between DM and quarks and gluons. Since the typical momentum transfer in direct detection, $q \sim \mu_{\chi N}v \ll \gev$, is relatively small, these fundamental interactions are matched onto effective interactions between the DM and nucleons and pions. The nucleon-level interactions are then taken as inputs to calculate the net interaction with nuclei~\cite{Lewin:1995rx}. For interactions induced by a massive scalar or vector mediator, the cross section can be written in the form~\cite{Goodman:1984dc,Jungman:1995df}
\beq
\frac{d\sigma_N}{dE_R} = \frac{m_N}{2\mu_{\chi N}^2v^2}\,\bar{\sigma}_N\,|F(E_R)|^2 \ ,
\label{eq:sigr}
\eeq
where $\bar{\sigma}_N$ is constant in $E_R$, and $F(E_R)$ is a nuclear form factor that goes to unity as $E_R\to 0$~\cite{Lewin:1995rx}.  

A general approach to DM interactions with nuclei is to organize the many possibilities in terms of a set of effective operators~\cite{Fan:2010gt,Fitzpatrick:2012ix,Hisano:2015bma}. In many cases of interest, the most important interactions can be classified as either spin-independent~(SI) or spin-dependent~(SD)~\cite{Goodman:1984dc,Lewin:1995rx}. For SI scattering,  the interaction can be approximately coherent over the entire nucleus, with the DM scattering off any of the constituent protons or neutrons leading to the SI effective cross section scaling like $\bar{\sigma}_N^{\rm (SI)} \propto A^2$, where $A$ is the number of nucleons in the nucleus. This can be compared to SD scattering, where the SD nuclear cross section scales as $\bar{\sigma}_{N}^{\rm (SD)} \propto \left\langle S_N \right\rangle^2$, where $\left\langle S_N \right\rangle$ is the nuclear spin expectation value, which can either be for proton spin ($\left\langle S_p \right\rangle$) or neutron spin ($\left\langle S_n \right\rangle$). Because $A \sim 100$ and $\left\langle S_n \right\rangle \sim 1$ for heavy nuclei typically employed in direct detection experiments, there has been greater sensitivity attained for SI DM interactions. 

\begin{figure}[ttt]
\begin{center}
 \includegraphics[width = .6\textwidth]{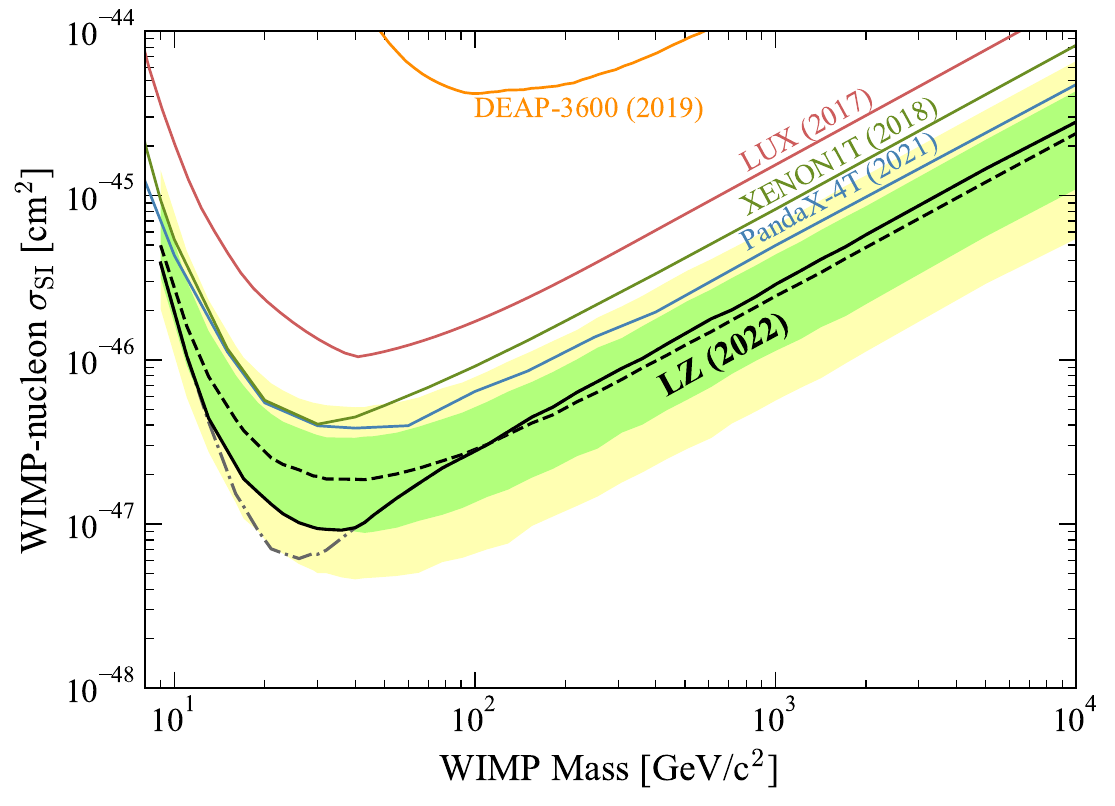}
 \caption{Direct detection bounds on WIMP-like DM with spin-independent scattering on nuclei from the LZ experiment together with other recent limits. Figure from Ref.~\cite{LZ:2022lsv}.
 } 
 \label{fig:silz}
\end{center} 
\end{figure}

Searches for DM-nucleus scattering are commonly presented in terms of a fixed per-nucleon cross section, $\sigma_{\chi n}$.  Since different experiments utilize different nuclear targets for the detection of DM, this allows them to be compared on an equal footing. For SI and SD scattering, the conversions are
\beq
\sigma_{\chi n}^{\rm (SI)} = \frac{1}{A^2}\frac{\mu_{\chi n}^2}{\mu_{\chi N}^2}\,\bar{\sigma}_N^{\rm (SI)} \  ~~~~{\rm and}~~~~~\sigma_{\chi n,p}^{\rm (SD)} = \frac{3}{4} \frac{J}{J+1} \frac{1}{\left \langle S_{n,p} \right \rangle^2} \frac{\mu_{\chi n}^2}{\mu_{\chi N}^2}\,\bar{\sigma}_N^{\rm (SD)} \ ,
\label{eq:si}
\eeq
where $\mu_{\chi n}$ is the DM-nucleon reduced mass, $J$ is the total angular momentum of the nucleus, and spin coupling factors have been omitted from $\sigma_{\chi n,p}^{\rm (SD)}$, which will depend on the DM model.

\begin{figure}[ttt]
\begin{center}
 \includegraphics[width = .45\textwidth]{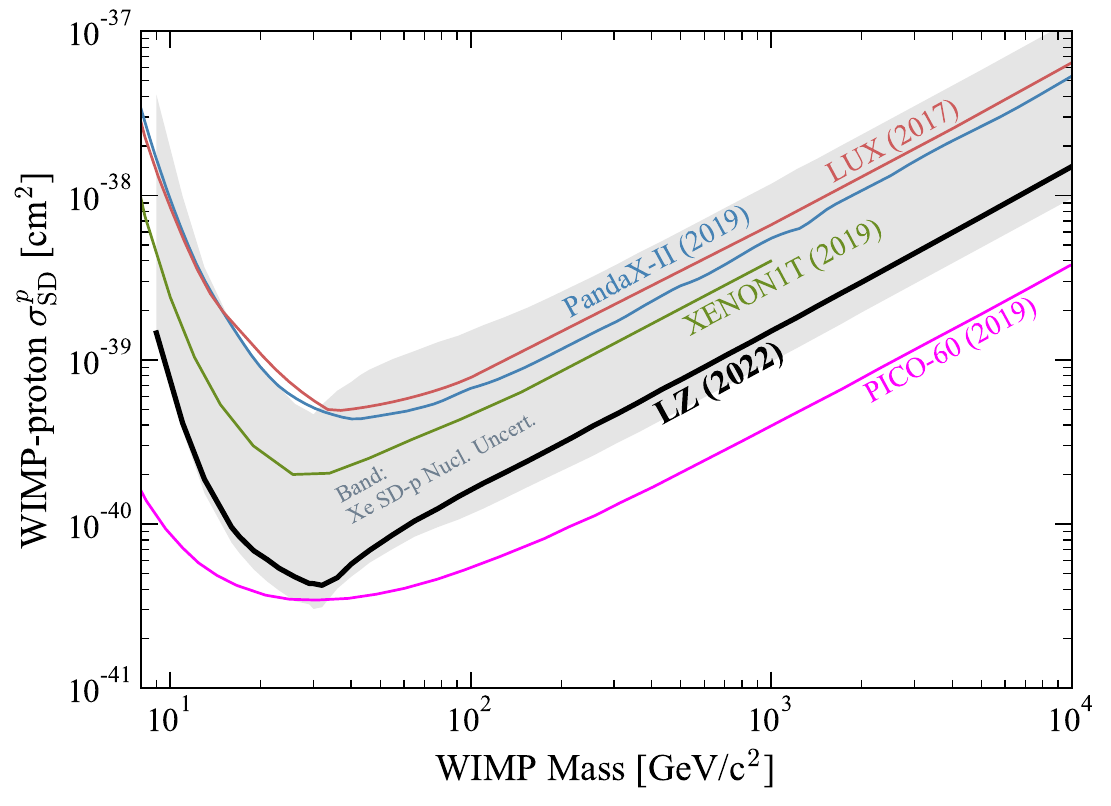}
 \includegraphics[width = .45\textwidth]{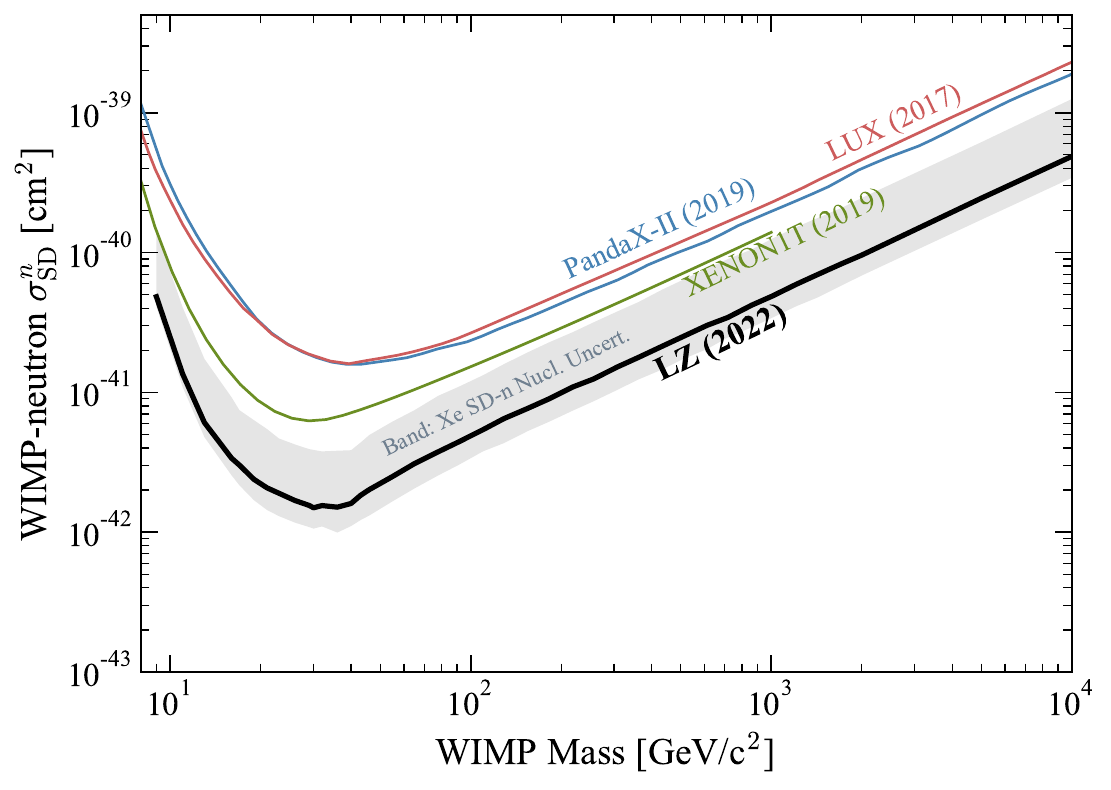}
 \caption{Direct detection bounds on WIMP-like DM with spin-dependent scattering on nuclei from the LZ experiment together with other recent limits. Figure from Ref.~\cite{LZ:2022lsv}.
 } 
 \label{fig:sdlz}
\end{center} 
\end{figure}

Direct searches for WIMP-like DM have grown rapidly in size and sensitivity over the past twenty years. For heavier DM with mass above $m_\chi \gtrsim 10\,\gev$ and predominantly SI interactions with nuclei, the most sensitive current experiments are based on large-scale liquid xenon targets and include LZ~\cite{LZ:2022lsv}, XENONnT~\cite{XENON:2023cxc}, and PandaX-4T~\cite{PandaX-4T:2021bab}. DEAP-3600 is the most sensitive liquid argon dark matter experiment~\cite{DEAPFirstYear,DEAP_NREFT}. In Fig.~\ref{fig:silz} we show the current world-leading bounds on heavier SI DM from the LZ experiment~\cite{LZ:2022lsv}. These experiments, together with PICO-60~\cite{Amole:2019fdf}, also provide the best current bounds on SD DM scattering on nuclei in this mass range, as shown in Fig.~\ref{fig:sdlz}.
The exclusion contours in Figs.~\ref{fig:silz} and \ref{fig:sdlz} follow similar shapes. In both figures the strongest exclusions on the normalized scattering cross sections are found for masses near $m_{\chi} \sim 50\,\gev$, where the DM and target nucleus masses are similar. For lower DM masses, the recoil energy falls quickly below the sensitivity thresholds of the experiments shown, while for larger masses the recoil energy converges to a constant value when $\mu_{\chi N} \approx m_N$. However, at high DM masses the number density of DM, $\rho_{\chi}/m_{\chi}$, decreases, resulting in a linear decrease in cross section sensitivity at high DM masses. 

\begin{figure}[ttt]
\begin{center}
 \includegraphics[width = .65\textwidth]{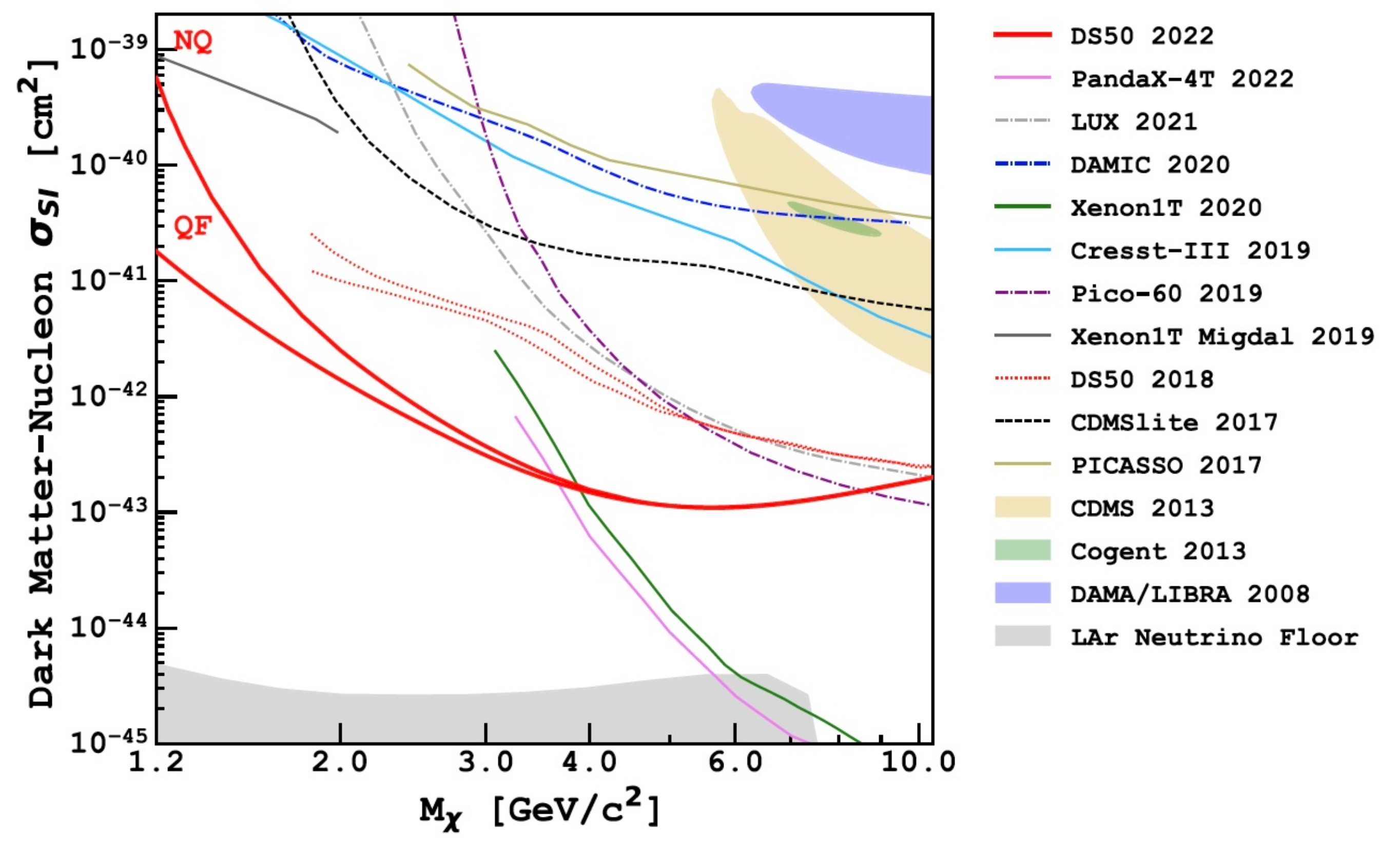}
 \caption{A compilation of recent spin-independent nucleon scattering searches for DM masses less than 10 GeV. Plot taken from \cite{DarkSide-50:2022qzh}.} 
 \label{fig:lmsi}
\end{center} 
\end{figure}

Nuclear scattering is a less efficient signal channel for DM with mass below $m_\chi \lesssim 10\,\gev$ because the energy transferred falls below the sensitivity threshold of the large-scale xenon detectors. Dedicated low-mass DM detectors with much lower detection thresholds such as SuperCDMS~\cite{Agnese:2016cpb} and NEWS-G~\cite{Gerbier:2014jwa,NEWS-G:2023qwh} are underway. Some additional sensitivity to lighter DM candidates can also be obtained from the Migdal effect, in which the recoil of a nucleus due to DM scattering liberates or displaces electrons~\cite{Ibe:2017yqa,Dolan:2017xbu}. This has been applied to extend the sensitivity of noble liquid detectors including  XENON1T~\cite{XENON:2019zpr} and DarkSide-50~\cite{DarkSide-50:2022qzh,DarkSide:2022dhx}, as shown in Fig.~\ref{fig:lmsi}.

Planned future experiments are expected to extend the sensitivity to WIMP-like DM by expanding to larger detection volumes while reducing backgrounds. The international community has converged on large-scale noble liquid detectors based on xenon or argon. The major plans with xenon are the XLZD/Darwin~\cite{DARWIN:2016hyl,Cooley:2022ufh} and PandaX-xT~\cite{PandaX:2024oxq} experiments with expected target masses on the order of 50~tonnes. With argon, the Global Argon Dark Matter Collaboration is currently building the DarkSide-20k~\cite{DarkSide-20k:2017zyg,Zani:2024ybb} experiment with a 20~tonne detection fiducial volume with plans underway to expand in the future to the ARGO project with volumes up to hundreds of tonnes~\cite{Galbiati:2018}. These experiments are expected to probe DM cross sections down to the so-called neutrino floor/fog where solar, supernova, and atmospheric neutrino scattering in the detector will present a challenging background to DM detection~\cite{Billard:2013qya,Ruppin:2014bra,Gaspert:2021gyj,OHare:2021utq,Akerib:2022ort}.

\subsubsection{Electron Scattering}

Dark matter below a GeV is difficult to detect in elastic nuclear scattering due to the mass mismatch with the target; from Eq.~\eqref{eq:er} this implies small energy transfers $E_R\sim m_\chi^2v^2/m_N$  that are hard to observe. Motivated by this challenge, several experiments have been developed to search for sub-GeV DM through its scattering with electrons~\cite{Dedes:2009bk,Kopp:2009et,Essig:2011nj,Graham:2012su,Essig:2015cda}. In this context, the electrons may be bound to individual atoms or delocalized within the target material, allowing for a broad range of kinematic responses to DM scattering~\cite{Lin:2019uvt}.

The total event rate per unit mass of detector material for electron scattering can be written in the general form~\cite{Lin:2019uvt,Trickle:2019nya}
\beq
R = \frac{1}{\rho_T}\lrf{\rho_\chi}{m_\chi}\int\!d^3v\,f_{\rm lab}(\vec{v})\,\Gamma(\vec{v}) \ ,
\eeq
where $\rho_T$ is the target mass density, and 
\beq
\Gamma(\vec{v}) = \frac{\pi\,\bar{\sigma}_e}{\mu_{\chi e}^2}\,\int\!\frac{d^3q}{(2\pi)^3}\,|F_{\rm DM}(\vec{q})|^2\,S(\omega_{\vec{q}},\vec{q}) \ .
\eeq
Here, $\bar{\sigma}_e$ is the DM cross section with a free electron evaluated at the reference scale $\alpha m_e$, $S(\omega_{\vec{q}},\vec{q})$ 
is the material response to momentum transfer $\vec{q}$ and energy $\omega_{\vec{q}} = \vec{q}\cdot\vec{v}-q^2/2m_\chi$, and $F_{\rm DM}(\vec{q})$ is a DM form factor. This latter quantity allows for the possibility that a mediator $A^\prime$ between DM and the electron is light relative to the squared momentum transfer, $m_{A^\prime}^2 \ll \vec{q}^2$. Given the normalization of $\bar{\sigma}_e$ above, we have $F_{\rm DM}(\vec{q}) \to 1$ for a heavy mediator with $m_{A^{\prime}} \gg \alpha\,m_e$, and $F_{\rm DM}(\vec{q}) \to (\alpha\,m_e/q)^2$ for a light mediator with $m_{A^{\prime}} \ll \alpha\,m_e$~\cite{Essig:2011nj}.

\begin{figure}[ttt]
\begin{center}
 \includegraphics[width = 0.8\textwidth]{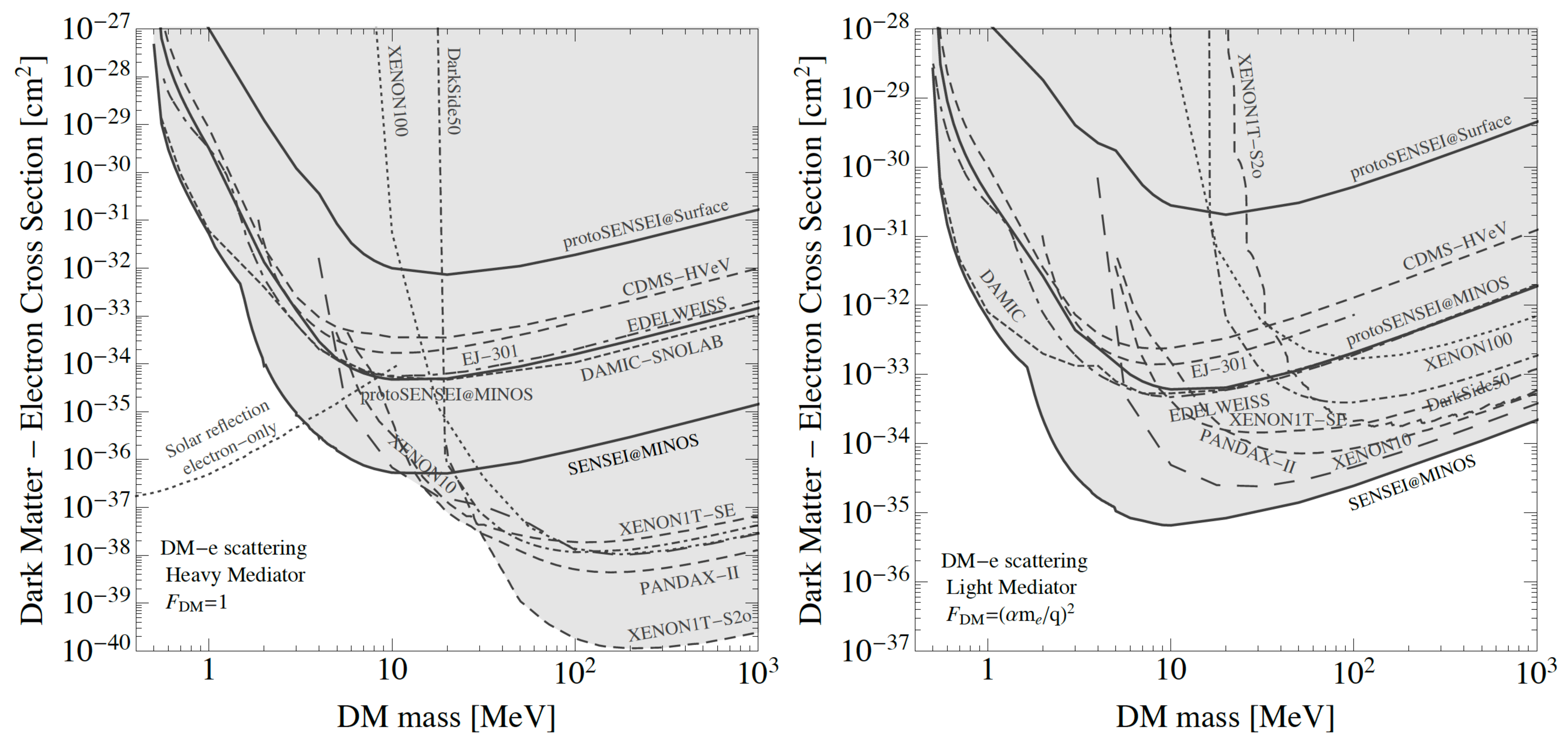}
 \caption{A compilation of recent electron scattering searches for DM masses less than a GeV. Plots taken from Ref.~\cite{Essig:2022dfa}.} 
 \label{fig:elec}
\end{center} 
\end{figure}

Searches to date for DM through electron scattering have been conducted primarily using ionization interactions in noble liquid detectors~\cite{Essig:2012yx,Essig:2017kqs,XENON:2016jmt, XENON:2021qze,Aprile:2019xxb, PandaX-II:2021nsg}, or with recoils in cryogenic semiconductors~\cite{Agnese:2018col,SuperCDMS:2020ymb,DAMIC-M:2023gxo, Aguilar-Arevalo:2019wdi,SENSEI:2020dpa,SENSEI:2023zdf}. In Fig.~\ref{fig:elec} we show a collection of recent bounds on DM electron scattering from Ref.~\cite{Essig:2022dfa}, with the left panel corresponding to a heavy mediator ($F_{\rm DM}=1$) and the right panel to a light mediator ($F_{\rm DM}=(\alpha m_e/q)^2$).  

Future searches based on electron scattering are expected to expand the sensitivity to sub-GeV DM across multiple fronts. In the MeV--GeV mass range, larger and cleaner implementations of cryogenic semiconductors such as SENSEI~\cite{SENSEI:2023zdf}, DAMIC-M~\cite{DAMIC-M:2023gxo}, SuperCDMS~\cite{SuperCDMS:2022kse}, and Oscura~\cite{Oscura:2022vmi} offer a promising and well-defined path for orders of magnitude improvements. The lower-mass sub-MeV mass region is more challenging. Here, recent work has investigated the detection of such very low-mass DM through collective excitations in novel materials due to electron (or nuclear) scattering~\cite{Hochberg:2015pha,Hochberg:2015fth,Hochberg:2016ajh,Schutz:2016tid,Knapen:2016cue,Bunting:2017net,Budnik:2017sbu,Hochberg:2017wce,Knapen:2017xzo,Kahn:2021ttr,Hochberg:2021yud,Essig:2022dfa}.

\subsubsection{Strongly-Interacting Particles}

While numerous cosmological considerations and DM models favor DM that is relatively weakly interacting, with $\sigma_{\chi n} \lesssim 10^{-39}~{\rm cm^2}$, it is also possible that DM is much heavier than the weak scale with a much larger cross section with SM particles~\cite{Carney:2022gse}. 
Mechanisms for the origin of such strongly-interacting particles in the early universe are discussed in Refs.~\cite{Witten:1984rs,Chung:1998rq,Giudice:2000ex,Oaknin:2003uv,Zhitnitsky:2006vt,Khlopov:2007ic,Hardy:2014mqa,DeLuca:2018mzn}. In many cases, these arise as highly composite objects made of many fundamental constituents. 

Very large cross sections can modify or hide the signals of DM at underground experiments. For sufficiently large interaction strengths, galactic DM will be slowed by scattering with the Earth's atmosphere and crust, and will not deposit a detectable amount of recoil energy in underground experiments through single scattering events~\cite{Starkman:1990nj,Erickcek:2007jv,Jacobs:2014yca,Bramante:2018qbc,Bramante:2018tos}. Instead, an ultraheavy, strongly-interacting DM species could scatter multiple times in a detector with very little deflection along its path owing to $m_\chi \gg m_N$. This motivates multi-scattering searches for heavy, strongly-interacting DM~\cite{Bramante:2018qbc,Bramante:2018tos}. 

\begin{figure}[ttt]
\begin{center}
 \includegraphics[width = 0.49\textwidth]{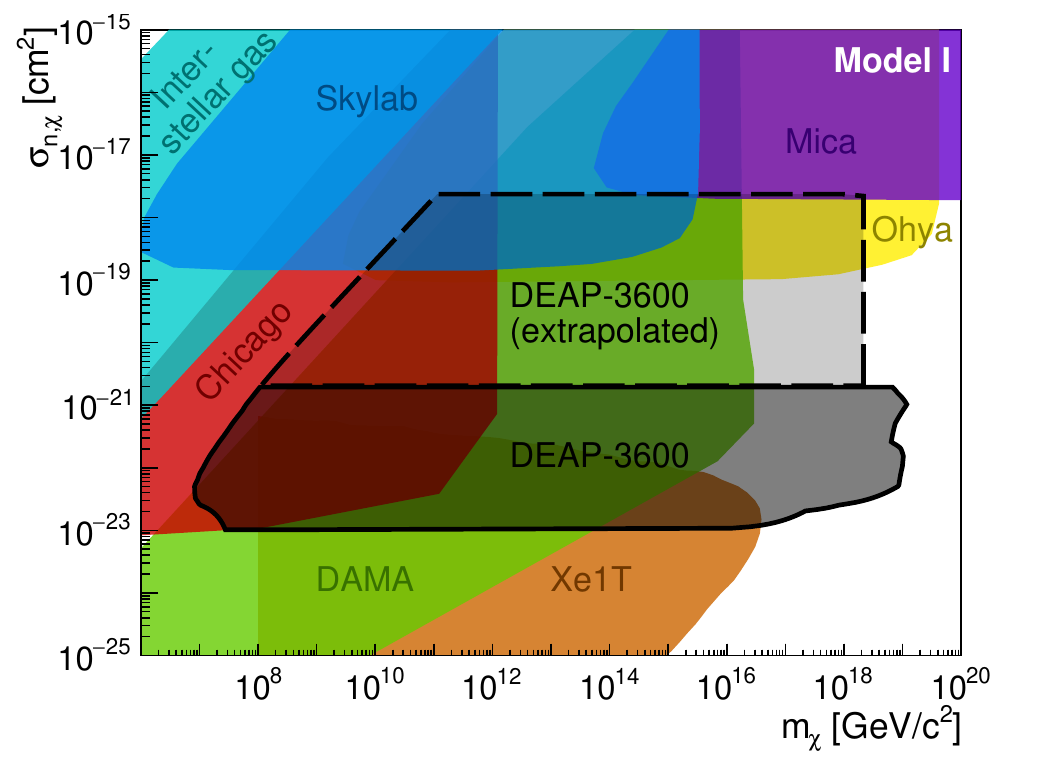}
 \includegraphics[width = 0.49\textwidth]{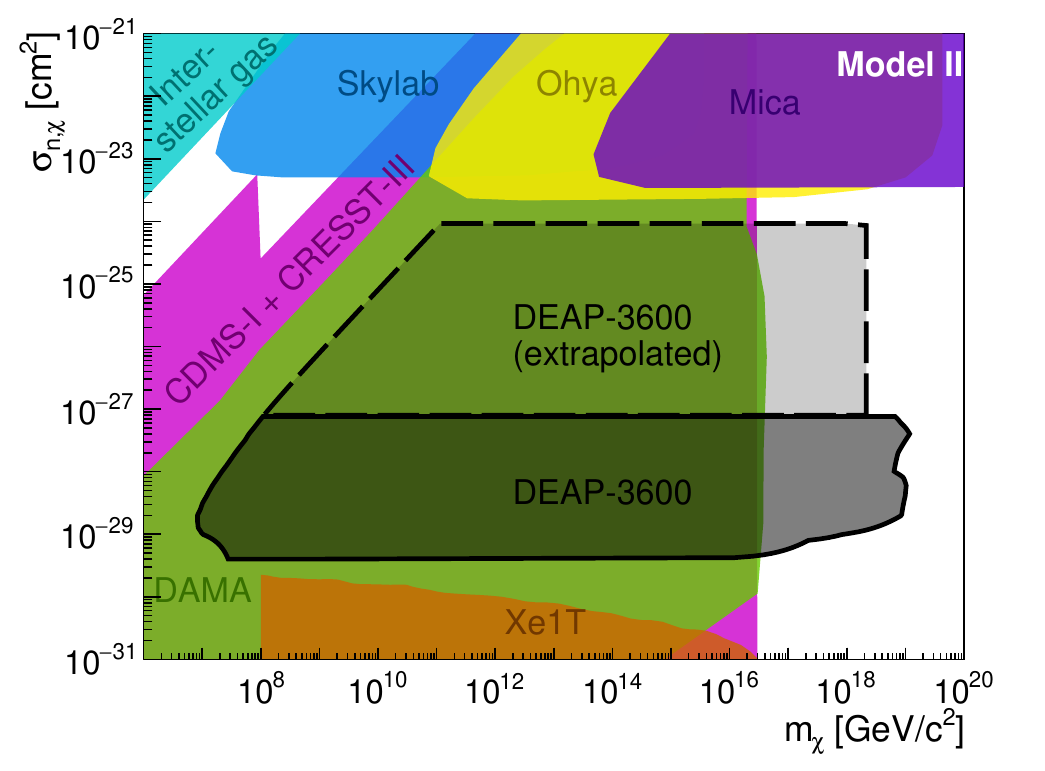}
 \caption{Compilation of bounds on heavy, strongly-interacting DM scattering on nuclei through for two (spin-independent) interaction models discussed in the text. Plots come from Ref.~\cite{DEAPCollaboration:2021raj}.} 
 \label{fig:totsir}
\end{center} 
\end{figure}

Searches for multiscattering and related events from heavy DM have been implemented recently by Majorana~\cite{Clark:2020mna}, DEAP-3600~\cite{DEAPCollaboration:2021raj}, XENON1T~\cite{XENON:2023iku}, and LZ~\cite{LZ:2024psa}. In Fig.~\ref{fig:totsir} we show a collection of bounds on strongly-interacting DM from Ref.~\cite{DEAPCollaboration:2021raj} assuming spin-independent~(SI) interactions. The two panels correspond to two models for translating the per-nucleon cross section to an effective cross section on nuclei. Model I is very conservative and simply identifies $\bar{\sigma}_{\chi N} = \sigma_{\chi n}$ in Eq.~\eqref{eq:sigr}. In Model~II, Eq.~\eqref{eq:sigr} is used but now with $\bar{\sigma}_N$ related to $\sigma_{\chi n}$ as in Eq.~\eqref{eq:si} implying a scaling with $A^4$, which sometimes implies a cross section exceeding nuclear size. If DM is a point particle, there are implications for how realistic point-like DM models are at such a large cross section~\cite{Digman:2019wdm}; on the other hand such large cross sections are obtainable for dark matter composites \cite{Acevedo:2021kly,Acevedo:2020avd}.

\subsubsection{Ultralight Boson Dark Matter}

Dark matter could be made of ultralight bosons ($m \ll \kev$) if they are wave-like in nature, with their energy density dominated by coherent field oscillations or a nearly zero-temperature condensate. As discussed above in Sec.~\ref{sec:candidates}, specific examples include the QCD axion, more general ALPs, ultralight dark photons, and fuzzy DM. These candidates are the targets of dedicated direct search experiments that rely on their specific wave-like properties, but they can produce novel signals in standard direct searches for WIMP-like DM as well.

\begin{figure}[ttt]
\begin{center}
 \includegraphics[width = 0.49\textwidth]{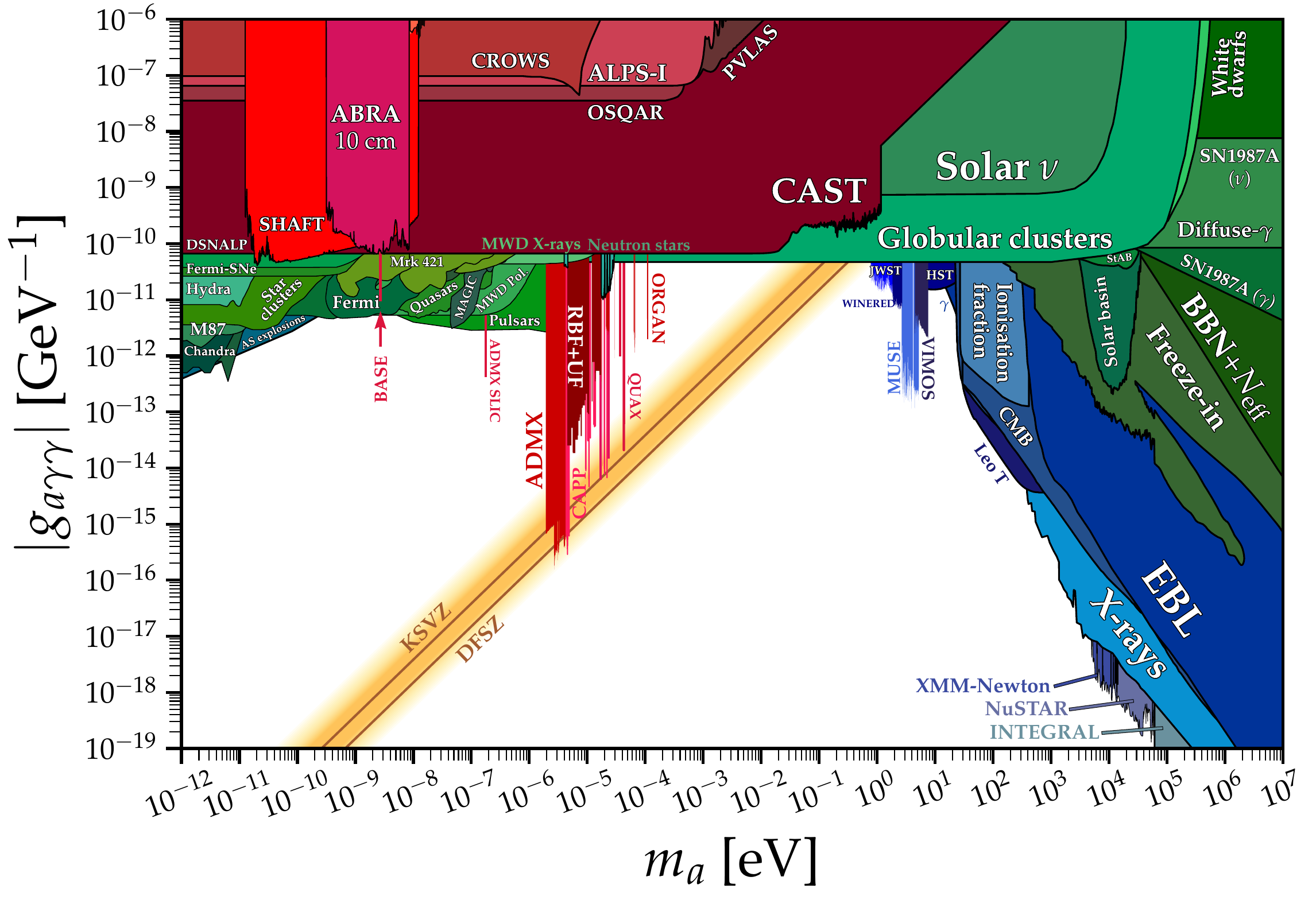}
 \includegraphics[width = 0.49\textwidth]{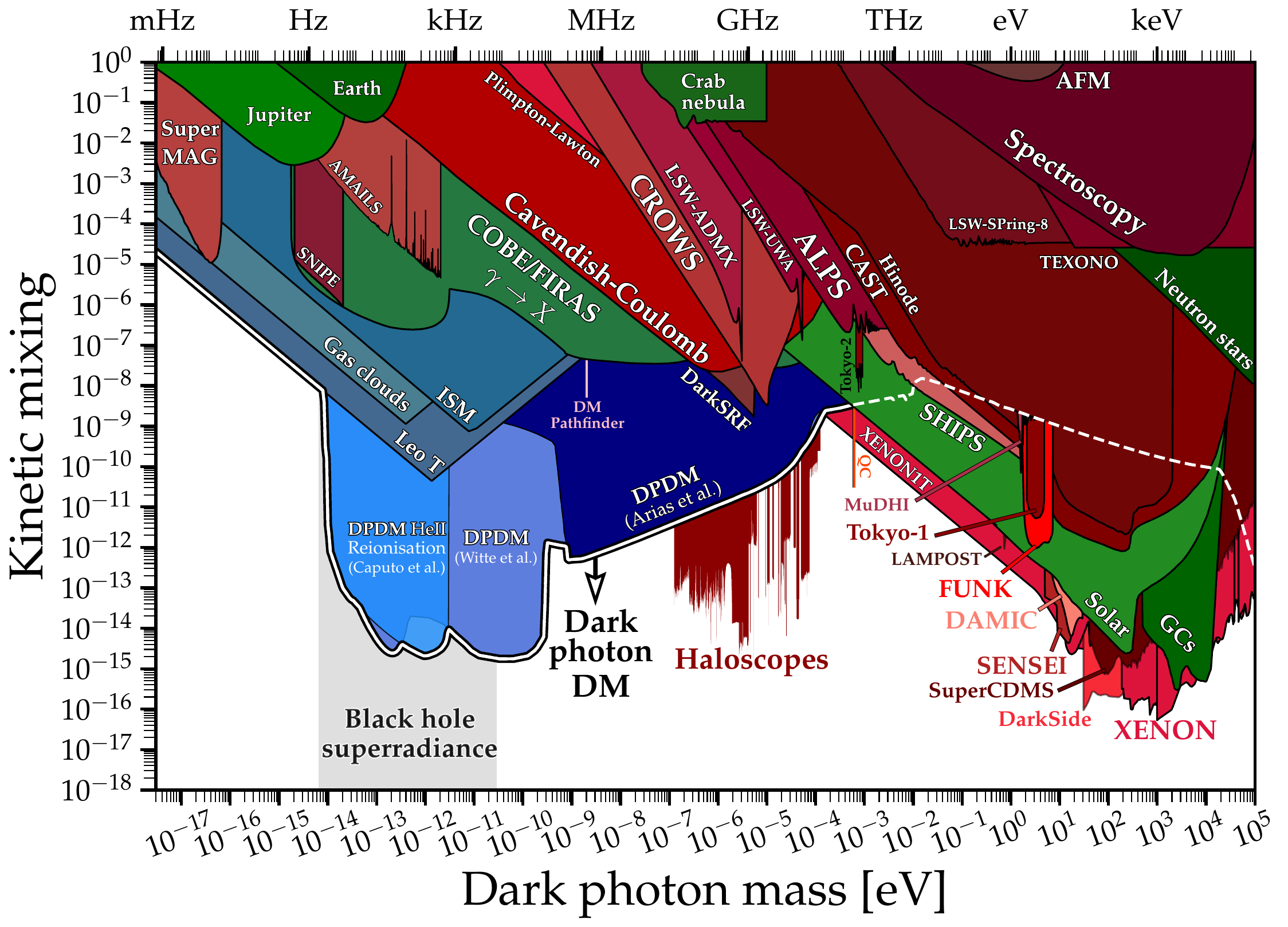}
 \caption{Compilation of laboratory, astrophysical, and cosmological tests of axions and ALPs~(left) and dark photons~(right). Axion bounds are given in terms of their effective coupling to photons $g_{a\gamma\gamma}$ while dark photon limits are given in terms of the kinetic mixing parameter. In both panels, red exclusions denote laboratory searches, green regions are excluded by astrophysics, and blue regions are ruled out by cosmology. Both figures are from Refs.~\cite{Caputo:2021eaa,AxionLimits}.
} 
 \label{fig:ohare}
\end{center} 
\end{figure}

Searches for axions and ALPs typically rely on a coupling to photons of the form of Eq.~\eqref{eq:agg}. This interaction can induce the conversion of an axion particle into an oscillating, low-frequency electric field in a resonant cavity when a transverse magnetic field is applied~\cite{Sikivie:1983ip,Graham:2015ouw,Du:2018uak,Zhong:2018rsr}.  Current bounds on axions and ALPs from such \emph{haloscopes} are shown in the left panel of Fig.~\ref{fig:ohare} together with a range of other indirect bounds~\cite{Caputo:2021eaa,AxionLimits}. Haloscopes can also be used to search for other types of ultralight bosons, such as dark photons~\cite{Caputo:2021eaa,AxionLimits}, shown in the right panel of Fig.~\ref{fig:ohare}. Additional methods for the detection of axion DM are considered in Refs.~\cite{Budker:2013hfa,Graham:2015ouw,TheMADMAXWorkingGroup:2016hpc,Ringwald:2016yge,Ouellet:2018beu,Irastorza:2018dyq,Salemi:2019xgl,Agrawal:2021dbo}. Related methods testing non-photon axion couplings to the SM may be sensitive to other types of ultralight bosons as well~\cite{Safronova:2017xyt,Antypas:2022asj,Chaudhuri:2014dla,Arvanitaki:2016fyj,Arvanitaki:2017nhi,Baryakhtar:2018doz,Knirck:2018ojz}.

Wave-like ultralight boson DM can also appear in direct detection experiments looking for the scattering of particle-like DM. Signals here come from the direct absorption of individual ultralight boson particles extracted from the background population making up the DM~\cite{Pospelov:2008jk,An:2014twa,Bloch:2016sjj,Hochberg:2016sqx}. This is analogous to the photoelectric effect, but with the energy absorbed corresponding to the ultralight boson mass. Searches by XENON1T~\cite{XENON:2019gfn}, XENONnT~\cite{XENON:2022ltv}, LZ~\cite{LZ:2023poo}, SENSEI~\cite{SENSEI:2023zdf}, COSINE-100~\cite{COSINE-100:2023dir} and SuperCDMS~\cite{SuperCDMS:2019jxx} place world-leading bounds on DM ALPs and dark photons over the mass range $m \sim$ eV--MeV.

\subsubsection{Directional Detection}

Directional DM detectors aim to measure both the energy and direction of a recoiling nucleus or electron in an underground detector. Due to the Sun's motion in the Milky Way, the peak DM flux from the Galactic halo is expected to arrive on Earth from the direction of the constellation Cygnus~\cite{Spergel:1987kx}. This directionality of the DM flux  causes a clustering of the nuclear or electron recoils in the detector around the same direction and creates a \emph{dipole feature} in the recoil rate. Detecting the dipole feature is a smoking gun signature for DM since no known backgrounds produce this directional signature. Several directional DM experiments  using different  detection strategies are either currently operating or are planned to start in the future~\cite{Mayet:2016zxu,DRIFT:2016utn,Shimada:2023vky,Santos:2011kf,Baum:2021jak,OHare:2021cgj,Vahsen:2020pzb,Amaro:2022gub}.

In addition to the dipole feature, there are other secondary directional signatures which can be searched for with the next generation of directional detectors. These consist of 
ring-like~\cite{Bozorgnia:2011vc} and aberration~\cite{Bozorgnia:2012eg} features in the directional recoil rate. The ring-like feature occurs for heavy enough DM particles and low enough recoil energies, where instead of the dipole, a ring of maximum recoil rate forms around the average arrival direction of DM particles. Aberration features
are changes in the recoil direction pattern and can provide additional information on the DM velocity distribution. These signatures require more events to be detected compared to the dipole feature.

Directional detection can also provide new insights into the nature of DM. In particular, directional signatures could be used for the determination of the spin of the DM particle~\cite{Catena:2017wzu,Catena:2018uae,Jenks:2022wtj}, or to confirm if DM is inelastic~\cite{Bozorgnia:2016qkh}. For example, the spin of the DM particles can impact the anisotropy patterns in the nuclear recoil directions, and the next generation directional experiments can distinguish between different DM spins~\cite{Catena:2017wzu,Catena:2018uae,Jenks:2022wtj}. On the other hand, for inelastic exothermic DM in which the DM particle down-scatters in mass after the collision with a nucleus, the mean recoil direction and the ring-like feature would be opposite to the average DM arrival direction~\cite{Bozorgnia:2016qkh}. Detecting such a unique directional signature would experimentally confirm this type of DM.

\subsubsection{Dependence of Direct Detection Signals on Halo Astrophysics}

An important point of consideration in the interpretation of the results from direct detection experiments is the accurate modeling of the Milky Way's DM halo. The DM density and velocity distribution in the Solar neighborhood are important inputs in the calculation of the predicted event rates in direct detection experiments. While the local DM density enters in the calculation of direct detection event rates as a normalization constant, the local DM velocity distribution enters in a more complicated way, through an integration over DM velocities. Determining the local DM velocity distribution is therefore important for an accurate interpretation of direct detection results. In particular, searches for inelastic DM can be quite sensitive to the highest velocity component of the DM distribution~\cite{Bramante:2016rdh}, and axion DM searches can be most sensitive to the lowest velocity DM component~\cite{Foster:2017hbq}.

The DM halo has both a smooth well-mixed component and DM substructure. The simplest description for the smooth DM distribution in our Galaxy is the {\it Standard Halo Model} (SHM)~\cite{Drukier:1986tm}. In this model, the DM halo is assumed to be a spherical, self-gravitating system in thermal equilibrium. The DM velocity distribution is an isotropic Maxwell-Boltzmann distribution, with the most probable speed equal to the local circular speed. This distribution is characterized by the local DM density $\rho_{\chi}$ and the velocity dispersion $v_0$, with typical fiducial values taken to be $\rho_{\chi} = 0.3\,\gev\,\mathrm{cm}^{-3}$ and $v_0  = 220\,\mathrm{km}/\mathrm{s}$. The true DM distribution may, however, be different from the SHM. 

Cosmological simulations of galaxy formation provide insight into the DM distribution in the Milky Way galaxy. Early DM-only~(DMO) N-body simulations were performed without baryons, and assumed that all the matter content in the Universe is in the form of collisionless DM. High resolution DMO simulations predict a DM velocity distribution that deviates significantly from the Maxwellian velocity distribution~\cite{Vogelsberger:2008qb, Kuhlen:2009vh}. However, the inclusion of baryons is necessary to make realistic predictions for the DM distribution from cosmological simulations. 

Within the last decade, realistic hydrodynamical simulations that include baryons have become possible. Many up-to-date hydrodynamical simulations have reached significant agreement with observations, and can reproduce important galaxy properties~\cite{Grand:2016mgo, Sawala:2015cdf, Schaye:2014tpa, FIRE2, Stinson:2012uh}. Although there is a large variation in the local DM velocity distribution from different hydrodynamical simulations, and a substantial halo-to-halo variation within a given simulation suite, one can still  find a number of common trends regarding the impact of baryons on the local DM distribution~\cite{Bozorgnia:2017brl}. In particular, studies using state-of-the-art hydrodynamical simulations find that including baryons in the simulations deepens the gravitational potential of the galaxy in the inner halo and shifts the peak of the local DM speed distribution to higher speeds. Baryons also make the halos more isothermal, and the local DM velocity distributions closer to the  Maxwellian functional form~\cite{Bozorgnia:2016ogo, Kelso:2016qqj, Sloane:2016kyi, Poole-McKenzie:2020dbo, Rahimi:2022ymu}. 

In addition to the smooth DM halo, any DM substructure in the Solar neighborhood could also affect the interpretation of direct detection results. Substructure can be in the form of DM subhalos or over-densities in the spatial distribution of DM, or in the form of velocity substructures such as DM streams or debris flows. A number of stellar substructures have been identified in the second data release of the Gaia satellite. The properties of the DM component of such substructures can be inferred using theoretical modelling and cosmological simulations. A recent example is the DM component of the {\it Gaia Sausage-Enceladus}, an anisotropic stellar population identified in Gaia data~\cite{Necib:2018iwb, Fattahi18, Evans:2018bqy, Bozorgnia:2019mjk}. DM from satellite galaxies such as the Large Magellanic Cloud (LMC) and the Sagittarius dwarf spheroidal galaxy can also impact the local DM distribution. 

Among the substructure in our local neighborhood, the LMC in particular has a significant impact on the local DM distribution and direct detection limits~\cite{Smith-Orlik:2023kyl, Besla:2019xbx, Donaldson:2021byu}. Due to its large relative speed with respect to the Sun, the LMC boosts the high speed tail of the local DM velocity distribution, and significantly shifts the direct detection limits towards lower cross sections and smaller DM masses. In particular, the LMC causes the exclusion limits on the DM-nucleon cross section set by a future xenon experiment to shift by more than five orders  of magnitude towards lower cross sections for a DM mass of $m_\chi \sim 5~\gev$~\cite{Smith-Orlik:2023kyl}. Such significant effects should be considered in the analysis of future data from  direct detection experiments. New astrophysical data and future high resolution cosmological simulations can ultimately quantify with high precision the effect of local DM substructure on direct detection results.

\subsection{Indirect Searches}

Indirect searches constitute a very diverse set of astrophysical probes for possible DM signals, spanning events that could originate from very early cosmology to those of the present epoch. These signals can be due to DM annihilation, decay, or accumulation in stars.  The decay or annihilation products might be visible through electromagnetic interactions, via the production of anomalous isotopes in the galaxy, or detectable by high-energy neutrino observatories.  The presence of DM in stars could change their structure or evolution away from that predicted by standard physics.  Nonstandard DM self-interactions or invisible decays could change the details of cosmological structure formation.  We review these various possibilities here.

\subsubsection{BBN, CMB, and the Cosmic Dawn}

In the standard thermal freezeout scenario, DM attains its relic abundance through annihilation into lighter particles, typically in the SM.  Even though after freezeout the rate of such reactions is below the Hubble rate, residual annihilations can lead to observable effects.  If the annihilation products interact electromagnetically, the energy injected into the cosmological plasma can modify the successful predictions for big bang nucleosynthesis~(BBN) or the CMB.  The effects are largest for light DM, since the density scales inversely to the DM mass.  An exception is ADM, which generically evades indirect constraints since annihilations are suppressed by the low abundance of the DM antiparticle.  Another exception is $p$-wave annihilating DM, whose annihilations are suppressed relative to those in the earlier universe at the time of freeze-out.

\begin{figure}[ttt]
\begin{center}
\includegraphics[width = 0.49\textwidth]{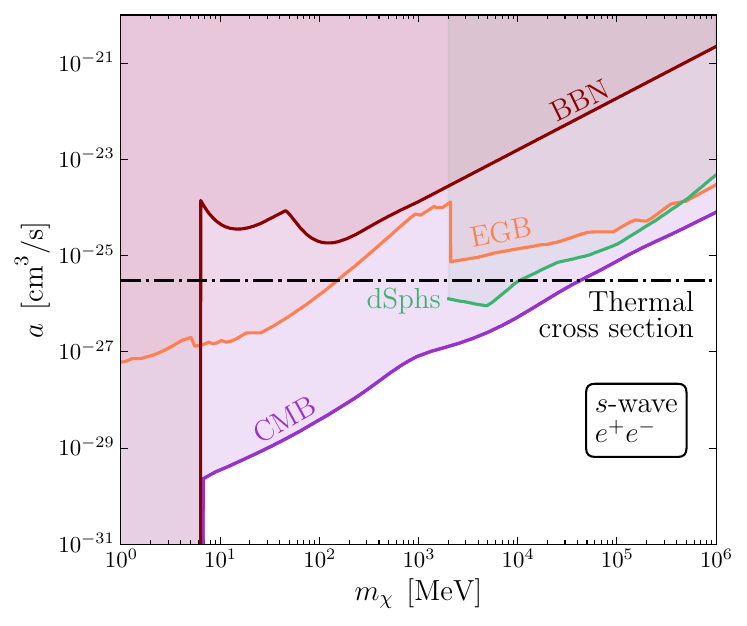}
 \includegraphics[width = 0.49\textwidth]{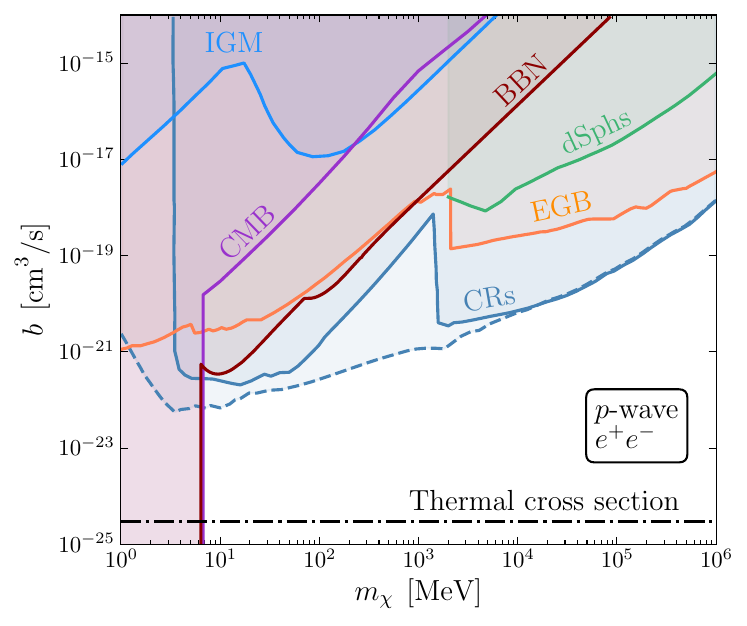}
 \caption{Various indirect bounds on the parameters 
in $\langle\sigma v_{\rm rel}\rangle = a + b\,v_{\rm rel}^2$ for 
DM annihilation  into $e^+e^-$ for $s$-wave~($b$ = 0, left) and 
$p$-wave ($a=0$, right) annihilations.  Figures from Ref.\ \cite{Depta:2019lbe}.} 
 \label{fig:bbn-cmb}
\end{center} 
\end{figure}

The current status of such bounds is summarized for annihilations into $e^+e^-$ in Fig.~\ref{fig:bbn-cmb}, taken from Ref.~\cite{Depta:2019lbe}. For $s$-wave annihilations, the strongest bound comes from the CMB~\cite{Slatyer:2015jla,Liu:2016cnk,Planck:2018nkj}, which limits thermal DM masses to be $\gtrsim 40\,$GeV. These bounds are much weaker for $p$-wave annihilation for masses above about $m_{\chi} \gtrsim 7\,\mev$. For DM masses below this, DM freeze-out completes during or after neutrino decoupling and the annihilation products heat the electromagnetic plasma relative to the neutrinos, thereby modifying the effective number of neutrino species $N_{\rm eff}$~\cite{Boehm:2012gr,Boehm:2013jpa,Nollett:2013pwa,Nollett:2014lwa}. Annihilations into hadrons give stronger BBN constraints than those into $e^+e^-$ \cite{Henning:2012rm,Kawasaki:2015yya}.  Exceptionally, models of fermionic DM with $p$-wave annihilations to light scalars can be strongly constrained by the CMB for parameters such that annihilation into DM bound states (which is $s$-wave) dominates~\cite{An:2016kie}. 

The recent detection of 21 cm absorption from the era of recombination also leads to limits on light $s$-wave DM annihilation \cite{DAmico:2018sxd,Cheung:2018vww,Liu:2018uzy}, which would heat the cosmic hydrogen gas and diminish the absorption trough. These constraints are somewhat stronger than  those from the CMB for certain mass ranges.

It is possible that DM is metastable on time scales longer than the age of the universe.  Analogous limits on the lifetime versus mass can be derived for decays that lead to electromagnetic energy deposition \cite{Diamanti:2013bia,Slatyer:2016qyl}.  The limit is typically of the order $\tau \gtrsim 10^{24}\,$s.

\subsubsection{Gravitational Probes -- Structure Formation}

Even if DM has no interactions with the SM, it may undergo decay into invisible particles, or interact with itself, which can both lead to observable consequences.  Ref.\ \cite{Poulin:2016nat} showed that the fraction $f$ of DM that is decaying times its decay width $\Gamma$ is bounded by $f\,\Gamma < 6\times 10^{-3}$\,Gyr$^{-1}$ from its gravitational effect on the CMB (see also Ref.~\cite{Gong:2008gi}).  Invisible DM decays can affect structure formation, reducing power at small scales and hence power in Lyman-$\alpha$ spectra.  This leads to somewhat weaker constraints, $\Gamma < 0.03\,$Gyr$^{-1}$~\cite{Wang:2013rha}.

Invisible DM decays were recently suggested as a solution to the tension between local measurements of the Hubble constant using supernovae as standard distance candles, versus the Planck determination based on the early expansion of the universe.  Ref.~\cite{Blinov:2020uvz} finds that the tension can be significantly reduced through the decays of a warm DM component with mass $1-40$\,eV, decaying with a lifetime of $10^2-10^4$ yr.

The standard $\Lambda$CDM picture assumes that DM does not interact significantly with itself.  This is borne out by observations of  galactic cluster mergers, most famously the Bullet Cluster, that leads to upper limits on the self-interaction cross section per DM mass, $\sigma/m \lesssim 1$\,cm$^2$/g~\cite{Randall:2007ph}. On the other hand, N-body gravitational simulations of structure formation in $\Lambda$CDM cosmology reveal a number of discrepancies with observations, known as the cusp-core problem, the missing satellites problem, and the too-big-to-fail problem; these have been reviewed in Ref.~\cite{Weinberg:2013aya}.  While it is still controversial whether these discrepancies can be resolved through standard baryonic physics effects and better measurements, it has been shown that DM self-interactions at a level close to the Bullet Cluster bound can ameliorate the problems, as has been comprehensively reviewed in Ref.~\cite{Tulin:2017ara}.

\subsubsection{Gravitational Waves}

The recent renaissance in gravitational wave astronomy \cite{LIGOScientific:2016aoc,KAGRA:2021vkt,EPTA:2023sfo} has brought with it new opportunities to discover the properties of dark matter through gravitational wave signatures. These include dark matter-associated gravitational waves produced in the early universe before recombination, sometimes associated with dark matter production, along with gravitational waves produced by dark matter or dark sectors at low redshift. In the latter scenario, dark matter may be black holes or other compact objects that give off gravitational waves during merger events, or may affect the gravitational wave signatures associated with Standard Model compacts stars and black holes.

Cosmologically sourced gravitational waves, often associated with the production of dark matter, have been studied since the advent of dark matter model building~\cite{Witten:1984rs,Grojean:2006bp,Das:2009ue,Schwaller:2015tja,Croon:2018kqn,Breitbach:2018ddu,Croon:2018erz,Bertone:2019irm,Amin:2019qrx,Croon:2019rqu,Marfatia:2020bcs,Guo:2020grp,Fornal:2020esl,Gavrilik:2020mjy,Fornal:2020ngq}. For example, this includes first order phase transitions~\cite{Schwaller:2015tja}, resonant scalar field dynamics producing gravitational waves associated with q-balls, dark matter mass generation, or ultralight dark matter~\cite{Kusenko:2008zm,Lozanov:2019ylm,Kitajima:2018zco,Bhoonah:2020oov}, and even gravitational waves associated with global symmetry breaking effects~\cite{King:2023ayw,King:2023ztb}. Many of these models predict a spectrum of high-frequency gravitational waves, where the detection of high frequency gravitational waves is an ongoing challenge~\cite{Domcke:2022rgu,Bringmann:2023gba,Aggarwal:2020olq}.

Dark matter that collects inside compact stars and transforms them into black holes would alter the expected population of merging neutron stars and stellar mass black holes, which in turn would change the gravitational wave signatures expected from merging compact objects (\textit{e.g.} a binary neutron star merger would become a black hole binary merger)~\cite{Bramante:2017ulk,Kouvaris:2018wnh,Takhistov:2020vxs,Tsai:2020hpi,Dasgupta:2020mqg,Steigerwald:2022pjo,Bhattacharya:2023stq}. This can occur via heavy asymmetric dark matter~\cite{Bramante:2017ulk,Kouvaris:2018wnh,Dasgupta:2020mqg} or primordial black holes~\cite{Bramante:2017ulk,Takhistov:2020vxs,Tsai:2020hpi} collecting inside and subsequently converting neutron stars into black holes. Separately, the waveforms of neutron star and black hole mergers can be altered by dark sector forces that introduce modifications to the potential between these objects while they merge~\cite{Alexander:2018qzg,Huang:2018pbu}.

Finally, dark matter may include a population of merging compact objects that produce gravitational waves detectable by ongoing or planned detectors. This includes the classic primordial black hole model~\cite{Zeldovich:1967lct,Hawking:1971ei}, whose expected merger rates and associated astrophysical signatures have recently been investigated in Refs.~\cite{Eroshenko:2016hmn,Ali-Haimoud:2017rtz}. It is also possible that dark matter has formed into ``clumps"~\cite{Diamond:2021dth} or dark sector compact stars~\cite{Hippert:2021fch,Hippert:2022snq}, which could yield associated merger gravitational waveform signatures~\cite{Giudice:2016zpa,Diamond:2021dth,Hippert:2022snq}, depending on the details of dark sector compact object formation and low redshift dynamics. The latter is particularly complicated for dissipative DM models like atomic dark matter, but recent simulations have started to enable predictions on the dark compact object distribution in these models~\cite{Roy:2023zar,Gemmell:2023trd}. Such dark compact objects may also be detected via gravitational microlensing~\cite{Winch:2020cju}, and in telescope surveys if they accumulate regular baryonic matter via a dark photon kinetic mixing~\cite{Curtin:2019ngc,Curtin:2019lhm,Howe:2021neq,Armstrong:2023cis}.

\subsubsection{Cosmic Gamma Rays and Radio Emission}

\begin{figure}[ttt]
\begin{center}
\centerline{\includegraphics[scale=0.7]{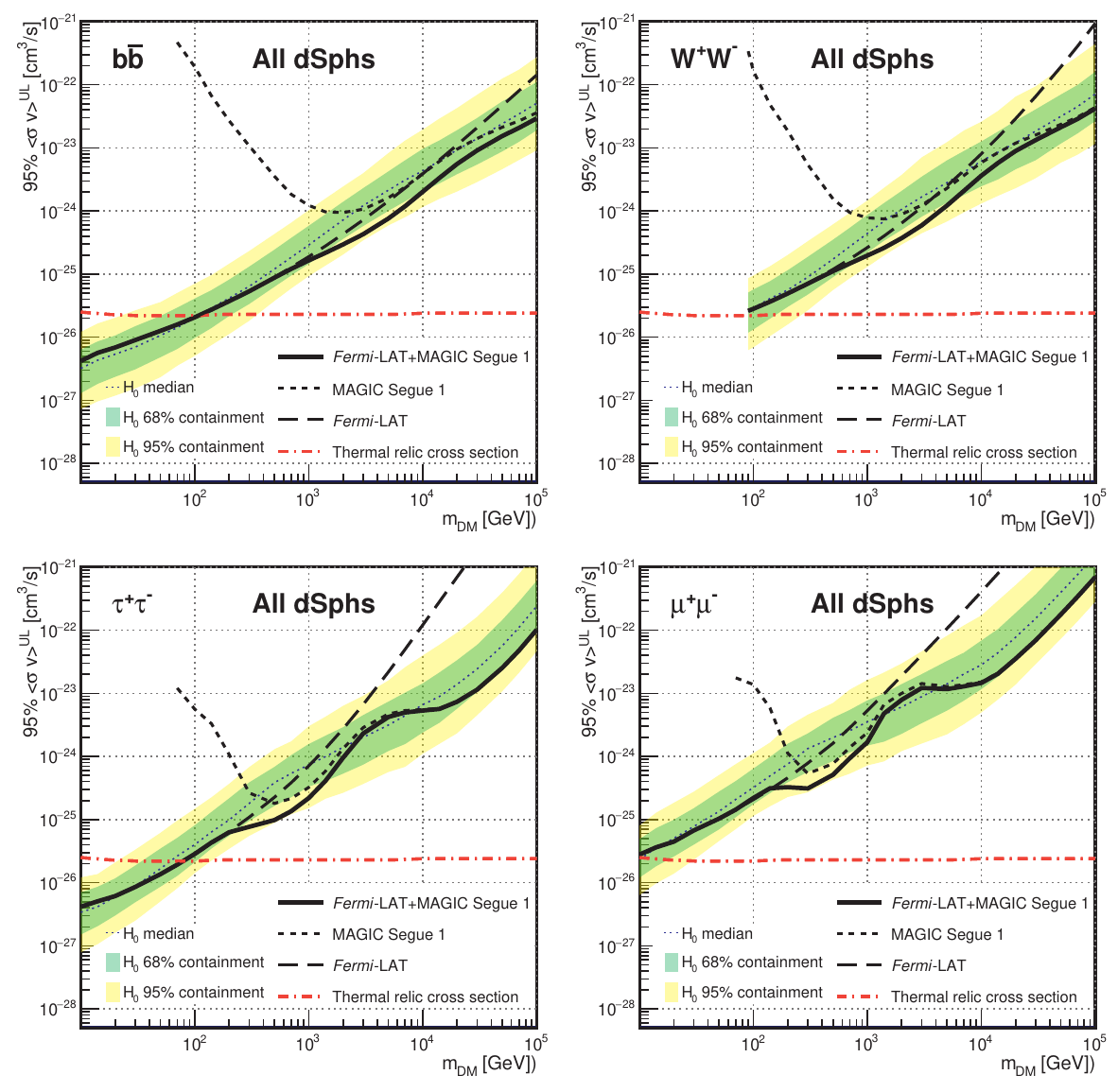}}
 \caption{Combined constraints from Fermi and MAGIC on DM annihilation from dwarf spheroidal galaxies, into $b\bar b$, $W^+W^-$, $\tau^+\tau^-$, $\mu^+\mu^-$ final states. Figures from Ref.~\cite{Ahnen:2016qkx}.}
\label{fig:fermi-magic}
\end{center} 
\end{figure}

Since DM is much more concentrated in galaxies than in the intergalactic medium, its residual annihilations (or decays) will be more easily detectable from our galaxy or others, in particular if the final state particles include or produce photons.  For WIMP DM, these would typically be gamma rays.  They can be detected from space by the Fermi Gamma-ray Space Telescope~\cite{Fermi-LAT:2009ihh,Hoof:2018hyn}, or from the ground by Cherenkov detectors such as HAWC~\cite{HAWC:2017mfa}, VERITAS~\cite{VERITAS:2017tif}, HESS~\cite{HESS:2018cbt}, MAGIC~\cite{MAGIC:2021mog}, and in the future by  CTA~\cite{CTAConsortium:2017dvg}.  In addition, charged particles emitted can produce synchrotron radiation that may be observed by radio telescopes.

In setting limits on DM annihilation or decay from such sources, the important uncertainties are the detailed density profile of the DM in the system of interest in the case of $s$-wave annihilation, and the modeling of the backgrounds from conventional astrophysical sources.  Dwarf spheroidal satellite galaxies of the Milky Way have been favored targets of observation because of their low baryon content, being DM dominated.  For these systems, the systematic uncertainty remains of how cuspy their DM density profiles are, which affects the predicted flux from DM annihilations. The limits also depend upon the assumed annihilation channels. Fig.~\ref{fig:fermi-magic} shows combined limits from Fermi and MAGIC on DM annihilation from 15 satellite galaxies (observed by Fermi) plus Segue~1 (observed by MAGIC) for several different annihilation channels~\cite{Ahnen:2016qkx}. DM annihilation directly into photons gives stronger limits, $\langle\sigma v\rangle \lesssim 3\times 10^{-29}$\,cm$^3$/s\;$(m_{\chi}/\gev)$~\cite{Ackermann:2015lka} assuming a standard NFW profile.  This work also studied the effect of varying the assumed DM halo profiles, leading to a variation of more than two orders of magnitude depending on the cuspiness.

For velocity-dependent DM annihilation models, such as $p$-wave, $d$-wave, and Sommerfeld-enhanced annihilation, the full DM velocity distribution of the source enters in the calculations of the annihilation signals. Usually simplified models for the DM velocity distribution is used to set  limits on the velocity-dependent cross section using the kinematic data from dwarf spheroidal galaxies. Recent studies using cosmological simulations of galaxy formation provide more accurate estimates of the DM velocity distribution and the expected velocity-dependent DM annihilation signals~\cite{Vienneau:2024xie, Piccirillo:2022qet, Blanchette:2022hir, Board:2021bwj}.

In addition to upper limits, there is evidence for excess gamma rays from the galactic center and possibly from neighboring M31 and Sagittarius dwarf spheroidal galaxy, with energies at the GeV scale \cite{Hooper:2011ti,Abazajian:2012pn,Calore:2014xka,TheFermi-LAT:2017vmf, Ackermann:2017nya,DiMauro:2021raz, Fermi-LAT:2022byn, Evans:2022zno}.  Although the excess itself is statistically quite significant, it is controversial as to whether the signal is better explained by DM annihilations or emission from a population of unresolved millisecond pulsars \cite{Gordon:2013vta,Bartels:2015aea}. Recently it has been argued that the statistical properties of the excess photons are more consistent with a point source origin (pulsars) than a smooth one (DM)~\cite{2011ApJ...738..181M,Lee:2014mza,Buschmann:2020adf}.  However this has been challenged in Refs.\ \cite{Leane:2019xiy,Leane:2020nmi,Leane:2020pfc,Cholis:2021rpp}, leaving the door open for the DM interpretation.  An example of a model that fits the excess while being consistent with dwarf constraints is DM in the mass range $40-60$\,GeV, annihilating to $b\bar b$ \cite{Ipek:2014gua,Cholis:2019ejx}.

\subsubsection{Cosmic Ray Leptons and Baryons}

Unlike photons, that point back to their sources, charged particles in cosmic rays are deflected by the galactic magnetic field. Nevertheless, their spectra can be predicted and deviations could indicate an exotic source from DM annihilations.  One such deviation, an excess of positrons versus electrons observed by the PAMELA experiment~\cite{Adriani:2008zr}, sparked intense interest in DM indirect detection via cosmic rays and led to an enormous number of theoretical explanations based on DM annihilation~\cite{Cirelli:2008pk,ArkaniHamed:2008qn} or decay~\cite{Yin:2008bs,Ibarra:2008jk,Chen:2008qs,Nardi:2008ix}.  While this excess has since been confirmed by Fermi-LAT~\cite{FermiLAT:2011ab} and AMS-02~\cite{Accardo:2014lma,Aguilar:2019owu}, DM explanations for it have become very challenging as theoretical predictions and observations from other experiments have improved~\cite{Meade:2009iu}, making proposed astrophysical origins such as pulsars more plausible~\cite{Hooper:2008kg,Yuksel:2008rf,Profumo:2008ms}.

Previously mentioned ground-based Cherenkov detectors have sensitivity to cosmic ray electrons, and space-borne instruments including Fermi-LAT~(Large Area Telescope)~\cite{Fermi-LAT:2009ihh}, AMS-02~\cite{AMS:2016oqu,AMS:2021nhj}, CALET~\cite{CALET:2017uxd,CALET:2022vro}, and DAMPE~\cite{DAMPE:2017cev}. The spectrum has been measured up to $E\sim 5$ TeV energies by HESS and DAMPE~\cite{Ambrosi:2017wek}.  The spectrum shows a break from single power-law behavior near 2.6\,TeV; this could be understood as coming from the nearby Vela pulsar~\cite{TheDAMPE:2017dtc}. The DAMPE spectrum has one anomalously high bin near $1.6\,$TeV that has prompted some DM enthusiasts to fit it to models of DM with this mass annihilating (or decaying) into leptons.  However the best-fit models require cross sections or decay rates that exceed bounds from the CMB or extragalactic background radiation~\cite{Yuan:2017ysv}, making the DM interpretation unlikely. 

The AMS-02 experiment on the International Space Station has the capability of measuring cosmic ray nuclei as well, which affords another probe of DM annihilation effects.  One measurement that has garnered attention is the antiproton spectrum \cite{AMS:2016oqu,AMS:2021nhj}, which was argued by several groups to have an excess over theoretical expectations in the range of $10-20$\,GeV~\cite{Cuoco:2016eej,Cui:2016ppb}, that could be explained by the same annihilating DM as needed for the GeV gamma ray excess~\cite{Cholis:2019ejx}. AMS-02 is also measuring heavier cosmic ray nuclei, and would be able to discriminate antinucleons, that could be produced infrequently in cosmic ray collisions or DM annihilations. The antideuteron $\bar d$ is of interest, being easier to produce than more complex antinuclei.  If enough $\bar d$'s could be observed to produce a spectrum, its shape would discriminate between a cosmic ray versus a DM origin \cite{PhysRevD.62.043003}.

\subsubsection{High Energy Neutrinos}

Dark matter annihilation to SM particles other than $e^\pm, \gamma$ necessarily yield a non-negligible neutrino flux. Unlike gamma rays and cosmic rays, these travel unimpeded through the interstellar/galactic media, pointing straight back towards their source. Unfortunately, their low interaction cross section also makes them hard to detect: in general, constraints from neutrinos on annihilation to charged particles tend to be the weakest. A recent combined analysis of DM annihilation in the Galactic Centre by ANTARES and IceCube~\cite{ANTARES:2020leh} puts limits on $\sv$ derived from neutrino fluxes consistently about two orders of magnitude weaker than gamma ray limits from dwarf spheroidal galaxies (Fermi-LAT+MAGIC~\cite{Ahnen:2016qkx}) and the galactic centre (HESS~\cite{Abdallah:2016ygi}) for DM masses between 50 GeV and 1 TeV. 

\begin{figure}[ttt]
    \centering
    \includegraphics[width = 0.7\textwidth]{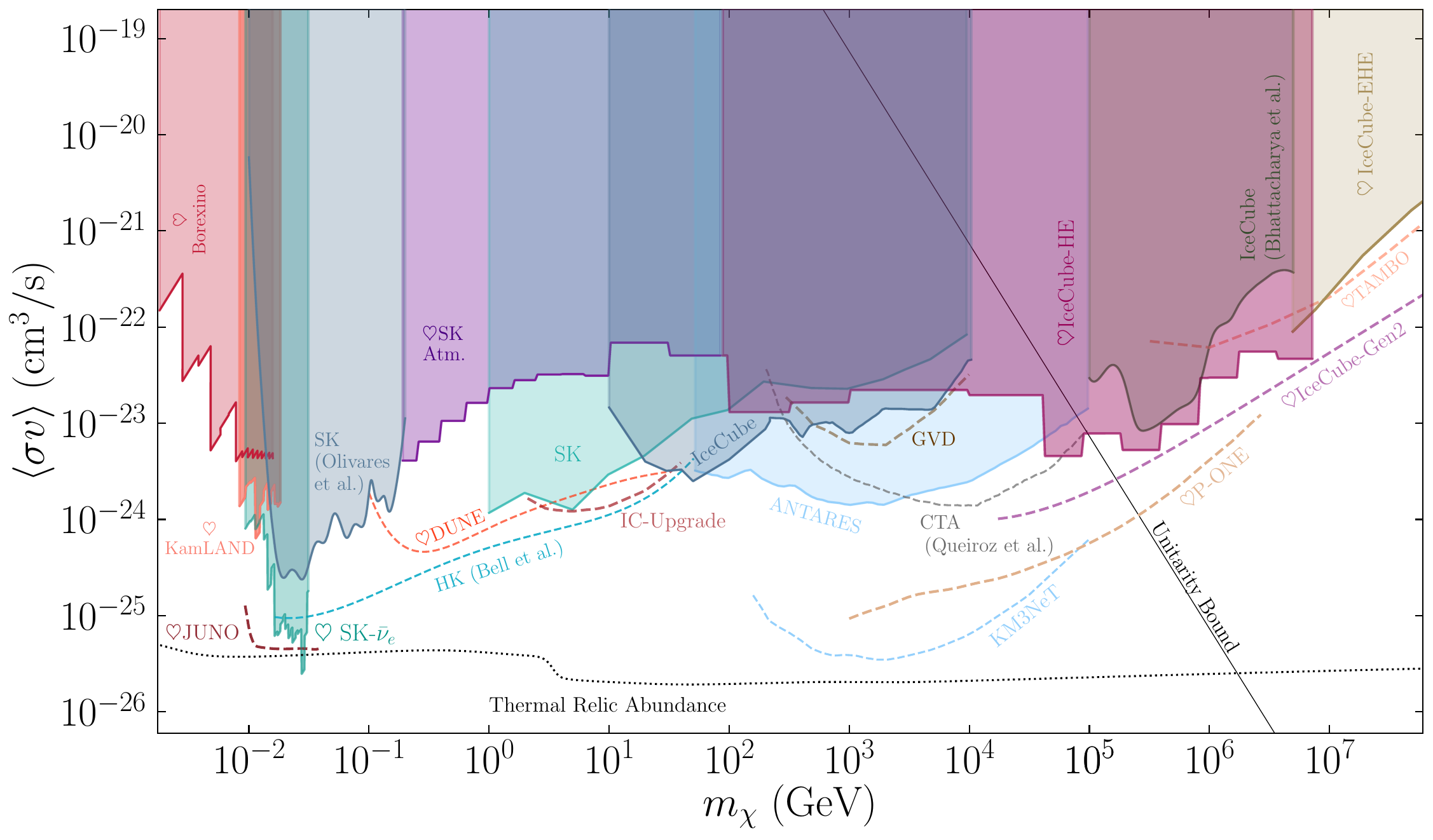}
    \caption{Limits on the DM annihilation cross section to neutrino-antineutrino pairs as a function of the DM mass $m_\chi$. Shaded regions correspond to current bounds, while dashed lines are projected sensitivities of future experiments. Figure (and references for each curve) from Ref. \cite{Arguelles:2019ouk}.}
    \label{fig:nunubarlimits}
\end{figure}

Dark matter could annihilate in part or completely to neutrino pairs. Since the $ \nu \bar \nu$ channel is the most difficult to constrain, a limit on annihilation to neutrinos provides a rough upper limit on the total annihilation rate to the SM~\cite{Beacom:2006tt,Yuksel:2007ac}. In Fig.~\ref{fig:nunubarlimits} we show the upper limit on DM annihilation to $ \nu \bar \nu$ pairs from a full set of experiments and analyses, from Ref.~\cite{Arguelles:2019ouk}. Interestingly, these constraints remain rather flat over a mass range spanning 15 orders of magnitude (see Fig.~4 of Ref.~\cite{Arguelles:2019ouk} for constraints that reach masses of $m_\chi \sim 10^{12}\,\gev$). This is due in part to the growing neutrino-nucleus cross section with $E_\nu$ (helping detection rates), and in part to the larger effective areas of high energy neutrino detectors, which are necessary because of the very low expected fluxes, $\phi \lesssim E_\nu^{-2}$, from conventional astrophysical sources. Limits on the lifetime of decaying DM can be obtained in a similar manner. IceCube~\cite{Aartsen:2018mxl,IceCube:2022clp} has placed an upper limit on the decay of DM particles between $10^{28}-10^{30}$ seconds, depending on the decay channel and the DM mass.

High-energy neutrinos can provide additional signatures of particle DM. Elastic scattering interactions between DM and neutrinos in the early universe can lead to changes in the matter power spectrum due to diffusion damping \cite{Boehm:2000gq,Escudero:2015yka,Stadler:2019dii}. Interestingly, introducing a similar DM-neutrino interaction for only a fraction $\sim 10^{-2}$ of the DM can yield a shift in the CMB acoustic peaks, possibly reconciling the $H_0$ tension between early (CMB) and late time (SNe Ia) observables. For light enough DM, scattering can also be observed via its effects on high-energy extragalactic neutrinos propagating through the Milky Way's DM halo. In this case, the absence of a significant deficit in the high energy neutrino flux observed at IceCube allows bounds to be placed on the interaction rate \cite{Arguelles:2017atb,Cline:2022qld,Ferrer:2022kei}. 

At higher energies still, recent observations by ANITA, a balloon-borne detector designed to detect ultra high-energy (UHE) neutrinos via the Askaryan effect in ice, indicate a small number of UHE events $\sim 0.5$ EeV~\cite{Gorham:2016zah,Gorham:2018ydl}. The observed upward direction of these events preclude a SM explanation, as the Earth is opaque to neutrinos at these energies. Many beyond Standard Model (BSM) explanations have been put forth, but most would lead to additional low-energy or high-zenith signals at IceCube. Nonetheless, it seems that decaying heavy DM could provide such a signal if the decay product can interact in the Earth \cite{Cline:2019snp,Hooper:2019ytr}. Given their sensitivity to the highest energy neutrinos, observatories such as ANITA and planned in-ice radio detectors like ARA and ARIANNA could be the best way to get at superheavy DM candidates \cite{Dudas:2020sbq}.

\subsubsection{The Sun and other Main Sequence Stars}

Scattering between DM and SM matter can lead to sizeable populations of DM accumulating in main sequence stars such as the Sun~\cite{Gould:1987ju}. If a scattering event brings the DM below the local escape velocity, the DM will become gravitationally bound. The population of DM in a star is set by \cite{Jungman:1995df}:
\begin{equation}
    \frac{dN(t)}{dt} = C(t) -2A(t) -E(t),
    \label{eq:dmpopsun}
\end{equation}
where $C(t)$ is the capture rate, $A(t) \propto N^2(t)$ is the self-annihilation rate and $E(t) \propto N(t)$ is the evaporation rate, negligible below DM masses of $m_{\chi} \sim 4$ GeV \cite{Gould:1987ju,Busoni:2017mhe}. If the annihilation cross section is large enough, Eq.~\eqref{eq:dmpopsun} will lead to a steady state solution, with the total annihilation rate set by the capture rate -- and thus proportional to the DM-matter elastic scattering cross section. For typical WIMP self-annihilation cross sections, the steady state population in the Sun will be small enough that electromagnetic annihilation products will not contribute in any detectable way to the Sun's heat budget. However, annihilation to any final state other than electrons or photons will lead to a substantial neutrino flux, with energies $E_\nu \sim m_{\chi}$. A GeV-scale neutrino flux from the Sun, detectable in experiments such as SuperKamiokande, IceCube and Antares, is thus considered a \emph{smoking gun} signal of WIMP-like DM. 

\begin{figure}[ttt]
    \centering
\includegraphics[width=0.9\textwidth]{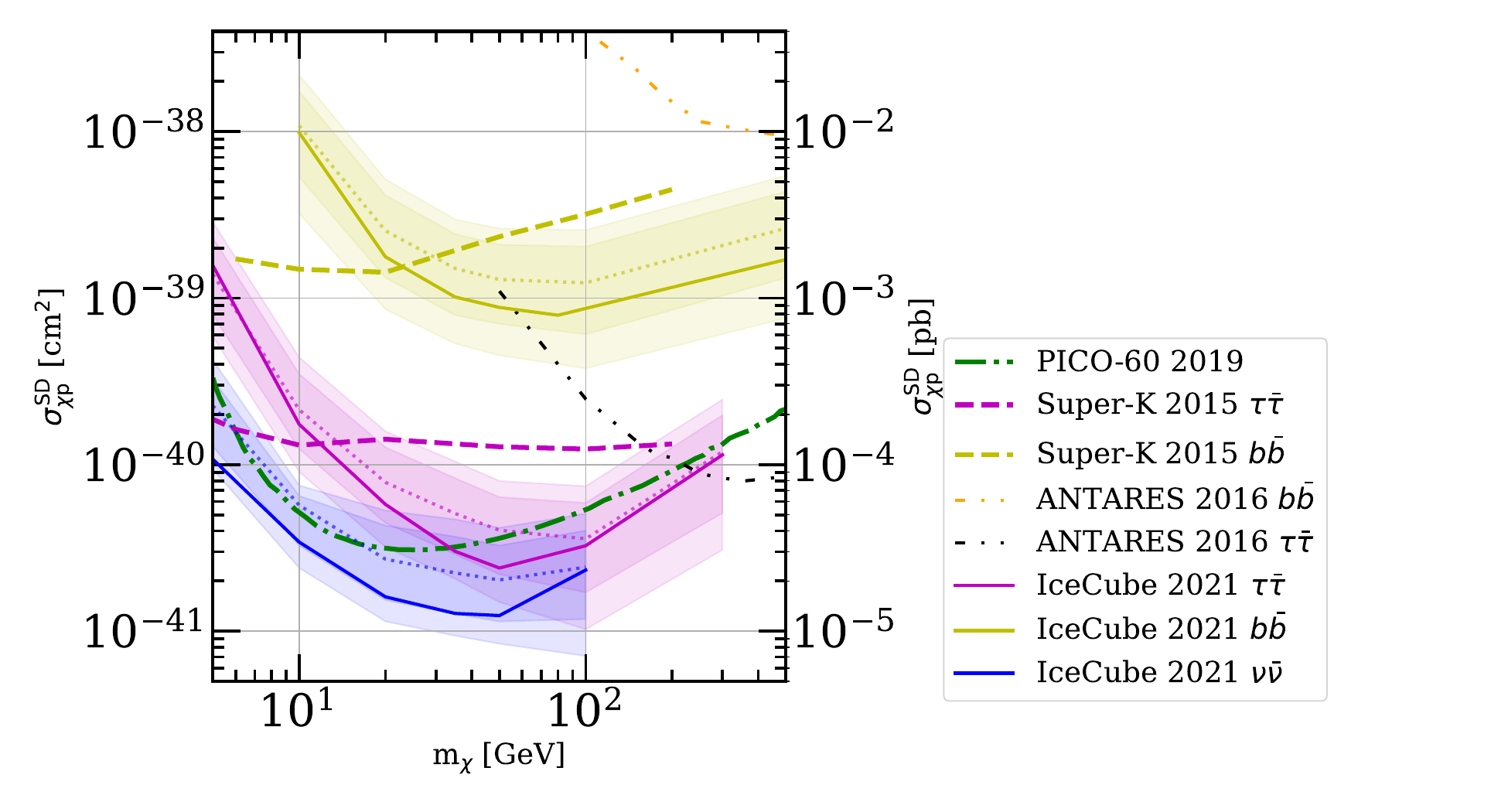}
    \caption{Upper bounds on the spin-dependent DM-proton elastic scattering cross section, $\sigma_{\chi p}^{SD}$, 
    from DM capture and annihilation in the Sun from IceCube~\cite{Choi:2015ara,IceCube:2021xzo}, Super Kamiokande~\cite{Choi:2015ara}, and ANTARES~\cite{Adrian-Martinez:2016gti} for various annihilation channels. Also shown are the leading bounds from direct detection from PICO~\cite{Amole:2019fdf}. Figure from Ref.~\cite{IceCube:2021xzo}.}
    \label{fig:SDneutrino}
\end{figure}

Because the steady state solution is set by the elastic scattering cross section, bounds on neutrino fluxes can be directly compared with the results of direct detection experiments. Since the Sun is composed mainly of hydrogen, it is a particularly sensitive probe of SD scattering. At present, some of the strongest bounds on SD DM-nucleon scattering are from the search for annihilation products from neutrinos by SuperKamiokande~\cite{Choi:2015ara}, ANTARES~\cite{Adrian-Martinez:2016gti}, and IceCube~\cite{Aartsen:2016zhm,IceCube:2021xzo}, depending on the annihilation channel. Current constraints on the SD DM-proton scattering cross section from this indirect search approach are shown in Fig.~\ref{fig:SDneutrino} and compared to the leading limits from direct detection from PICO-60~\cite{Amole:2019fdf}. It should be noted that even though they are present in very small quantities ($\ll 2\%$), heavier elements can contribute significantly to the total capture rate. This is especially true for certain DM-nucleon operators \cite{Catena:2015uha}. 

In some DM models, such as ADM, the annihilation rate can be suppressed such that sizeable populations (up to $10^{-10} {\rm M}_\odot$) can accumulate in the Sun. In this case, the DM can act as an efficient heat conductor, as its mean free path can be quite large~\cite{Press:1985ug,Nauenberg:1986em,Gould:1989hm}. This can lead to hydrostatic equilibrium in the Sun with a shallower thermal gradient and a lower central temperature (without affecting the total radius and $pp$ luminosity), leading to lower $^8$B and $^7$Be neutrino fluxes. Changes in stellar structures can also be sensitively measured by helioseismology. Interestingly, changes by certain models of DM to the pressure and sound speed profiles can alleviate the strong and yet-unexplained tension between standard solar models and helioseismologically-measured sound speed profiles, small frequency separations and convective zone boundary~\cite{Vincent:2014jia,Vincent:2015gqa,Vincent:2016dcp}.

The same mechanism can lead to observable changes in other main sequence stars. Near the Galactic Centre, where the DM density is expected to be much higher, annihilation and heat conduction by captured DM in stars can lead to changes in the conditions for thermal and hydrostatic equilibrium, affecting their evolutionary tracks on the Hertzprung-Russel~(HR) diagram~\cite{Scott:2008ns,Zentner:2011wx,Iocco:2012wk}. In stars with slightly higher masses than the Sun, even modest amounts of ADM can suppress the formation of convective cores, an effect that would be visible via astroseismology in a few hundred nearby stars from sensitive telescopes such as Kepler~\cite{Casanellas:2012jp,Casanellas:2015uga,Brandao:2015lra}.

\subsubsection{Neutron Stars and White Dwarfs}

\begin{figure}[ttt]
\begin{center}
\includegraphics[width=0.42 \textwidth]{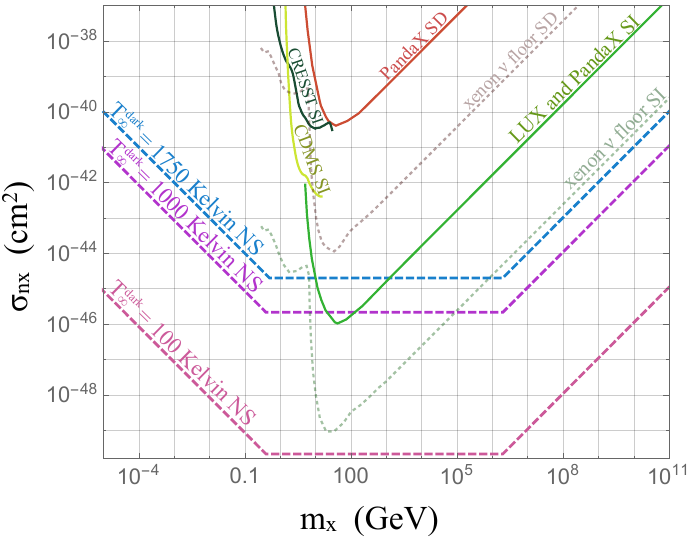}
~~\includegraphics[width=0.51 \textwidth]{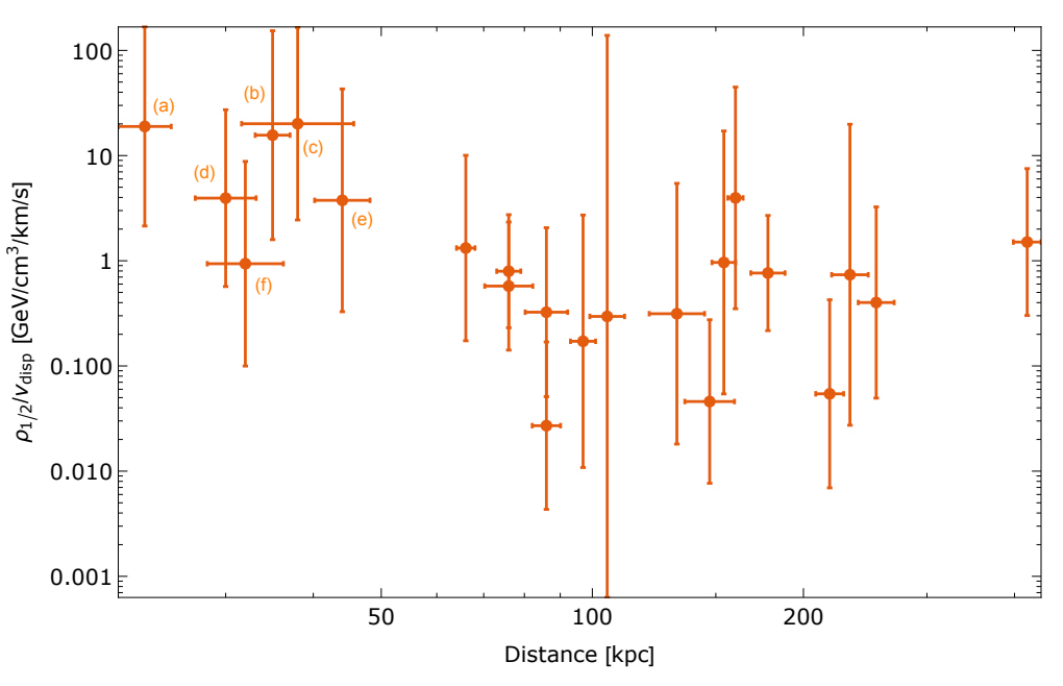}
\caption{Left: Projected limits on the DM-neutron scattering cross section from kinetic heating of neutron stars by DM scattering,
taken from Ref.\ \cite{Baryakhtar:2017dbj}. Right: Potential dwarf galaxy targets for DM searches using white dwarf temperatures, where dwarf galaxies in the upper left are closest to Earth ($x$-axis) with high DM density normalized to velocity dispersion ($y$-axis). The most suitable targets include (a) Segue I, (b) Segue II, (c) Willman 1, (d) Reticulum II, (e) Coma Berenices, (f) Ursa Major II as studied in
Ref.\ \cite{Krall:2017xij} from which the figure was obtained.
}
\label{fig:nskh}
\end{center} 
\end{figure}

Dark matter accumulating in cold astrophysical objects like neutron stars and white dwarfs can give potentially observable effects by heating them via DM annihilation, leading to surface temperatures higher than expected from standard stellar cooling~\cite{Bertone:2007ae,Kouvaris:2007ay,Kouvaris:2010vv,deLavallaz:2010wp,McCullough:2010ai,Bramante:2017xlb}, especially in the region of the galactic center where higher DM densities would produce the largest effects. In the case of neutron stars, projected excess temperatures from DM heating remain below current observational sensitivities. However, upcoming instruments including the James Webb Space Telescope~\cite{Gardner:2006ky}, the Thirty Meter Telescope~\cite{Skidmore:2015lga}, or the European Extremely Large Telescope~\cite{neichel2018overview} are expected to make such measurements.  Even in the absence of annihilation, kinetic heating can occur by DM-nucleon or DM-electron scattering~\cite{Baryakhtar:2017dbj,Raj:2017wrv,Bell:2018pkk,Bell:2019pyc,Acevedo:2019agu,Joglekar:2019vzy,Joglekar:2020liw,Anzuini:2021lnv}.  The projected constraints, shown in the left panel of Fig.~\ref{fig:nskh}, can surpass those from direct detection, and especially in theories of DM where the signal in direct searches is suppressed such as sub-GeV DM~\cite{Dasgupta:2020dik} or DM connecting to the SM through a pseudoscalar mediator~\cite{Coffey:2022eav}.

On the other hand, observations already exist of temperatures of white dwarfs in the globular cluster M4, deduced from their luminosities~\cite{McCullough:2010ai,Dasgupta:2019juq}.  In principle, these can give very strong limits on DM-nucleon scattering, improving on direct limits over a wide range of DM masses.  However such constraints depend strongly upon the assumed ambient DM density $\rho_{\chi}$ in the vicinity of the white dwarf. 
For atomic dark matter, Ref.~\cite{Curtin:2020tkm} shows that white dwarf cooling constraints supply the strongest bounds even for highly  subdominant atomic dark matter mass fractions, in either a halo or disk configuration in our galaxy. 
%
For WIMP-like dark matter, the bounds presented in Refs.~\cite{McCullough:2010ai,Dasgupta:2019juq} are based on heating white dwarfs above $T \sim 3000$ Kelvin, which requires that white dwarfs to capture most of the DM flux while sitting in a background DM density of $\rho_{\chi} \sim 10^3 \,$GeV/cm$^3$. Hence if the DM density is much lower than $10^3 \,$GeV/cm$^3$ in M4, like in the solar neighborhood where it is $\rho_\chi \sim 0.3\,\gev/\text{cm}^3$, these M4 white dwarf observations would provide no constraint.  Ref.~\cite{Krall:2017xij} identifies several nearby dwarf galaxies as promising targets for future white dwarf searches, that could yield limits that are less subject to the uncertainty in $\rho_{\chi}$. These dwarf galaxies have more reliably measured DM densities, shown in the right panel of Fig.~\ref{fig:nskh}. In this respect, future constraints from neutron stars are promising, since even in an environment with $\rho_{\chi}=0.3\,\gev/\text{cm}^3$, the maximum saturated temperatures can be well above the expected values, making a discovery possible.

\begin{figure}[ttt]
\begin{center}
 \centerline{
   \includegraphics[width=0.49 \textwidth]{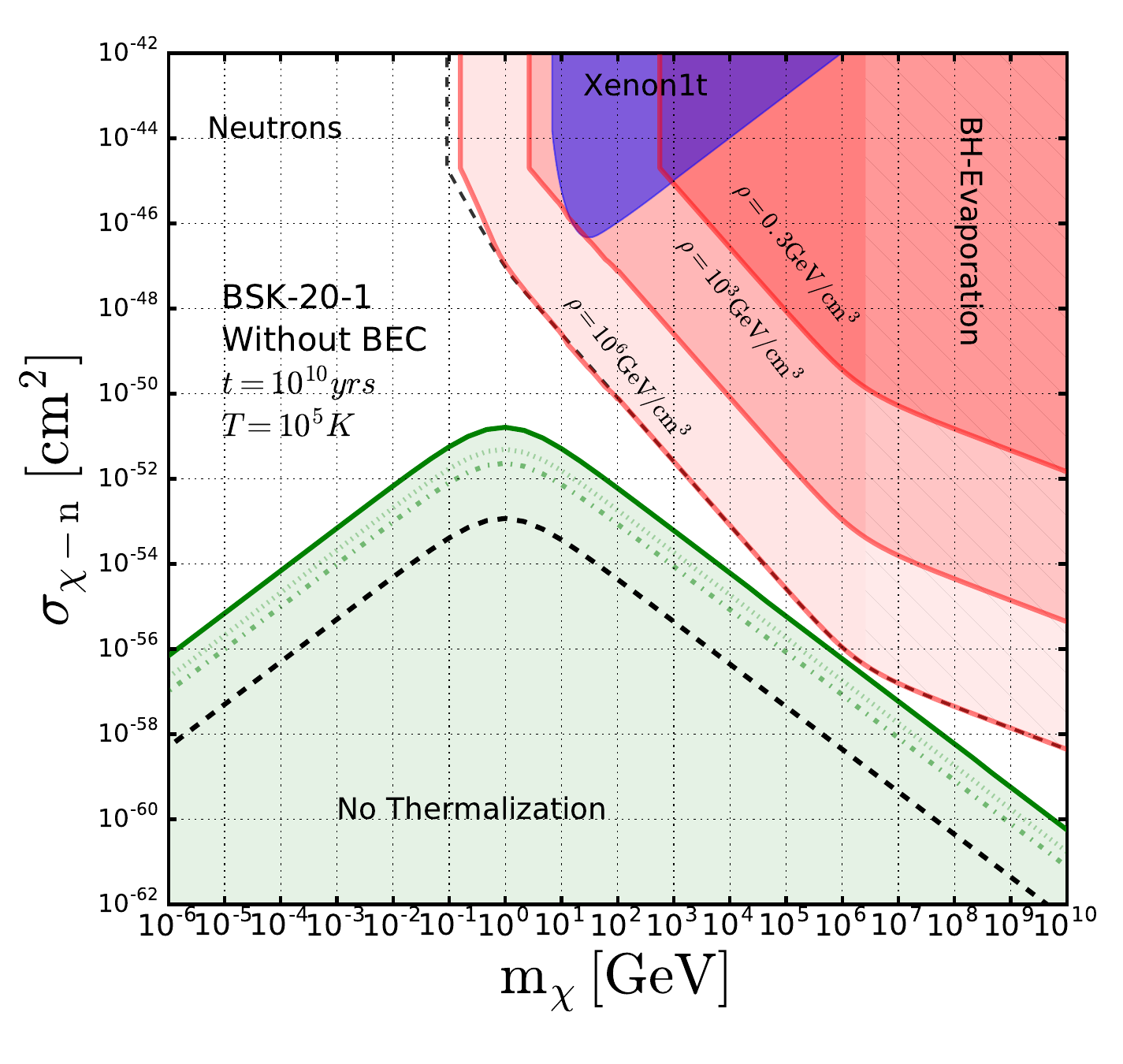}
   \includegraphics[width=0.49 \textwidth]{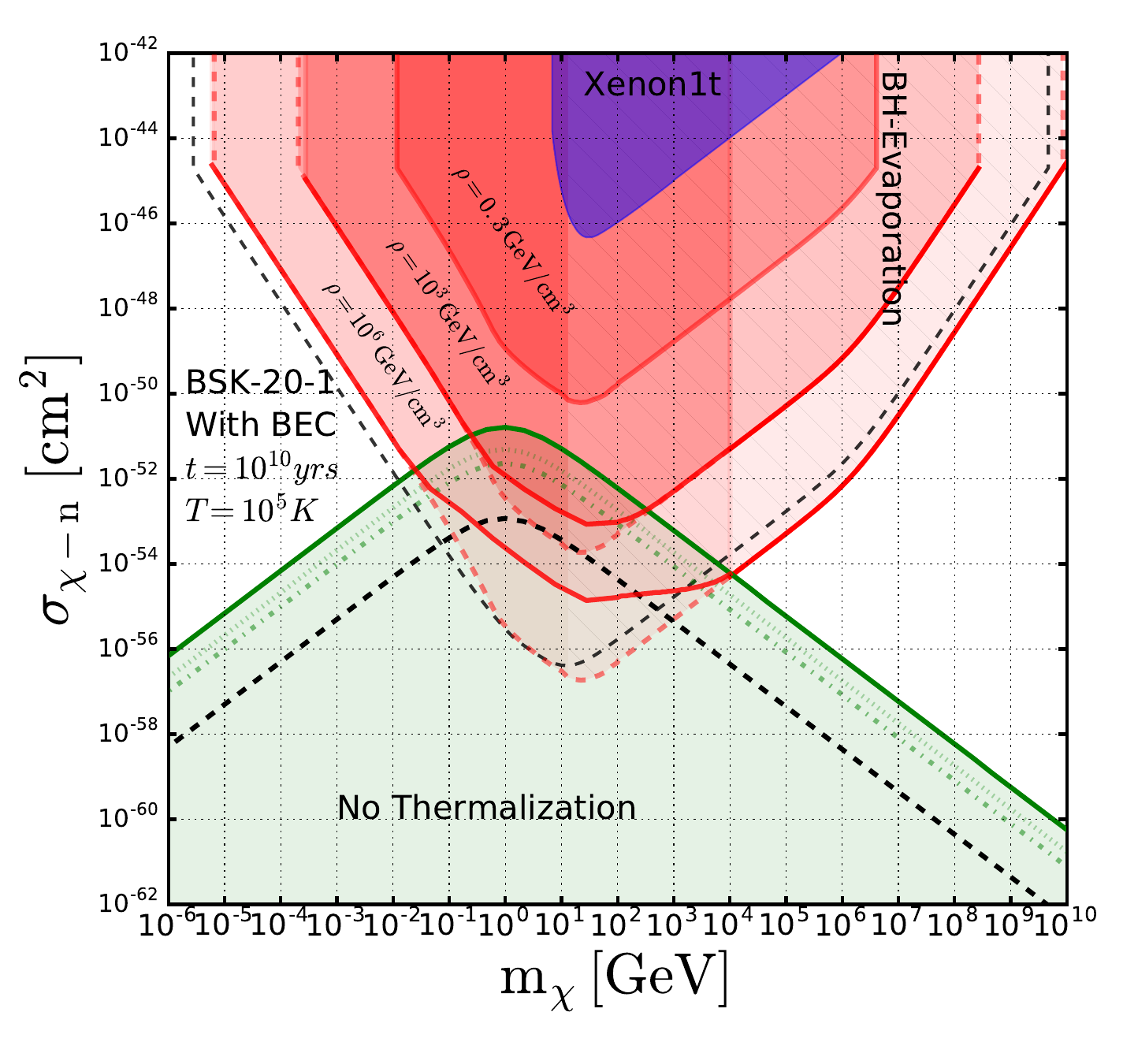}
}
\caption{Neutron star constraints on bosonic DM scattering with neutrons, from Ref.\ \cite{Garani:2018kkd}.  The limits depend upon whether the formation of a DM Bose-Einstein condensate is taken into account (right) or not (left), and also the ambient DM density surrounding the neutron star (different red contours).  Current direct detection limits from XENON1T (blue)
are shown for comparison.}
\label{fig:nscollapse}
\end{center} 
\end{figure}

Observations of neutron stars can provide even stronger bounds for the specific case of DM with a very low self-annihilation cross section. This class includes ADM as the key example. For such candidates, the lack of self-annihilation leads to a strong suppression of other signals of indirect detection. However, this feature also implies that DM captured by a star through scattering with nucleons or leptons leads to an accumulation of DM in the star.  If sufficient DM accumulates, it can alter the evolution of the star, even leading to its collapse~\cite{PhysRevD.40.3221,Bertone:2007ae,McDermott:2011jp,Bramante:2013hn,Bell:2013xk,Bramante:2015dfa,Garani:2018kkd}.  The most dramatic effects are reserved to bosonic DM with a very small quartic self-coupling~\cite{Bramante:2013hn,Bell:2013xk}, since the degeneracy pressure of fermionic DM resists collapse of the star; see Ref.~\cite{Zurek:2013wia} for a review.   In the case of fermionic DM, only very heavy DM above the PeV scale~\cite{Bramante:2017ulk,Gresham:2018rqo} can induce collapse.

Recent limits on bosonic DM with suppressed self-annihilation based on the observation of old neutron stars are shown in Fig.~\ref{fig:nscollapse} for DM scattering with neutrons.  Similar limits are also found for scattering with protons and even muons.  Moreover, although the corresponding bounds on fermionic DM are weaker, they can surpass those from direction detection for superheavy DM~\cite{Bramante:2017ulk,Garani:2018kkd}. Additional astrophysical methods have been proposed to search for ADM imploding neutron stars, based on the location of neutron star mergers in galaxies~\cite{Bramante:2017ulk}. In addition, heavy ADM can also be constrained using non-observation of white dwarf explosions \cite{Bramante:2015cua,Acevedo:2019gre}. For asymmetric bosonic DM, there are also constraints on the self-interaction cross section, which can be stronger than Bullet Cluster bounds, since self-interactions in conjunction with scattering on baryons increases the capture efficiency~\cite{Guver:2012ba}.

\subsubsection{Astrophysical Uncertainties in Indirect Detection}

As with direct detection, astrophysical uncertainties can limit what we may infer on the particle nature of DM from indirect detection observables. For example, consider  constraints that can be placed on the annihilation of DM within the Milky Way halo into neutrino pairs, using data from neutrino telescopes (a subset of the data presented in Fig.~\ref{fig:nunubarlimits}). The expected flux is
\begin{equation}
    \frac{d\Phi_{\nu+ \bar{\nu}}}{d\Omega dE_\nu} = \frac{1}{4 \pi}
    \frac{\sv}{\kappa m_\chi^2}  \frac{1}{3}  \frac{dN_\nu}{dE_\nu} J(\Omega),
    \label{eq:galaxyAnnRate}
\end{equation}
where the $J$-factor parametrizes the projection of the astrophysical DM density distribution $\rho_\chi(\vec r)$:
\begin{equation}
 J(\Omega) \equiv   \int_{\mathrm{l.o.s.}}\!dx\;  \rho_\chi^2(x, \Omega) \ . 
 \label{eq:Jfactordef}
\end{equation}
Because of the $\rho^2_{\chi}$ dependence, small uncertainties in the DM distribution in the galactic centre can translate to very large effects on the expected flux -- and thus prospects for exclusion or detection. While the local DM density is relatively well-constrained, the matter distribution becomes more strongly baryon-dominated for smaller galactocentric distances. Our knowledge of DM density distributions comes from N-body simulations. Recent state-of-the-art hydrodynamical simulations provide estimates for the DM density profile of Milky Way-like galaxies~\cite{Board:2021bwj, Calore:2015oya, Schaller:2015mua}. The part of the DM profile that matters the most for indirect searches is also the most difficult to resolve: nearing the centres of halos, profiles vary strongly within sub-resolution volumes, and the DM behavior within galactic central regions remains an open and active question of debate.  

Recent work compiled and binned rotation curve data from a large number of observations of stars in the Milky Way~\cite{Benito:2019ngh}. By combining the data in a statistically consistent way, the authors were able to produce a likelihood function that can be used to evaluate how well a given set of parameters describing an NFW-like profile describe the kinematic data. These likelihoods can be combined to produce likelihoods on integrated $J$-factors, or can be used directly in analyses, such as the one performed in Ref.~\cite{Arguelles:2019ouk}.

\subsection{Collider Searches}
\label{collider-searches}

Being uncharged and stable, DM created in a collider typically escapes the detectors without leaving a trace. Even so, since the DM carries away energy and momentum, an apparent imbalance in momentum will be seen if visible objects are also created in the collision. The characteristic signature of DM at (hadron) colliders is therefore missing (transverse) momentum, and its associated magnitude called the missing transverse energy~(MET). More generally, new physics associated with DM can lead to an even broader range of collider signals.

Collider searches at the LHC and other machines are highly complementary to direct and indirect probes of DM. Due to the high-energy nature of the collisions, there is no kinematic barrier to creating lighter DM with $m_{\chi}\lesssim 10\,\gev$ that is more challenging to observe in direct detection. Colliders can also make DM with primarily $p$-wave annihilation or that is asymmetric, whose indirect detection signals are thus strongly suppressed. And should a DM candidate be found through direct and indirect detection, even in the form of a canonical, SI scattering, $s$-wave annihilating WIMP, confirmation of its existence and a full characterization of its mass and interactions would benefit greatly from collider measurements.

Just like other DM tests, collider searches can be conducted with varying degrees of model dependence. Ultraviolet~(UV) complete theories like SUSY allow us to study the full phenomenology of the DM candidate and its interactions with SM and BSM particles.  While SUSY has not turned up yet at the LHC, this approach continues to play an important role since these searches often have the greatest sensitivity to specific DM candidates by exploiting all possible signatures in detail. However, given our ignorance of the specific new physics giving rise to DM, it is now also standard practice to organize collider DM searches using \emph{Simplified Models} containing only a few essential ingredients~\cite{Abdallah:2015ter}.  This is a useful way to classify and interpret collider searches, but in some cases the approach can be too simple and miss qualitatively important signatures such as long-lived particles~(LLPs) or other exotic collider signals related to DM.

In this section we present the current state of DM collider searches in the context of simplified models. We also discuss more general signals motivated by DM, such as LLPs and lepton jets, and how they can be searched for at the LHC and beyond.  While most of the focus of this section is on searches connected to the LHC program, we also discuss briefly searches for lighter dark sectors at B-factories and fixed-target experiment.

\subsubsection{Simplified Models of Dark Matter}

Most interactions of WIMP-like DM relevant for direct detection can be parametrized in terms of a low-energy effective field theory~(EFT) that connects the quantum field describing DM directly to nucleons and electrons~\cite{Bai:2010hh,Fox:2011fx,Fan:2010gt,Fitzpatrick:2012ix,Hisano:2015bma}. However, this nearly model-independent EFT approach breaks down for collider searches when the relevant collision energies approach or surpass the masses of the mediator particles that connect DM to the SM, since now the mediators can play an important role in the interaction or be created directly. Simplified models of DM attempt to retain a large degree of the model independence of the EFT approach while also capturing the key dynamical features of more complete UV theories with regards to DM production and LHC signatures.  In turn, simplified models can be mapped to the EFT operator space probed by direct detection experiments.

Simplified models for DM production at the LHC can be divided into two classes: i) those in which the mediator is a singlet under the SM gauge charges as well as the symmetry that stabilizes DM~\cite{Boveia:2016mrp}; ii) those where the mediator is charged under both sets of symmetries~\cite{Aaboud:2019yqu}. Broadly speaking, the former leads to LHC production processes with the mediator in the $s$-channel giving rise to both SM resonances from the mediator and DM pair production signatures, while the latter gives rise to $t$-channel processes which always include two DM particles in the final state. We present here a few simple examples of both categories that demonstrate the most important features of this approach and its complementarity with direct detection searches.

\begin{figure}[ttt!]
\begin{center}
\begin{tabular}{cc}
\includegraphics[width=0.47\textwidth]{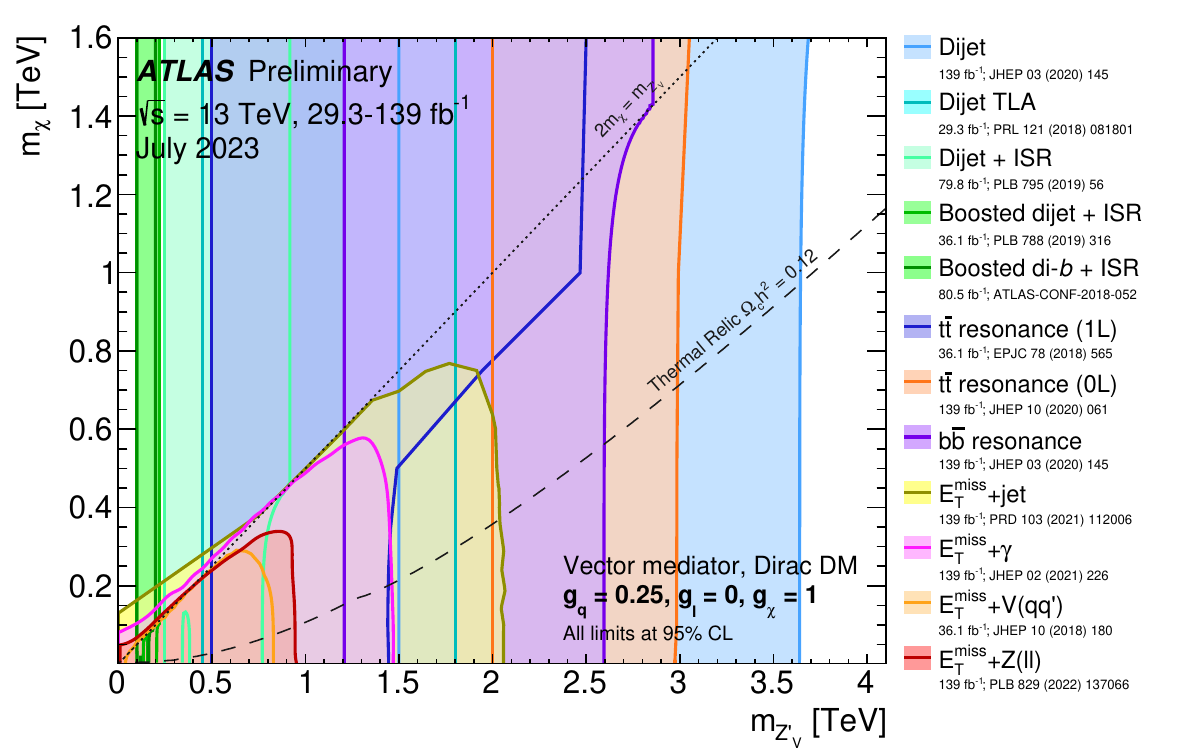}
&
\includegraphics[width=0.47\textwidth]{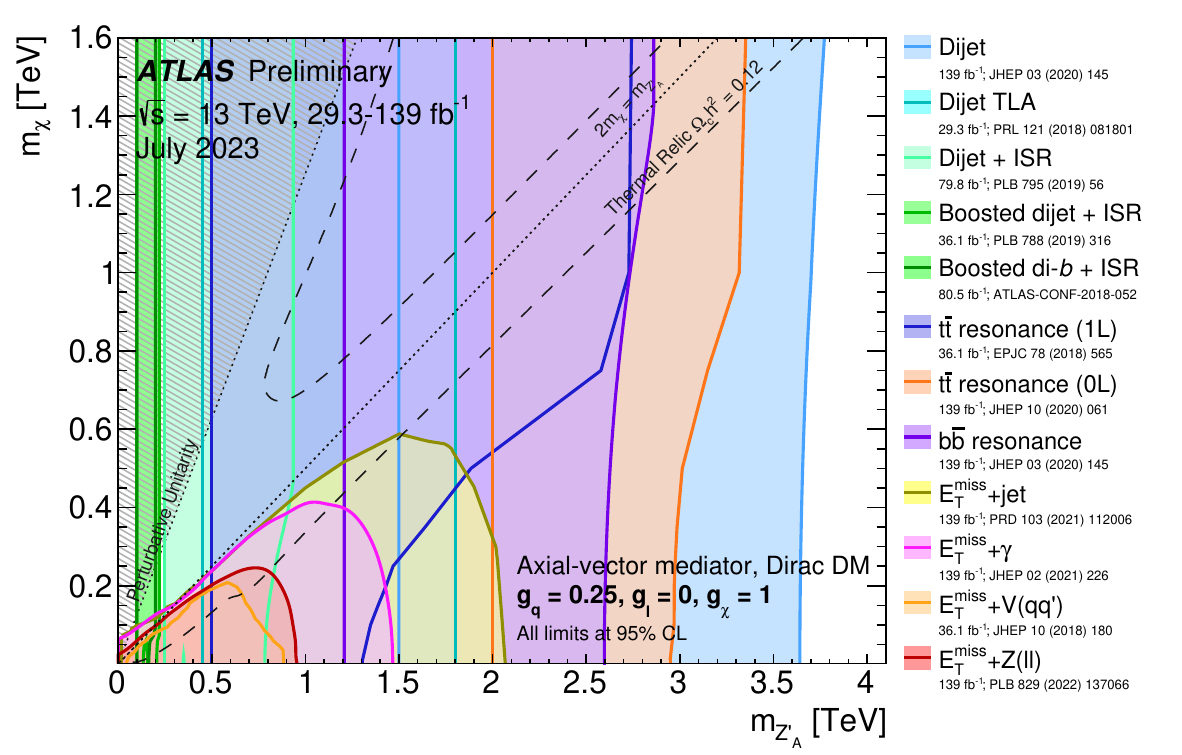}
\\
\includegraphics[width=0.47\textwidth]{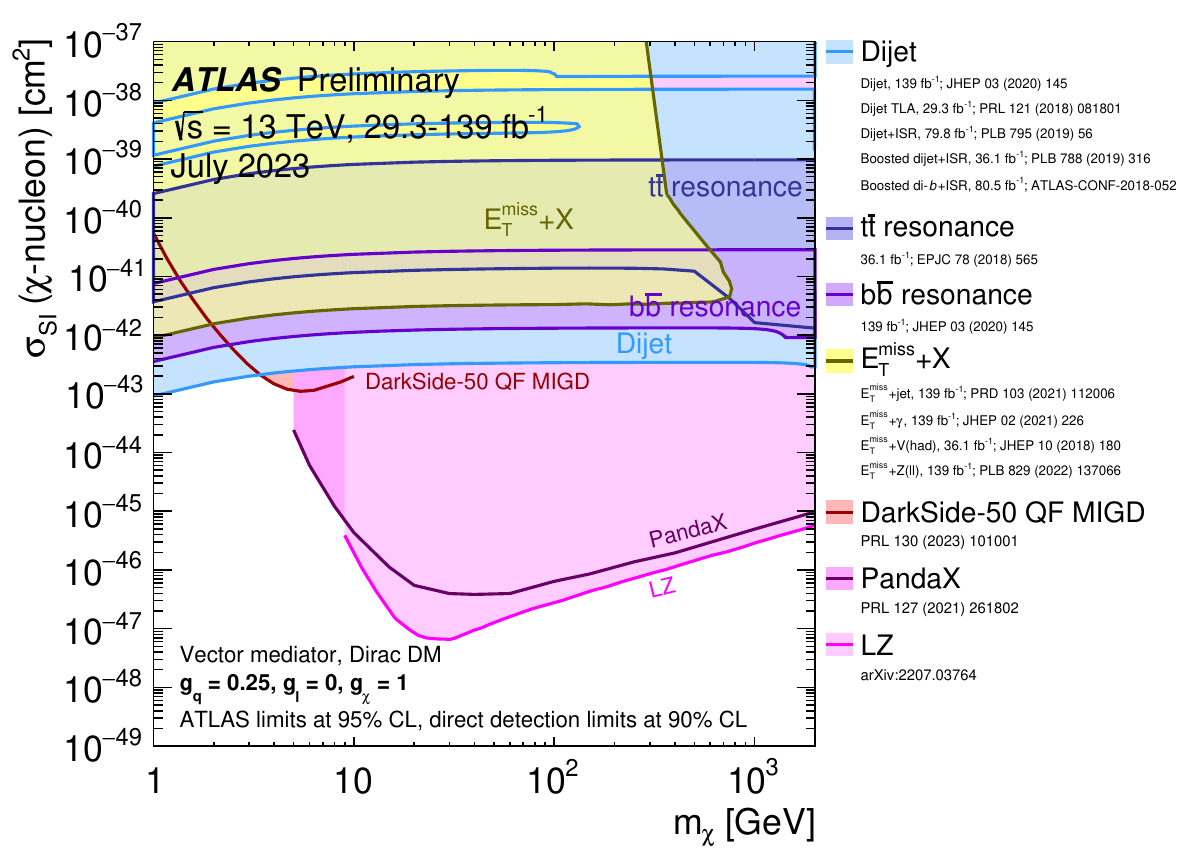}
&
\includegraphics[width=0.47\textwidth]{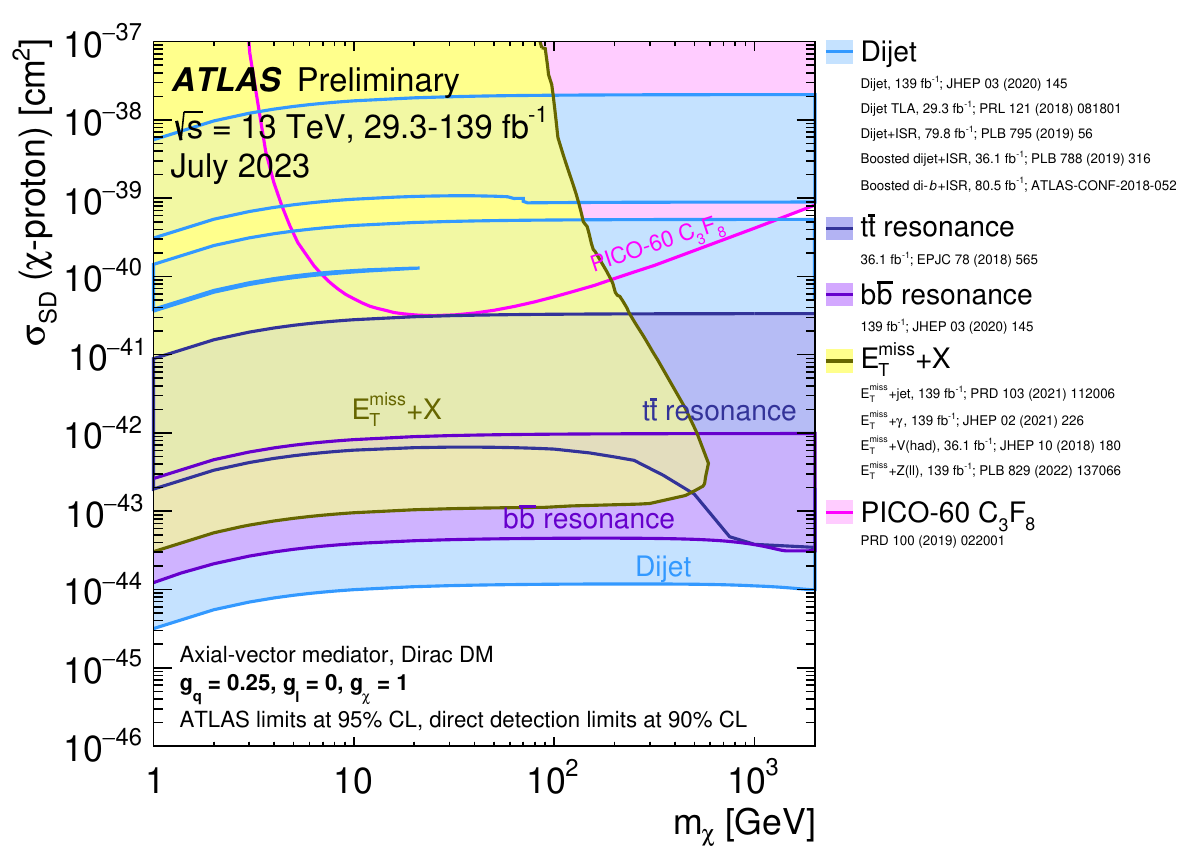}
\end{tabular}
\end{center}
\caption{
ATLAS exclusion bounds on quark- and lepton-universal (axial) vector mediator models, with indicated couplings $g_q, g_\ell$. Left: leptophobic vector mediator. Right: axial vector mediator with equal lepton and quark couplings. Top: collider bounds in the mediator-DM mass plane $(m_{Z'_V}, m_\chi)$.  Bottom: collider bounds compared to limits from direct detection obtained by matching the simplified models to the low-energy EFT of DM-nuclear scattering. 
All figures from Ref.~\cite{ATL-PHYS-PUB-2023-018}. 
}
\label{fig:atlasdm}
\end{figure}

In \emph{$s$-channel simplified models}, the DM particle $\chi$ is assumed to couple to a new mediator state that also connects to the SM. To be concrete, the DM is often assumed to be a (Dirac) fermion. It connects to the SM through a mediator with coupling $g_\chi$ to DM, and $g_q$ and $g_l$ couplings to quarks and leptons. The mediator is typically taken to be a vector or axial vector $Z'$ with (usually) flavour-universal couplings $g_{q,l}$. For scalar or pseudoscalar mediators, more complete theories are typically used.

This portal produces two main types of new signals at colliders: i) DM pair production in association with additional SM radiation such as jets or one or more of $\gamma/W/Z/h$ leading to a characteristic MET signal $X$+MET; ii) SM production via resonant $s$-channel production of the $Z'$ mediator in $q \bar q$ collisions leading to dijet or dilepton mass peaks. A wide variety of SM resonance and MET searches therefore have sensitivity to this simplified model parameter space.  In Fig.~\ref{fig:atlasdm} we show a summary of ATLAS constraints on vector portal simplified models together with a comparison to current exclusions from direct detection~\cite{ATL-PHYS-PUB-2023-018}.  Resonance searches give the strongest bounds here, while the most important $X$+MET bounds typically come from mono-jet searches. The exclusions are relatively independent of the DM mass as long as the $Z' \to \bar \chi \chi$ decay is kinematically allowed.  

The bounds of Fig.~\ref{fig:atlasdm} also demonstrate two specific advantages of collider searches compared to direct detection. First, the sensitivity of collider searches does not decrease with lighter DM mass, while direct detection experiments that rely on nucleon scattering typically lose sensitivity for $m_{\chi} \lesssim 10\,\gev$. Second, direct detection sensitivity to SD scattering is typically much weaker than for the SI. These processes correspond to vector and axial vector mediators, respectively, and LHC searches have similar sensitivities to both. Thus, lighter or SD DM candidates are particularly amenable to collider tests. Let us emphasize, however, that the comparison between collider bounds and direct detection shown in Fig.~\ref{fig:atlasdm} assumes fixed couplings close to unity. For lighter mediators, typically below the weak scale, the same DM-nucleon cross sections can arise for much smaller vector couplings to the SM. This has the effect of greatly weakening or even eliminating the constraints from high-energy colliders. An interesting implication of this point is that the discovery of lighter or SD DM in direct detection would point towards the existence of a dark sector with a new force mediator~\cite{Fox:2011pm}.

Simplified models with scalar or pseudoscalar mediators can also be constructed. In this case, avoiding large flavor-violating effects typically requires that the SM-mediator couplings be proportional to the SM Higgs Yukawa couplings, $g_{q,\ell} y_{q, \ell}$. This arises naturally if the new scalar is in some way connected to the Higgs boson, and thus scalar or pseudoscalar portals are representative of many BSM scenarios with extended scalar sectors such as the NMSSM~\cite{Ellwanger:2009dp} and families of 2HDMs~\cite{Branco:2011iw} with DM candidates. On account of the flavour structure of the scalar portal, the strongest collider bounds are typically found in the  $t\bar{t}$+MET channel~\cite{Aaboud:2019yqu}. The connection of the scalar portal to the Higgs boson also points towards the exciting possibility of exotic Higgs decays involving DM or its mediators~\cite{deFlorian:2016spz}. 

Turning next to \emph{$t$-channel} models, the mediator is assumed to carry the same charge that stabilizes the DM.  As the name suggests, this implies that the mediator can only contribute to collider processes in the $t$ channel. These scenarios are motivated by SUSY, where the supersymmetric version of the fermion-higgs or fermion-gauge interactions give rise to couplings between a neutral DM candidate, a SM fermion, and a sfermion carrying the quantum number that stabilizes DM as well as gauge charge.  Unlike $s$-channel mediators, these scenarios only give rise to missing energy signatures at colliders, either via associated DM-mediator production from a gluon-quark initial state, or mediator pair production and subsequent decays down to DM pairs.  As with $s$-channel models, bounds from collider searches are highly complementary to direct DM searches.

\subsubsection{Dark Matter and Long-Lived Particles}

While the DM simplified model framework is a useful way to organize and interpret collider searches at the LHC, it is important to keep in mind that the underlying theory of DM could be much more complicated and generate even more exotic collider signals. 
Such signals arise in many UV-complete models of DM, and in some cases they are directly connected to the mechanism that creates the DM relic density in the early universe.  A specific example in many theories of DM are LLPs.

Long-lived particles refer to new species with characteristic decay lengths that are large relative to the dimensions of a (LHC) collider detector. The production of neutral or charged LLPs is motivated both from bottom-up and top-down perspectives, with many UV-complete theories that address the electroweak hierarchy problem, baryogenesis, neutrino masses, and DM predicting their existence at the LHC~\cite{Curtin:2018mvb}. Searches for LLPs are a rapidly growing priority in the ATLAS, CMS and LHCb research programs~\cite{Alimena:2019zri}. They have also motivated dedicated LHC far detectors like MATHUSLA~\cite{Chou:2016lxi,Alpigiani:2018fgd,Curtin:2018mvb,MATHUSLA:2020uve,Curtin:2023skh}, FASER~\cite{Ariga:2018zuc, Ariga:2018uku,FASER:2022hcn}, and MoEDAL-MAPP~\cite{Pinfold:2009oia,Acharya:2014nyr,MoEDAL-MAPP:2022kyr}, which greatly extend the LLP sensitivity of the LHC near detectors. The LLP search program at the LHC is still in its early stages but maturing rapidly, with promising discovery potential across large regions of motivated signature space.

If a DM candidate is discovered at the LHC in the form of excess missing energy, a crucial question is whether it really is the source of DM or just a new long-lived neutral particle. Dedicated LLP detectors could play a crucial role in distinguishing these possibilities~\cite{Curtin:2018mvb}, and possibly test LLP lifetimes up to $\tau \sim 0.1\,\mathrm{s}$. While such a lifetime is very short compared to the age of the universe, it is an important benchmark since (visible) decays slower than this are very strongly constrained by primordial nucleosynthesis~\cite{Kawasaki:2004qu,Jedamzik:2006xz}.

Beyond testing the stability of a potential DM candidate, LLP searches can also elucidate a broad range of DM models in which the DM particle is connected to the SM through a long-lived mediator. In this class of models, sometimes referred to as hidden valleys~\cite{Strassler:2006im,Han:2007ae}, the main production mode of DM at the LHC can be through the decay of the mediator to DM + SM~\cite{Canetti:2012vf,Canetti:2012kh,Co:2015pka,DAgnolo:2017dbv}. In other scenarios, DM is produced in conjunction with a heavier, long-lived partner~\cite{TuckerSmith:2001hy,Izaguirre:2015zva}, and the decay of the accompanying LLP can serve as a tag to identify the DM itself.

\subsubsection{Dark Sectors at Colliders}

As discussed in Sec.~\ref{sec:dsector}, DM could be part of a larger dark sector containing several new particles and force mediators. Such dark sectors can lead to a wide range of exotic collider signals at the LHC as well as at lower energy machines. 

Lighter dark sectors below the weak scale are typically probed most efficiently in lower energy collisions of enormous intensity~\cite{Pospelov:2008zw,Bjorken:2009mm}. These features are illustrated for a dark photon decaying mainly to the SM in the left panel of Fig.~\ref{fig:dpbounds} and primarily to DM in the right panel. In both cases, the strongest bounds come from $B$-factories down to about $m_{A'} \sim 300\,\mev$~\cite{Batell:2009yf,Essig:2009nc}, and from fixed-target experiments at even lower masses~\cite{Essig:2009nc}.  Expanded collider searches for dark sectors of various types are planned or underway. These include Belle~II~\cite{Essig:2013vha,Dolan:2017osp,Kou:2018nap,Duerr:2019dmv} as well as fixed-target experiments such as NA62~\cite{Dobrich:2018ezn}, SHiP~\cite{Alekhin:2015byh}, and LDMX~\cite{Akesson:2018vlm}. 

When the DM mass is near the weak scale or above, or connects to the SM primarily through heavier connectors, the full energy reach of the LHC can become important for testing a related dark sector. This is the premise behind the hidden valley paradigm~\cite{Strassler:2006im,Han:2007ae}, in which massive connectors between the dark and visible sectors are created in LHC collisions and decay to dark sector states. Exotic collider signals can arise if some of these dark states decay back to the SM. A specific example is SUSY connected to a dark $U(1)'$ gauge bosons and its associated superpartners~\cite{Strassler:2006qa,ArkaniHamed:2008qp}. If the dark sector has a characteristic mass below the weak scale, supersymmetric cascades down to the DM can be accompanied by the emission of strongly boosted dark vector bosons. If these decay back to visible particles, the dark vectors will give rise to so-called lepton jets consisting of multiple highly-collimated leptons (and possibly additional hadronic activity)~\cite{ArkaniHamed:2008qp,Baumgart:2009tn,Cheung:2009su}.  These can be prompt or displaced, and have been searched for by ATLAS~\cite{Aad:2015sms} and CMS~\cite{Sirunyan:2018mgs}. More complicated dark sectors can give rise to even more exotic collider signatures such as disappearing jets~\cite{Schwaller:2015gea}, disappearing charged tracks~\cite{Mahbubani:2017gjh}, and more.

\section{Summary and Conclusions \label{sec:conclusions}}

Evidence for dark matter~(DM) from astrophysics and cosmology is overwhelming. Our entire understanding of how visible matter in the universe condensed to form the pattern of galaxies we observe relies on the existence of DM. Unfortunately, despite a broad range of theoretical proposals and experimental searches, the underlying nature of DM remains a mystery. 

In this review we have presented the leading theoretical proposals for DM and we have summarized the most promising ways to test them. Potential DM candidates scan an enormous range of masses, from $\sim 10^{-19}\,\text{eV}$ to many times heavier than the Sun, with connections to visible matter through a range of interactions, from the strong to the weak to gravity to new dark forces. Experimental searches for DM are just as diverse, and include direct laboratory searches, indirect searches in astrophysics observations, and attempts to create and measure DM in high-energy collisions.

While we have not yet identified DM beyond its universal gravitational influence on ordinary matter, the efforts to do so have provided an enormous amount of information about what DM is not. In turn, this guides present and future search techniques. Enormous progress is being made in these efforts, both on the theoretical and experimental sides, with key contributions coming from the Canadian subatomic physics community. The coming years promise to be an exciting period with a strong likelihood of new fundamental discoveries.

\begin{acknowledgments}
The authors gratefully acknowledge funding from the Natural Sciences and Engineering Council of Canada~(NSERC), and the Canada First Research Excellence Fund through the Arthur B. McDonald Canadian Astroparticle Physics Research Institute. TRIUMF receives federal funding via a contribution agreement with the National Research Council~(NRC) of Canada.
\\${}$\\
Competing interests: The authors declare there are no competing interests.
\\
No data is associated with this work.
\end{acknowledgments}


\bibliography{CJP2_DMTheory}

\end{document}